\documentclass{aa}
\pdfoutput=1
\usepackage{graphicx}
\usepackage{amsmath}

\usepackage{units}
\usepackage{natbib}
\usepackage{upgreek}
\usepackage{mathtools}

\usepackage[usenames,dvipsnames]{color}
\usepackage{ulem}

\usepackage{hyperref}
\usepackage{datatool}
\usepackage{booktabs}
\usepackage{siunitx}
\usepackage{pgfplotstable}
\usepackage{longtable}

\usepackage{txfonts}

\hypersetup{colorlinks = true,
	    citecolor = blue,
	    linkcolor = blue}

\newcommand\Tstrut{\rule{0pt}{2.6ex}}         
\newcommand\Bstrut{\rule[-0.9ex]{0pt}{0pt}}   

\newcommand{\micron}{$\upmu$m}

\newcommand{\arcsecond}{$^{\prime\prime}$}
\newcommand{\sqarcsecond}{$^{\prime\prime}$ $\times$ $^{\prime\prime}$}

\newcommand{\msun}{M$_\odot$}
\newcommand{\lsun}{L$_\odot$}
\newcommand{\hii}{\textsc{Hii} }
\newcommand{\mal}{$\times$ }

\newcommand{\hms}[3]{#1$^\text{h}$#2$^\text{m}$#3$^\text{s}$}
\newcommand{\ra}{$\upalpha_\text{J2000}$}
\newcommand{\dec}{$\updelta_\text{J2000}$}

\begin{document}
%
\title{The physical and chemical structure of Sagittarius B2}
\subtitle{I. Three-dimensional thermal dust and free-free continuum modeling on \unit[100]{au} to \unit[45]{pc} scales}

\titlerunning{3d structure of Sgr B2}

\author{A.~Schmiedeke\inst{1}
    \and P.~Schilke\inst{1}    
    \and Th.~M\"oller\inst{1}
    \and \'A.~S\'anchez-Monge\inst{1}
    \and E.~Bergin\inst{2}
    \and C.~Comito\inst{1}
    \and T.~Csengeri\inst{3}
        \and D.C.~Lis\inst{4, 5}
    \and S.~Molinari\inst{6}
    \and S.-L.~Qin\inst{1, 7}
    \and R. Rolffs\inst{1}
}

\institute{
	I. Physikalisches Institut, Universit\"at zu K\"oln, Z\"ulpicher Stra\ss e 77, D-50937 K\"oln, Germany\\email: schmiedeke@ph1.uni-koeln.de
	    \and
    Department of Astronomy, The University of Michigan, 500 Church Street, Ann Arbor, MI 48109-1042, USA
        \and
    Max-Planck-Institut f\"ur Radioastronomie, Auf dem H\"ugel 69, D-53121, Bonn, Germany
    \and
    LERMA, Observatoire de Paris, PSL Research University, CNRS, Sorbonne Universit\'es, UPMC Univ. Paris 06, F-75014, Paris, France  
    \and
     California Institute of Technology, Pasadena, CA 91125, USA
    \and
    INAF - Istituto di Astrofisica e Planetologia Spaziali, via Fosso del Cavaliere 100, I-00133, Roma, Italy
    \and
    Department of Astronomy, Yunnan University, and Key Laboratory of Astroparticle Physics of Yunnan Province, Kunming, 650091, China
}

\date{Received ; Accepted}

  \abstract
   {
   We model the dust and free-free  continuum emission in the high-mass star-forming region Sagittarius B2.
   }
   {We want to reconstruct the three-dimensional density and dust temperature distribution, as a crucial input to follow-up studies of the gas velocity field and molecular abundances. 
   }
   {We employ the three-dimensional radiative transfer program RADMC-3D to calculate the dust temperature self-consistently, provided a given initial density distribution. This density distribution of the entire cloud complex is then recursively reconstructed based on available continuum maps, including both single-dish and high-resolution interferometric maps covering a wide frequency range ($\upnu$ = \unit[40]{GHz} - \unit[4]{THz}). The model covers spatial scales from \unit[45]{pc} down to \unit[100]{au}, i.e. a spatial dynamic range of  10$^5$.}
   {We find that the density distribution of Sagittarius B2 can be reasonably well fitted by applying a superposition of spherical cores with Plummer-like density profiles. In order to reproduce the spectral energy distribution, we position Sgr B2(N) along the line of sight behind the plane containing Sgr B2(M). 
   We find that the entire cloud complex comprises a total gas mass of \unit[8.0 $\times$ 10$^6$]{\msun} within a diameter of \unit[45]{pc}. This corresponds to an averaged gas density of \unit[170]{\msun pc$^{-3}$}. We estimate stellar masses of \unit[2400]{\msun} and \unit[20700]{\msun} and luminosities of \unit[1.8~$\times$~10$^{6}$]{\lsun} and \unit[1.2~$\times$~10$^{7}$]{\lsun} for Sgr B2(N) and Sgr B2(M), respectively. We report H$_2$ column densities of \unit[2.9~$\times$~10$^{24}$]{cm$^{-2}$} for Sgr B2(N) and \unit[2.5 $\times$ 10$^{24}$]{cm$^{-2}$} for Sgr B2(M) in a \unit[40]{\arcsec} beam. For Sgr B2(S), we derive a stellar mass of \unit[1100]{\msun}, a luminosity of \unit[6.6~$\times$~10$^{5}$]{\lsun} and a H$_2$ column density of \unit[2.2~$\times$~10$^{24}$]{cm$^{-2}$} in a \unit[40]{\arcsec} beam. We calculate a star formation efficiency of \unit[5]{\%} for Sgr B2(N) and \unit[50]{\%} for Sgr B2(M). This indicates that most of the gas content in Sgr B2(M) has already been converted to stars or dispersed.}
  {}


   \maketitle

%

\section{Introduction}\label{sec:intro}
Apart from in-situ measurements in the Solar System, all information gained in astrophysics is deduced from the analysis and interpretation of radiation received with ground or space-based telescopes. Gas and dust, in between the source and the telescope, influences the radiation. So analyzing the radiation received from an astrophysical object not only provides information about the source, but also about the medium in between the object and the observer.  Radiative transfer is thus one of the most fundamental phenomena in astrophysics.

As summarized by \citet{Steinacker2013}, three-di\-men\-sional dust radiative transfer calculations are essential to make progress in many fields of astronomy. Dust grains modify the radiation field in many objects such as protoplanetary disks, evolved stars, reflection nebulae, supernova remnants, molecular clouds, the interstellar medium, galaxies, galactic nuclei, and the high-redshift universe.

\begin{figure}[t]
    \centering
    \includegraphics[width=0.45\textwidth]{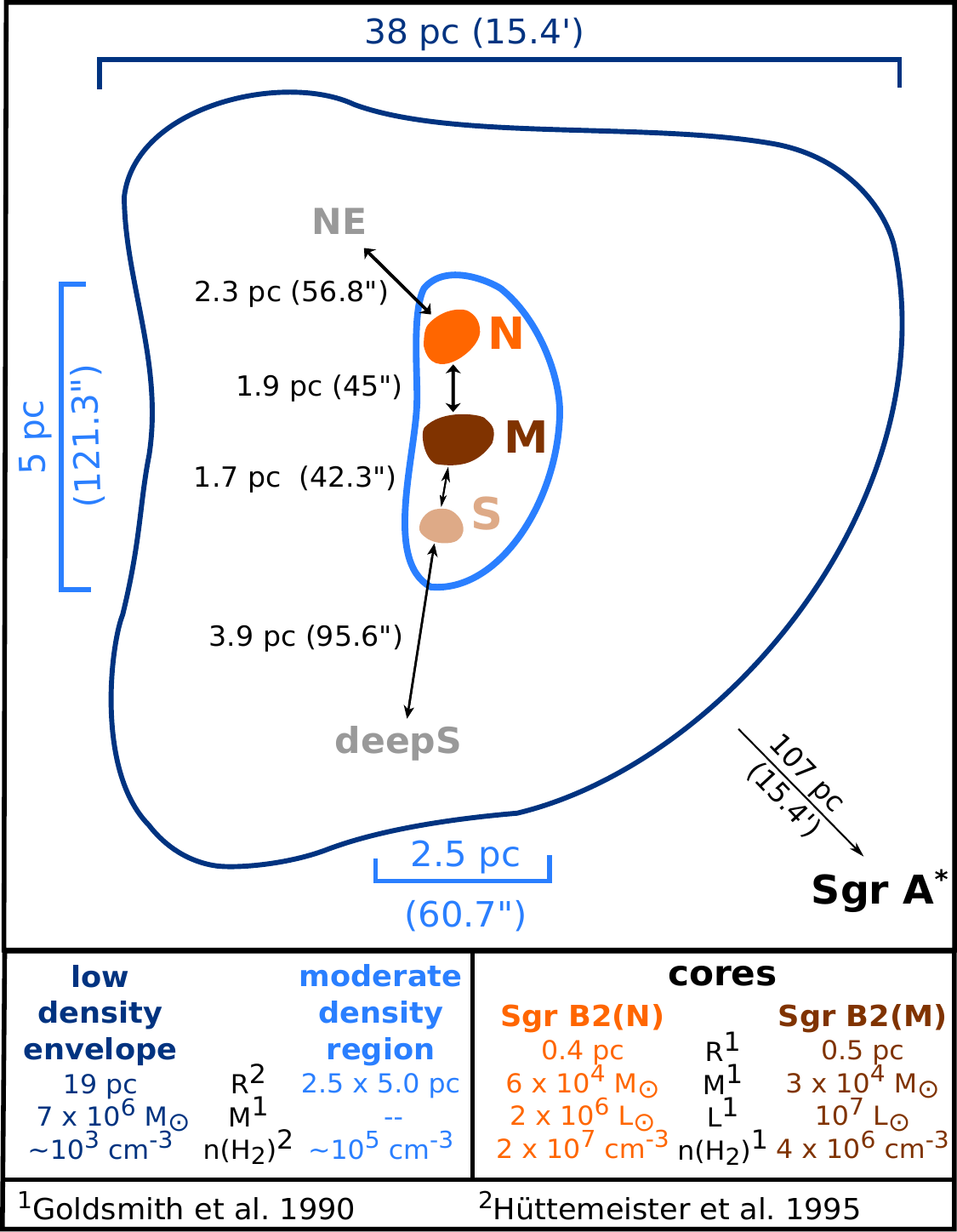}
    \caption{Sketch of the Sgr B2 region, adapted from \citet{Huettemeister1995}.}
    \label{fig:sketch}
\end{figure}

In this paper, we focus on the modeling of the dust and free-free continuum emission of the high-mass star forming molecular cloud Sagittarius B2 (hereafter Sgr B2) by applying detailed three-dimensional radiative transfer modelling.

Sgr B2 is one of the most massive molecular clouds in the Galaxy. It is located at a distance of \unit[8.34$\pm$0.16]{kpc} \citep{Reid2014}\footnote{In this paper, we assume a distance to Sgr B2 of \unit[8.5]{kpc}.} and has a projected distance of \unit[107]{pc} (or \unit[43.4]{\arcmin}) from Sgr A$^\ast$, the compact radio source associated with the supermassive black hole located at the Galactic Center. 
\citet{Huettemeister1993a} distinguish three different parts in Sgr B2: (i) a low density envelope, (ii) a moderate density region extended around (iii) local hotspots, which are the most compact, densest molecular regions (see Fig. \ref{fig:sketch}). The envelope measures \unit[38]{pc} (or \unit[15.4]{\arcmin}) in diameter \citep[][corrected for distance]{Scoville1975} and has a gas mass of \unit[7 $\times$ 10$^6$]{\msun} \citep{Goldsmith1990}. The average H$_2$ density n(H$_2$) $\sim$ \unit[10$^3$]{cm$^{-3}$},  and H$_2$ column density N(H$_2$) $\sim$ \unit[10$^{23}$]{cm$^{-2}$}, are relatively low compared to the central part of the region. The moderate density region extends over \unit[2.5]{pc} $\times$ \unit[5.0]{pc} around the local hotspots. Its density and H$_2$ column density are higher, n(H$_2$) $\sim$ \unit[10$^5$]{cm$^{-3}$} and N(H$_2$)~$\sim$~\unit[10$^{24}$]{cm$^{-2}$}. Embedded in this intermediate region are the local hotspots. At least three of them are sites of active star formation \citep{Gordon1993}. These three  sources are historically named according to their relative location in an equatorial coordinate system: Sgr B2(N)(orth), Sgr B2(M)(ain), and Sgr~B2(S)(outh).  They are positioned along a north-south line. In projection, Sgr~B2(M) is located \unit[1.9]{pc} (or \unit[45]{\arcsec}) south of Sgr~B2(N) and Sgr~B2(S) is located \unit[1.7]{pc} (or \unit[42.3]{\arcsec}) south of Sgr~B2(M), see Fig. \ref{fig:sketch}. These cores have sizes of \unit[$\sim$ 0.5]{pc}, H$_2$ densities of \unit[$\sim$ 10$^7$]{cm$^{-3}$}, and column densities of \unit[10$^{25}$]{cm$^{-2}$} \citep{Huettemeister1995}.  Different works, e.g. \cite{Goldsmith1990, Etxaluze2013},  have derived the gas mass of the two cores Sgr B2(N) and SgrB2 (M) in the ranges of \unit[6 -- 25 $\times$ 10$^4$]{\msun} and \unit[3 -- 23 $\times$ 10$^4$]{\msun}, respectively. However, they used different radii (\unit[0.4 -- 1.6]{pc}), which makes a comparison of the results impossible.

Sgr~B2(N) and Sgr~B2(M) are sites of active massive star formation and comprise a plethora of (ultra-compact) \hii regions, X-ray sources associated with \hii regions, X-ray sources with no radio or IR counterparts (Takagi, Murakami \& Koyama 2002), dense cores, embedded protostars, and molecular masers \citep{Goicoechea2004}. More than 70 \hii regions have been detected in the whole Sgr B2 cloud complex \citep[see Fig. \ref{fig:largeScaleMap} ; ][]{Mehringer1993, Gaume1995, dePree1998}.

In the observed dust continuum maps, we see an extension of the cloud complex to the north-east. Following the historical naming scheme, we name this region Sgr B2(NE) throughout this paper. This component is located in projection \unit[2.3]{pc} (or \unit[56.8]{\arcsec}) north-east of Sgr B2(N). Another extension is visible towards the south of the cloud. We will name this extension Sgr B2(deepS)outh throughout this paper. This component is located in projection \unit[3.9]{pc} (or \unit[95.6]{\arcsec}) south of Sgr B2(S), see Fig. \ref{fig:sketch} .

The modeling of the continuum emission of the Sgr B2 complex presented here provides us with the three-dimensional model of the structure (density distribution) of this region. For this multiwavelength, multiscale data is crucial to properly constrain the structure. In the next step this will enable us to model the line shapes (kinematics) and thus constrain molecular gas properties such as the gas velocity field,  molecular abundances, etc.

The paper is organized as follows: In Sect.\ \ref{s-observations} we introduce the observational dataset used throughout the paper. The modeling approach is presented in detail in Sect.\ \ref{s-modeling} . This is followed by the application of the modeling approach to Sgr B2 and the discussion of the results in Sect.\ \ref{s-analysis} . Finally the paper is concluded in  Sect.\ \ref{s-conclusions}.  In Sect.\ \ref{s-figures}, we present additional figures and in Sect.\ \ref{s-tables} we tabulate the setup parameters of all models. In Sect.\ \ref{app:derivationHii} we derive physical properties of \hii regions.


\section{Observations and data reduction}
\label{s-observations}

Multiwavelength, multiscale data is crucial to properly constrain the structure of Sgr B2. Towards the hot cores Sgr B2(N) and Sgr B2(M), the Herschel/HIFI spectral surveys provide the continuum information from the sub-mm up to the far-infrared regime. High-resolution interferometric maps towards both hot cores obtained with the Submillimeter Array (SMA) and the Very Large Array (VLA) provide the necessary spatial resolution on small scales. To cover the large-scale structure, we use dust continuum maps obtained within the surveys ATLASGAL and HiGAL, described in detail below. A summary of the data is presented in Table \ref{tab:obsSummary}. Fig.\ \ref{fig:freqScales} provides an overview of the employed datasets and the spatial scales they cover.

\begin{table*}
  \caption{Summary of observational data.}
  \label{tab:obsSummary} 
  \centering
  \begin{tabular}{|l || r | r | r | r | r | r |}
  \hline
		&          &             & 		 & center coordinates	&  		 & \Tstrut \\
  Telescope     & $\upnu$  & $\uplambda$ & resolution    & RA, DEC (J2000)	& map size	 & incl. sources \\
                & [GHz]    & [$\upmu$m]  & [\arcsecond]  & [17:47:s, -28:m:s] 	& [\sqarcsecond] & \Bstrut \\
  \hline
  
  \textbf{large-scale} & & & & & & \Tstrut \Bstrut \\
  
    VLA 			& 23.1 & 13000  & 0.27 \mal 0.23 \tablefootmark{1} & 20.166, 23:04.76  & 143 \mal 143\hspace{.2cm}  & Sgr B2 \\
    APEX \tablefootmark{2} 	& 345  & 870	& 19.2 \hspace{.2cm} & 19.943, 23:01.62 & 1100 \mal 1100 \tablefootmark{3} & Sgr B2 \\    
    Herschel \tablefootmark{4} 	& 600  & 500    & 42.5 \hspace{.2cm} & 19.639, 22:57.77 & 1100 \mal 1100 \tablefootmark{3} & Sgr B2 \\
				& 857  & 350	& 30.3 \hspace{.2cm} & 19.791, 22:59.68 & 1100 \mal 1100 \tablefootmark{3} & Sgr B2 \\    
				& 1200 & 250	& 23.4 \hspace{.2cm} & 19.939, 23:01.64 & 1100 \mal 1100 \tablefootmark{3} & Sgr B2 \\    
    Herschel \tablefootmark{5} 	& 4283 & 70	& 10.4 \hspace{.2cm} & 20.046, 23:02.89 & 1100 \mal 1100 \tablefootmark{3} & Sgr B2 \\
    
   \textbf{small-scale}    & & & & & & \Tstrut \Bstrut \\
   
    VLA  & 40.8	&  7000	& 0.15 \mal 0.10 \tablefootmark{6} 	& 19.902, 22:17.8 	& 24 \mal 24	& Sgr B2(N) \\ 
	 & 	& 	& 0.15 \mal 0.10 \tablefootmark{6} 	& 20.202, 23:05.3 	& 		& Sgr B2(M) \\
	 & 40.8	&  7000 & 0.049 \mal 0.079 \tablefootmark{7}	& 20.115 23:04.0 	& 10 \mal 10 	& Sgr B2(M) \\
    SMA	 & 342	& 874	& 0.37 \mal 0.22 \tablefootmark{8} 	& 19.883, 23:18.4 	& 16 \mal 16	& Sgr B2(N) \\
	 & 	& 	& 0.37 \mal 0.22 \tablefootmark{8} 	& 20.158, 23:05.0 	& 		& Sgr B2(M) \\
	 
   \textbf{single}& & & & & & \Tstrut \\
   \textbf{pointing}& & & & & & \Bstrut \\ 

   Herschel \tablefootmark{9}	& 480 -- 1250	& 625 -- 240	& 44.9 -- 17.2	& 19.88, 22:18.4 	& ---	& Sgr B2(N) \\
				&		& 		& 		& 20.35, 23:03.0 	& ---	& Sgr B2(M) \\
				& 1410 -- 1910	& 213 -- 157	& 15.3 -- 11.3	& 19.88, 22:18.4 	& ---	& Sgr B2(N) \\
			        & 		& 		&		& 20.35, 23:03.0 	& ---	& Sgr B2(M) \\
  \hline
 \end{tabular}
 \tablefoot{\tablefoottext{1}{The VLA was in the DnCnBnA hybrid array configuration.}
	    \tablefoottext{2}{LABOCA instrument. This map has been combined with the Planck map.}
	    \tablefoottext{3}{The coverage of these maps is beyond the extent of Sgr B2. We have thus extracted cutouts.}
            \tablefoottext{4}{SPIRE instrument.}
            \tablefoottext{5}{PACS instrument.}
            \tablefoottext{6}{The VLA was in the BnA hybrid array configuration.}
            \tablefoottext{7}{The VLA was in the A array configuration.}	   
	    \tablefoottext{8}{The SMA was in the compact and in the very extended array configuration. Both data sets have been combined.}
	    \tablefoottext{9}{HIFI instrument.}}
 \end{table*}

 \subsection{Herschel / HIFI}

The Herschel / HIFI guaranteed time key project HEXOS \citep[Herschel / HIFI observations of EXtraOrdinary Sources;][]{Bergin2010} includes full line surveys of Sgr B2(N)  towards \ra~=~\hms{17}{47}{19.88}, \dec~=~\ang{-28;22;18.4} and Sgr B2(M) towards \ra~=~\hms{17}{47}{20.35}, \dec~=~-\ang{28;23;03.0}, covering the frequency ranges of \unit[480 -- 1250]{GHz} and \unit[1410 -- 1910]{GHz}. The corresponding half-power beam widths are \unit[44.9 -- 17.2]{\arcsec} and \unit[15.3 -- 11.3]{\arcsec}, respectively.

The spectral scans have been calibrated with HIPE version 10.0 \citep{Roelfsema2012}. The resulting double-sideband (DSB) spectra were reduced with the GILDAS CLASS\footnote{http://www.iram.fr/IRAMFR/GILDAS} package. Basic data reduction steps included removal of spurious features or otherwise unusable parts of the spectra. The continuum emission was subtracted from the DSB scans by mostly zero-th, first- or more rarely second-degree polynomial fitting. The continuum-subtracted DSB data were deconvolved \citep[sideband separation through pure $\chi^2$ minimization; ][]{Comito2002} to provide an equivalent single-sideband (SSB) spectrum for each HIFI band. 

Since a single full HIFI line survey is actually made up of 14 independent line surveys with seven different local oscillators (HIFI LO bands 1a through 7b), inconsistencies in the continuum level between HIFI bands were expected and indeed observed. A linear least squares fit of the subtracted continuum values as a function of local oscillator (LO) frequency provided a reliable --- because unaffected by spectral features --- parametrization of the continuum variation across each HIFI band, which was then folded back into each continuum-subtracted SSB spectrum. 

Finally, the overall continuum was rendered self-consistent in two steps: the first adjustment consisted of an additive factor for each band, to achieve a zero-continuum level for the observed saturated absorption features. This is based on the absorption of molecules with a high dipole moment. These molecules are located in foreground, i.e. low density, clouds along the line-of-sight towards Sgr B2. Thus they will have a very low excitation temperature. The second adjustment required a multiplicative factor, in order for the continuum values in overlap regions between bands to be consistent with each other.

For Sgr B2(N), the additive factors range between -1.08 and \unit[0.48]{K}, with a median of \unit[0.05]{K}; the multiplicative factors range between 0.82 and 1.42, with a median of 1.00. For Sgr B2(M), the additive factors range between -0.87 and \unit[+0.34]{K}, with a median of \unit[0.28]{K}; the multiplicative factors range between 0.86 and 1.42, with a median of 1.03.

\subsection{Submillimeter Array (SMA)}
Sgr B2 has been observed with the SMA in the com\-pact and very extended configurations. The observations were carried out on June 11, 2010 using seven antennas and on July 11, 2010 using eight antennas. The phase tracking centers were \ra~=~\hms{17}{47}{19.883}, \dec~=~\ang{-28;22;18.4} for Sgr B2(N) and \ra~=~\hms{17}{47}{20.158}, \dec~=~\ang{-28;23;05.0} for Sgr B2(M). 
The data reduction and results are described in detail in \citet{Qin2011}. The absolute flux scale is estimated to be accurate to within 20\%. Both sources were observed in double-sideband mode, and covered rest frequencies from \unit[342.2 to 346.2]{GHz} and from \unit[354.2 to 358.2]{GHz}. The line-free channels have been used to reconstruct the continuum image. We have re-imaged the continuum maps with almost uniform weighting, resulting in a beam of \unit[0.37]{\arcsec} $\times$ \unit[0.22]{\arcsec}, position angle (hereafter: PA) =~\unit[17.8]{$^\circ$} with a rms of \unit[$\sim$ 26]{mJy/beam}.

\subsection{Very Large Array (VLA)}

To constrain the location and physical parameters of the \hii regions in Sgr B2, we used several different data sets obtained with the VLA.

\subsubsection{\cite{Gaume1995} map at 1.3 cm covering Sgr B2}\label{sssec:vla_gaume}

\cite{Gaume1995} observed Sgr B2 with the VLA in three configurations, BnA, CnB, and DnC, between February 1989 and October 1989. The central pointing position is \ra~=~\hms{17}{47}{20.166}, \dec~=~\ang{-28;23;04.76}. The data set and its calibration is described in detail in \cite{Gaume1995}. We used an image of the combined DnCnBnA data, resulting in a beam of \unit[0.27]{$''$} \mal \unit[0.23]{$''$} (HPBW), PA~=~\unit[70]{$^\circ$}, with an rms of \unit[$\sim$ 0.38]{mJy/beam}.

\subsubsection{\cite{Rolffs2011a} maps at 7 mm covering Sgr B2(N) and Sgr B2(M)}\label{sssec:vla_rolffs}

Sgr B2 has been observed with the VLA in the BnA hybrid configuration at \unit[40.7669]{GHz}, corresponding to \unit[7.5]{mm} (project AR687). The phase tracking centers were \ra~=~\hms{17}{47}{19.902}, \dec~=~\ang{-28;22;17.8} for Sgr B2(N) and \ra~=~\hms{17}{47}{20.202}, \dec=~\ang{-28;23;05.3} for Sgr B2(M).  Sgr B2(M) was observed on 2009 January 31, and Sgr B2(N) on February 1.  The data set and its calibration is described in detail in \cite{Rolffs2011a}. 
The continuum was fitted using line-free channels. We have re-imaged the continuum map with almost uniform weighting, resulting in a beam of \unit[0.15]{$''$} $\times$ \unit[0.10]{$''$} (HPBW), PA~=~\unit[52.8]{$^\circ$}, with a rms of \unit[$\sim$ 0.9]{mJy/beam}.

\subsubsection{\cite{dePree1998} map at 7 mm covering Sgr B2(M)}\label{sssec:vla_depree}

\cite{dePree1998} observed Sgr B2(M) with the VLA using the A configuration on 1996 December 12, and 1997 January 17. The central pointing position is \ra~=~\hms{17}{47}{20.115}, \dec~=~\ang{-28;23;04.02}. The data set and its calibration is described in detail in \cite{dePree1998}. The image we used has a spatial resolution of \unit[0.049]{$''$} \mal \unit[0.079]{$''$} (HPBW), PA~=~\unit[11]{$^\circ$}, with an rms of \unit[$\sim$ 0.5]{mJy/beam}.

\subsection{APEX / ATLASGAL project}
Carried out using the Large APEX BOlometer CAmera \citep[LABOCA; ][]{Siringo2009}, the APEX Telescope Large Area Survey of the Galaxy \citep[ATLASGAL; ][]{Schuller2009} covers the full inner Galactic Plane at \unit[870]{\micron} with a resolution of \unit[19.2]{\arcsec} and an rms below \unit[70]{mJy/beam}. Each position of the inner Galatic plane has been mapped twice with different scanning directions to avoid striping, using the on-the-fly mapping technique. The pointing accuracy is of the order of \unit[4]{\arcsec} and the flux calibration uncertainty is lower than \unit[15]{\%}. A detailed description of the data reduction is given in  \citet{Csengeri2014}. This map was recently cross-calibrated using the data of the Planck mission and the large-scale structure that is filtered out during the processing of the LABOCA data has been added back into the map (Csengeri et al. 2015, submitted). We use this improved map in our study.

\subsection{Herschel / HiGAL project}

The Herschel Hi-GAL survey \citep{Molinari2010} provides photometric mid-IR observations at \unit[70]{\micron} and \unit[160]{\micron} using PACS \citep{Poglitsch2010}, and sub-milimeter observations at  \unit[250]{\micron}, \unit[350]{\micron}, and \unit[500]{\micron} using SPIRE \citep{Griffin2010}. Sgr B2 was observed as part of the Field 0 observations covering the Galactic Center (OD 481; obsids 1342204102, 1342204103). The observations were carried out in PACS/SPIRE parallel mode with a fast scanning speed of \unit[60]{\arcsec s$^{-1}$}. The data reduction is described in detail in \citet{Traficante2011}. This dataset was cross-calibrated using the data of the Planck mission on the long-wavelength side and the data from the IRAS mission on the short wavelength side. The angular resolutions at the five wavelengths, listed by increasing wavelength, are \unit[10.4]{\arcsec}, \unit[13.6]{\arcsec}, \unit[23.4]{\arcsec}, \unit[30.3]{\arcsec}, and \unit[42.5]{\arcsec} \citep{Traficante2011}. We note that the \unit[160]{\micron} PACS map is saturated towards Sgr B2(M) and N and is thus not used in our study.


\section{Modeling Procedure}

\label{s-modeling}
\begin{figure}[t]
    \centering
    \includegraphics[width=0.48\textwidth]{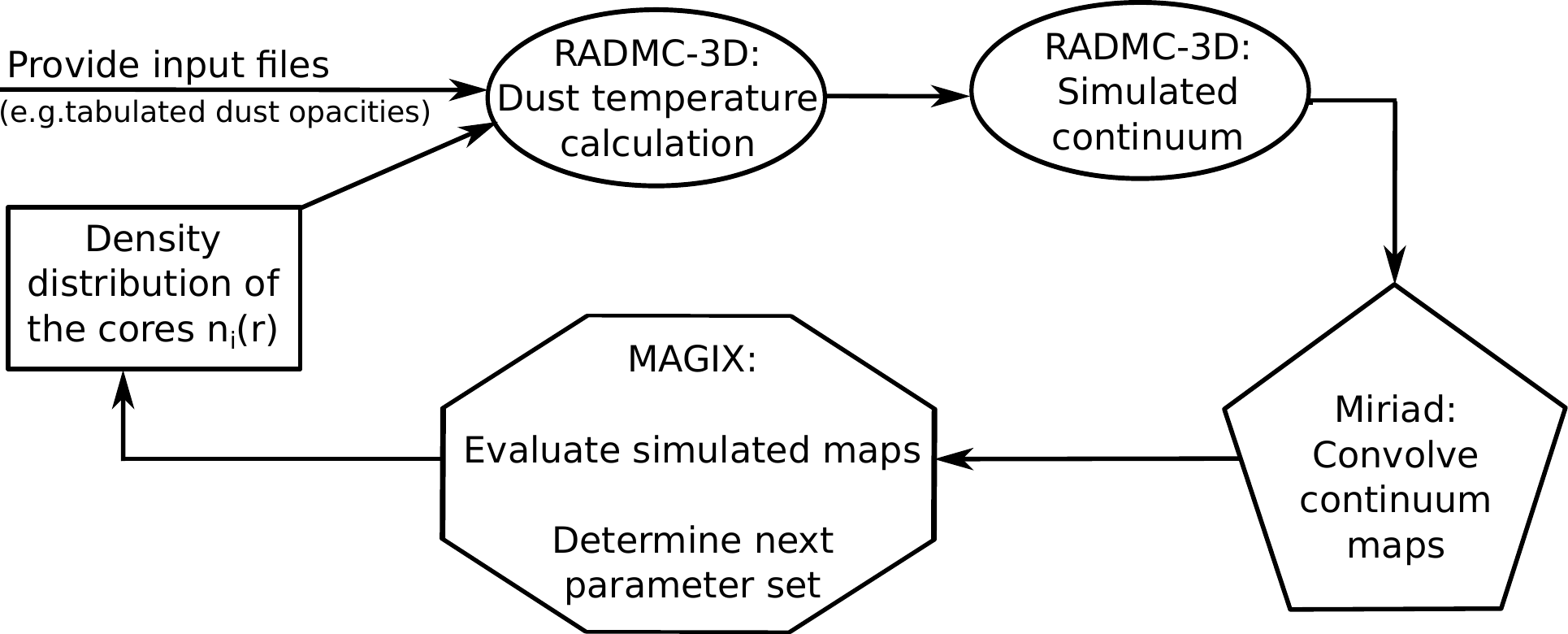}
    \caption{Flowchart of the modeling framework pandora.}
    \label{fig:flowchart}
\end{figure}

We have implemented a framework called \textit{ pandora} that follows the flowchart shown in Fig. \ref{fig:flowchart} for the continuum modeling. As part of this framework, we employ the three-dimensional, publicly available radiative transfer program \textsc{RADMC-3D} \citep[][version 0.39]{Dullemond2012}, Miriad for the post-processing, and MAGIX \citep{Moeller2013} for the optimization of the model input parameters. We describe the model setup and physical parameters in the following subsections. \textsc{RADMC-3D} is written in Fortran90 and allows the user to setup their models in a non-invasive way, i.e. via a separate subroutine. In the following subsections we describe different parts of the setup of our model.

\subsection{Coordinate system}

We use a positively right-handed cartesian coordinate system. This means that the x-axis points west on the sky, the y-axis points to the north and the z-axis points towards the observer. The origin of this system is located at the center of the model, which is chosen to be Sgr B2(M), i.e. \ra~=~\hms{17}{47}{20.172}, \dec~=~\ang{-28;23;04.58}.

\subsection{Grid refinement} 

We want to recover the large dynamic range in spatial resolution from \unit[38]{pc}, i.e. the diameter of the envelope \citep[][corrected for distance]{Scoville1975} down to \unit[100]{au} (small scale structure around the hot cores and \hii regions). If we would attempt this with a cartesian grid, i.e. a grid where the elements are unit cubes,  this would require 800 billion cells. This is not only computationally unfeasible but also unneccessary when e.g. the density distribution is flat.

We use the adaptive mesh refinement technique \citep[originally presented by][]{Berger1984, Berger1989} to locally increase the spatial resolution of our numerical radiative transfer simulation. The initial grid consists of $11^3$ cells. We employ the RADMC option to include a tree-based AMR method \citep[][and references therein]{Khokhlov1998}. This means that, on a cell-by-cell basis, parent cells are refined into children cells resulting in a recursive tree structure. The resulting grid provides high resolution where needed based on the refinement criteria. 

In our approach, two criteria are used to check if a cell needs to be refined: (i) the presence of a dust core center within the cell and (ii) the dust density distribution within the cell. Since we cover a large dynamical range, it is crucial to make sure that small clumps are recognized by the refinement routine. Thus as long as the cell size is larger than twice the radius of a dust core, this cell is refined if the core center is located within the cell. A cell is also refined into eight sub-cells when 

\begin{equation}
    \frac{\vert n_\text{center} - n_i \vert}{n_\text{center}} > \upepsilon,
\end{equation}

\noindent where $n_\text{center}$ is the density at the cell center and $n_i$ is the density at the center of the faces or the corners of the cell and $\upepsilon$ is the allowed maximum density difference, which is set to \unit[10]{\%}. 

The grid is refined to level 13, i.e. at least one initial cell has been refined 13 times. This results in a minimum cell size of \unit[100]{au} and in total \unit[$\sim$68 million]{final cells}.

\subsection{Heating sources}\label{ssec:heating}

\textsc{RADMC-3D} provides several methods to include luminous sources. We choose the option to manually specify the individual luminous sources and assume that all stars are point sources, i.e. their radius is not taken into account. We use a two-step approach. 
In the first step we account for observed early-type high-mass stars by including the known \hii regions \citep[][see Table \ref{tab:sgrb2_hii} for a complete list of all parameters]{Mehringer1993, Gaume1995, dePree1998}. Due to a revised distance of the Galactic Center we estimated the corresponding parameters, see Sect.\ \ref{subsec:hii}. We specify their position and calculate the luminosity and temperature from the Zero-Age-Main-Sequence (hereafter: ZAMS) type using Table 5 from \citet{Vacca1996}. Here we assume that each \hii region is ionized by a single star. We use the \textsc{RADMC-3D} option to assume simple blackbody spectra for each star, by specifying their blackbody temperature. This accounts for stars down to the B0 spectral type. 

In the second step we take later spectral types, i.e. stars which cannot produce \hii regions detectable with current observations, with stellar masses between \unit[0.01]{\msun} and \unit[$\sim$19.0]{\msun}, into account. We do this as follows: Based on the gravitational potential, our algorithm randomly determines positions for the new stars. For this, we follow the procedure explained in the appendix of \citet{Aarseth1974}. A luminosity, which is randomly drawn from the initial mass function (IMF) of \citet{Kroupa2001}, is assigned to each star.

We do this as follows: Assuming a spherical symmetric star cluster, we specify the radius of this cluster. This radius is determined using a radial histogram of the distributed \hii regions (see Fig. \ref{fig:starClusterRadius}).
From Kroupa's IMF we then calculate the fraction $f_\text{HM}$ of the stars in the high mass regime (i.e. between M$_1$ and M$_2$; M$_1$ $<$ M$_2$) to the mass of all stars (i.e. between M$_0$ and M$_2$; M$_0$ $<$ M$_1$ $<$ M$_2$). The lowest mass boundary M$_0$ is provided, the mass boundaries M$_1$ and M$_2$ are set by the high-mass stars that were distributed within the cluster in the first step described above. The total stellar mass of the cluster M$_\text{cluster}$ is then calculated from the ratio of the sum of the stellar mass of all high-mass stars distributed between M$_1$ and M$_2$ to the ratio $f_\text{HM}$. We then randomly draw stars between M$_0$ and M$_1$ from the IMF until the total stellar mass of the cluster, M$_\text{cluster}$, is reached.

The model can contain several of these star clusters. After the stars are sprinkled separately within each of these clusters, the entire envelope is treated as a cluster itself and the procedure is repeated once more, i.e. necessary stars are sprinkled everywhere except in the previously mentioned star clusters until the calculated total stellar mass of the computational domain is reached.

We then convert the stellar mass $M$ to luminosity  $L$ using the parameterized mass-luminosity relation from \citet{Griffiths1988}:
\begin{equation}
    \log_{10}\left(\frac{L}{L_\odot}\right) = 
    \begin{dcases}
       0.006 + 4.16 \log_{10}\left(\frac{M}{M_\odot}\right),\\
       \hspace{2.5cm} 0.682 < \log_{10}\left(\frac{M}{M_\odot}\right)\ <  0.461\\
       0.370 + 3.51 \log_{10}\left(\frac{M}{M_\odot}\right),\\
       \hspace{3.7cm} \log_{10}\left(\frac{M}{M_\odot}\right) > 0.461
    \end{dcases}
\end{equation}
\citet{Griffiths1988} provide two sets of boundaries. One is non-continuous but has a better $\upchi^2$ value whereas the other set, which we choose to use, provides a continuous behaviour of the mass-to-luminosity relation while having a slightly worse $\upchi^2$ value.  

We only include luminosity in terms of ZAMS luminosity. The total luminosity determined in the literature, however, includes contributions from accretion luminosity, as well as from high-mass stars without an \hii region \citep{Hosokawa2009}. We thus note that the luminosity calculated with our approach is a lower limit. We determine the corresponding effective temperatures of the stars using the fundamental stellar parameters (spectral type, luminosity, effective temperature) compiled by \citet{Straizys1981}. Since we need to provide an effective temperature for each star, the limitation of the tabulated data sets the lower mass limits of the star clusters to \unit[0.2]{\msun}.

\subsection{Dust density distribution}\label{ssec:dustdistr}

The overall density structure is obtained by the superposition of the density profiles of all dust cores, i.e. in overlap regions, the density simply adds up. In each cell $j$, the density is determined as
\begin{equation}
    n_j = \sum_{i = 1}^{N} n_{i, j}(\pmb{r})
\end{equation}
where i is the index of the dust cores and N is the number of cores. Following \cite{Qin2011}, we use a modified Plummer-like profile to model the SMA dust density cores as well as the large-scale envelopes. For a subset of our models, we find it useful to introduce elongated density structures (see Sect.\ \ref{s-analysis}). We define them as follows:
\begin{equation}\label{eq:plummer}
  n_i(\pmb{r}) = \frac{n_c}{\left(1+\vert\pmb{r}\vert^2\right)^{\upeta/2}}
\end{equation}
where $n_c$ is the central density given in H$_2$ cm$^{-3}$ and $\pmb{r}$ is given by the Euclidean norm, including scaling factors
\begin{equation}
    \vert\pmb{r}\vert = \sqrt{\left(\frac{r_x}{r_{0, x}}\right)^2 + \left(\frac{r_y}{r_{0,y}}\right)^2 + \left(\frac{r_z}{r_{0,z}}\right)^2}
\end{equation}
where $r_{x,y,z}$ are the components of $r$ and $r_{0,x}$, $r_{0,y}$, and $r_{0,z}$ set the size of the cluster core in each of the three principal axes. A spherical symmetric distribution is obtained by setting $r_0=r_{0,x}=r_{0,y}=r_{0,z}$. The density distribution is flat inside the radius $r_0$ and approaches a power-law with an exponent $\upeta$ at $r\gg r_0$.


\subsection{\hii regions} \label{subsec:hii}

In our model, we consider \hii regions as Str\"omgren spheres, i.e. as a fully ionized, spherical regions of uniform electron density with no dust. There are more than 70 \hii regions known in Sgr B2 \citep{Mehringer1993, Gaume1995, dePree1998}. 

We obtained the high-resolution interferometric maps presented in \cite{Gaume1995} and \cite{dePree1998} and performed the following analysis steps. \hii regions come in many different shapes, which makes them hard to fit with Gaussians. So in order to match them as best as possible with spheres, we masked everything below three times the rms in the continuum maps and by eye inspection adjusted circles to enclose the observed \hii regions. We then integrated the enclosed flux and assuming optical thin emission and a homogeneous, non-expanding \hii region, we calculated the number electron density $\mathrm{n_e}$, the emission measure $\mathrm{EM}$, and the number of ionizing photons $\mathrm{\dot N_\mathrm{i}}$ using the following formulas. 

\begin{equation}
    \left(\frac{\text{EM}}{\text{pc cm}^{-6}}\right) 
    = 3.217 \times 10^{7} 
	\left(\frac{F_\nu}{\text{Jy}}\right)  
	\left(\frac{T_e}{\text{K}}\right)^{0.35} 
	\left(\frac{\nu}{\text{GHz}}\right)^{0.1}
	\left(\frac{\theta_\text{source}}{\text{arcsec}}\right)^{-2}
\end{equation}

\begin{align}\label{eq:electronDensity}
     \left(\frac{n_e}{\text{cm}^{-3}}\right)
     = {} & 2.576 \,\times\, 10^{6} \, 
	    \left(\frac{F_\nu}{\text{Jy}}\right)^{0.5}  
	    \left(\frac{T_e}{\text{K}}\right)^{0.175} 
	    \left(\frac{\nu}{\text{GHz}}\right)^{0.05} \nonumber \\
       & \left(\frac{\theta_\text{source}}{\text{arcsec}}\right)^{-1.5}
	    \left(\frac{D}{\text{pc}}\right)^{-0.5} 
\end{align}

\begin{equation}
    \left(\frac{\dot N_\mathrm{i}}{\text{s}^{-1}}\right)
    = 4.771 \times 10^{42} \times
	    \left(\frac{F_\nu}{\text{Jy}}\right)  
	    \left(\frac{T_e}{\text{K}}\right)^{-0.45} 
	    \left(\frac{\nu}{\text{GHz}}\right)^{0.1}
	    \left(\frac{D}{\text{pc}}\right)^2
\end{equation}
Here $F_\upnu$ is the flux density of the \hii region, $T_\mathrm{e}$ is the electron temperature, $\upnu$ is the frequency, $\mathrm{D}$ is the distance to the source, and $\uptheta_\text{source}$ is the angular diameter of the \hii region. A derivation of these formulas can be found in the appendix \ref{app:derivationHii}.

Deriving the number electron density at a certain frequency from the observed flux density assuming optical thin emission will underestimate the number electron density if the \hii region is actually optically thick. This underestimate will then lead to the intensity of the (optically thick) \hii region being underpredicted by the simulation compared to the observed data at this frequency. To account for this discrepancy, we compare the synthetic intensities obtained from our model setup (see Sect.\ \ref{ssec-images}) with the observed intensities. From the deviation of the intensity levels, we identify the \hii regions which must be optically thick. We then iteratively adjust the flux density values and recalculate the number electron density $\mathrm{n_e}$, the emission measure $\mathrm{EM}$, and the number of ionizing photons $\mathrm{\dot N_\mathrm{i}}$ until we obtain a resonably good match between the observed and synthetic intensities. As our main goal is to model the dust emission here we need to derive the free-free contribution from the \hii regions. Hence we did not aim at a very sophisticated  model of the \hii regions, which would have to include a different geometry and a density structure.

The \hii regions detected by \citet{Mehringer1993} were included according to the parameters stated therein. However, \citet{Mehringer1993} assumed a source distance of \unit[7.5]{kpc}. We have thus re-calculated the radii and the parameters stated above for a distance of \unit[8.5]{kpc}. But we have not corrected the values to account for the optical depth effects. Thus we note that these values are a lower limit. All parameters, including the radii and the precessed coordinates, are listed in Table \ref{tab:sgrb2_hii}.

\subsection{Dust temperature calculation}\label{ssec:Tdust}

The dust temperature is calculated self-consistently by \textsc{RADMC-3D} using the Monte Carlo method of \cite{Bjorkman2001}, with various improvements, such as the continuous absorption method of \citet{Lucy1999}. Initially, all cells have a dust temperature equal to zero. To determine the dust temperature, RADMC-3D first identifies all sources of luminosity, i.e.\ the total amount of energy available. This amount is divided into photon packages,  which are separately emitted by the individual stars one after another. As the photon packages move through the grid, they interact with the dust present (scattering, absorption) causing them to change their direction or wavelength. In the case of absorption, the photon package is immediately re-emitted in a different direction with another wavelength according to \cite{Bjorkman2001}. However, the luminosity fraction each photon package represents remains the same. Whenever a photon package enters a cell, it increases the energy of that cell and thus also increases the temperature of the dust. Photon packages never get lost, they can only escape the model through the outer edge of the grid. After the last photon package has escaped the grid, the dust temperature is obtained. In summary the dust temperature of each cell is basically the sum of the energy fractions of each photon package passing through the cell. In total we use 10 million photon packages.

The calculated temperature is an equilibrium dust temperature, since it is assumed that the amount of energy each dust grain acquires and re-radiates stays the same. For most cases, this is presumably a very good approximation, because the heating and cooling timescales for dust grains are typically very short, compared to any time-dependent dynamics of the system \citep{Choudhury2015}. Thus transiently heated small grains are not important within the wavelength range considered here.

Note, we only consider stars as heating sources (see Sect. \ref{ssec:heating}). The calculated dust temperature in the outer parts of the envelope is on average \unit[$12-15$]{K}.  Modified blackbody fitting of SCUBA data \citep{Pierce-Price2000} and Herschel data \citet{Etxaluze2013} yields average dust temperature values for Sgr B2 (along the line-of-sight) of \unit[$\sim$ 20]{K}. Apart from line-of-sight effects in the modified blackbody fitting, which do not influence the modeling, part of this discrepancy is most likely due to heating by sources not considered here, e.g. cosmic rays or coupling with the turbulently heated gas.

We furthermore make use of the Modified Random Walk (MRW) method that is implemented in \textsc{RADMC-3D} in the simplified form described in \cite{Robitaille2010}. This method prevents a photon package from getting trapped in high-density regions by predicting where the photon will go next. This allows \textsc{RADMC-3D} to make one single large step of the photon package, saving the computation time for the otherwise necessary hundreds or thousands of absorption or scattering events.

\subsection{Images and post-processing}\label{ssec-images}

We use one dust species throughout the model and do not include scattering events. We use the tabulated dust opacity from \citet{Ossenkopf1994} for dust without grain mantles and no coagulation, as found to best fit the Sagittarius B2 region by \citet{Rolffs2011}. Including the free-free emission we cover the frequency range from \unit[40]{GHz} up to \unit[4]{THz}. Continuum maps at various wavelengths of interest are produced with \textsc{RADMC-3D}.

In order to compare the synthetic maps with real obervations, telescope-dependent post-processing is necessary.  For this purpose we employ the data reduction package Miriad \citep{Sault1995}. In case of interferometric observations, we fold the simulated maps with the uv-coverage. The imaging is then performed for both maps (synthetic and observed) with the same imaging parameters. In case of single dish observations, the synthetic maps are convolved with the beam of the telescope at that frequency, which we assumed to be Gaussian. Final maps are then produced for the respective telescopes at various wavelengths in intensity units (Jy/beam).

\subsection{Fitting procedure}

Three-dimensional modelling intrinsically has many free parameters. For this initial study, we have focused on deriving the density field for a fixed dust setup and, apart from by-hand adjustments to account for optical depths effects, fixed \hii regions. Still, the amount of free parameters is large. We have 20 dust core components and 9 components for the large-scale envelope. For each component we modified, in order of increasing importance, the density exponent, a stellar heating source and accordingly a stellar cluster, the position along the line of sight, the displacement in right ascension and declination, the radius and the central density. In total this amounts to approximately 140 free parameters, many of them degenerated. 

Concerning the position along the line-of-sight we unfortunately only have limited constraints available for certain sources to fix the three-dimensional structure on all scales. For optically thin radiation, there is no information on the line-of-sight structure.  Optically thick radiation comes from a photospheric surface where the opacity exceeds unity.  Thus, maps from different wavelengths, with different opacities tracing different radii, can give some constraints on the relative positions of sources along the line-of-sight, but in general it is not possible to fix the absolute positions along the line-of-sight.
The more observational data is available the better the full three-dimensional structure could be deduced. We have thus only varied the line-of-sight position in a very limited way. More will be possible when fitting individual molecular lines, because multiple lines from many isotopologues of the same species provide many more surfaces with the opacity exceeding unity, allowing a tomography of the source.

Exploring the complete parameter space of all remaining free parameters is prohibitive in terms of computing time.  Thus, we used a hybrid approach to constrain the other parameters mentioned. First, we derive a good guess by varying parameters by hand, and judging the quality visually ($\upchi^2$-by-eye). This enabled us to fix the exponent, the presence of stellar heating source/star clusters and the displacement in right ascension and declination.
We then employed the model optimizer MAGIX \citep{Moeller2013} to iteratively search for the best solution, i.e. the parameter set with the lowest reduced $\upchi^2$ value. For this run, we use the Genetic Algorithm, leaving the central densities and the radii as free parameters. The contribution from the \hii regions (e.g. number electron density) was kept fixed during the entire fitting procedure.

We fit the \unit[874]{GHz} SMA maps, \unit[480 -- 1280 and 1440 -- 1900]{GHz} HIFI spectral scans as well as the large scale maps from ATLASGAL at \unit[850]{\micron}, HiGAL SPIRE at \unit[500, 350 and 250]{\micron} and HiGAL PACS at \unit[70]{\micron}.

In total, we have run approximately 2 $\times$ 10$^4$ different models (multi-core). The runtime for a single run varies between 0.5 and 6.0 hours, depending on the density structure, which, e.g. affects the number of final cells in the grid.


\section{Analysis and Discussion} \label{s-analysis}

Within the scope of this paper, we will focus the analysis on the regions for which a wealth of data is available, namely Sgr B2(N),  Sgr B2(M), and partly Sgr B2(S). We included Sgr B2(NE) and Sgr B2(deepS) in the model. However, we did not attempt to fit these components, since the available dataset is scarce.

\subsection{Large scale dust continuum}

\begin{figure*}[ht]
    \centering
    \includegraphics[width=.98\textwidth]{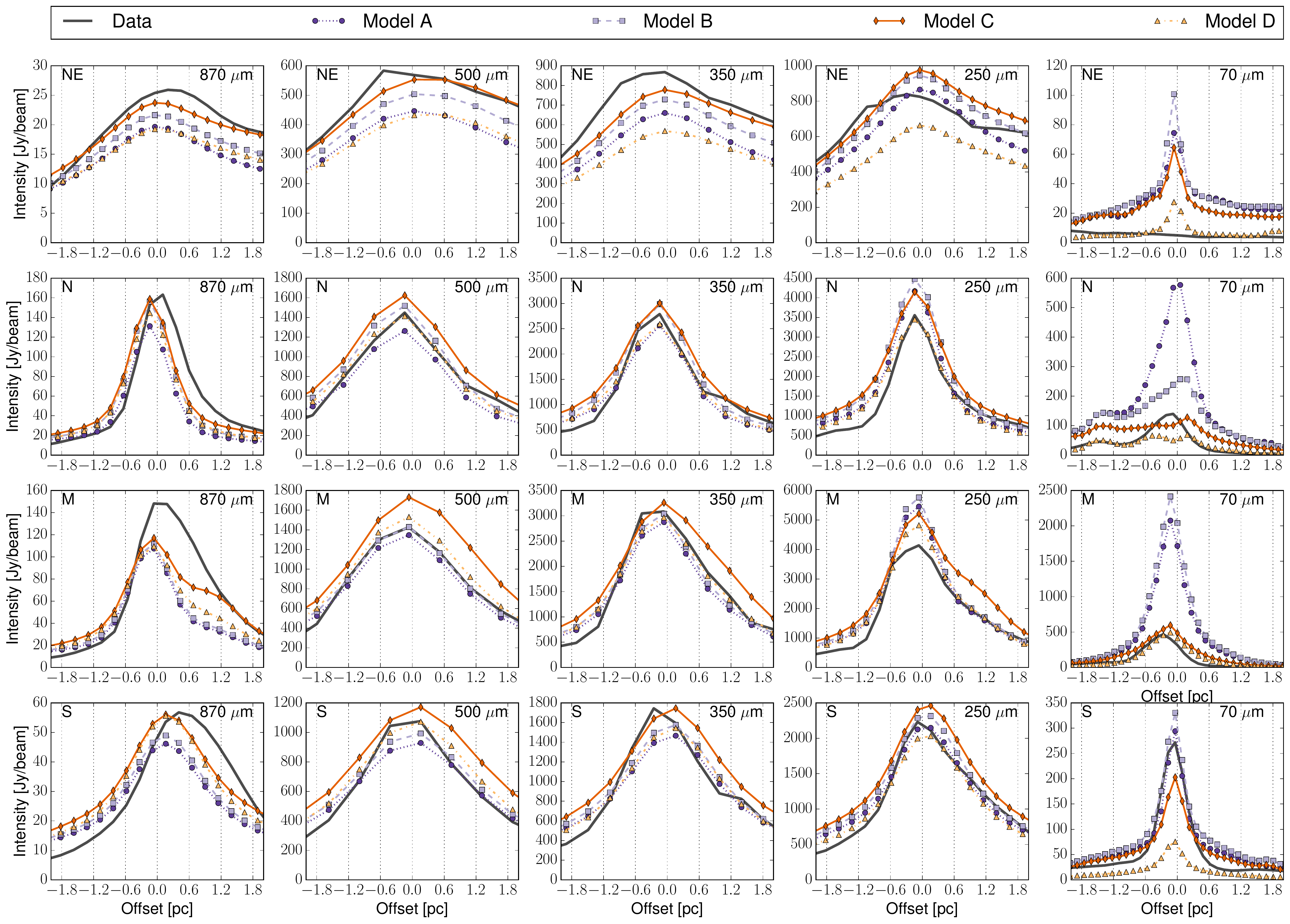}
    \caption{Intensity cuts along the reversed right-ascension axis at the constant declination of the envelope components of Sgr B2(NE), Sgr B2(N), Sgr B2(M), and Sgr B2(S) from top row to bottom row. The data is plotted in solid black. The wavelength decreases from left to right: 870 \micron, 500 \micron, 350 \micron, 250 \micron, and 70 \micron. Model A is denoted by the circular markers, Model B by the squared markers, Model C by the diamonds shaped markers and model D by the triangular markers.}\label{fig:cut_largeScale}
\end{figure*}

\begin{figure*}[t]
    \centering
    \includegraphics[width=0.98\textwidth]{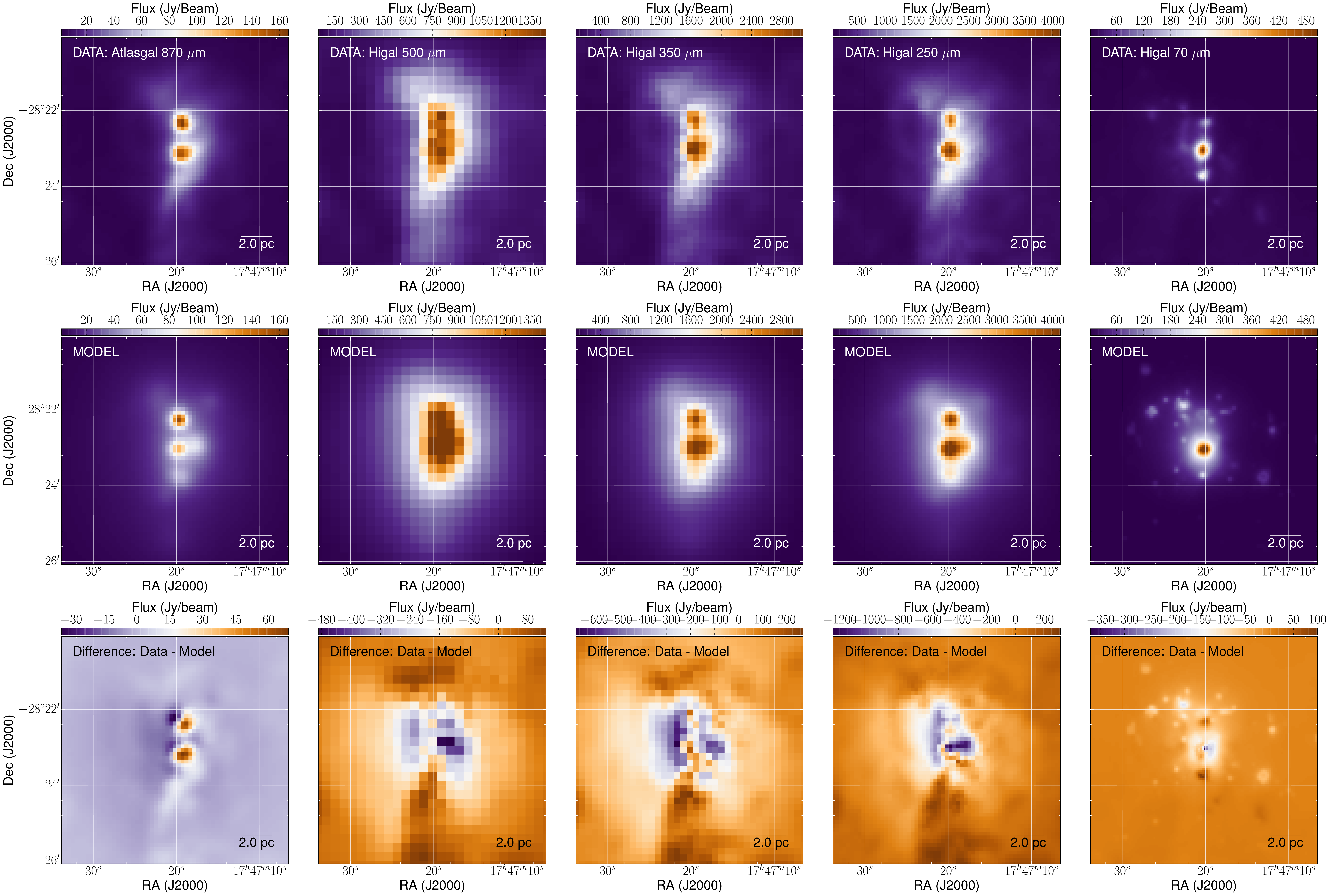}
    \caption{Sgr B2, large scale continuum maps (Model C). From left to right: ATLASGAL \unit[870]{\micron}, Hi-GAL \unit[500]{\micron}, \unit[350]{\micron}, \unit[250]{\unit}, and \unit[70]{\micron}. First row: Data. Second row: Simulation. Third row: Difference between data and simulation.}\label{fig:singleDishMaps}
\end{figure*}

To recover the large scale structure of the envelope, as visible in the ATLASGAL and Hi-GAL intensity maps, we superimpose many density components having profiles with varying exponents (see Sect. \ref{ssec:dustdistr}). While fitting the large-scale dust continuum maps, we noted the following problem. For a model where the density profile for each component is spherically symmetric and all components are located in the z=0 plane (Model A), the \unit[70]{\micron} intensity was always overestimated and the \unit[870]{\micron} intensity was underestimated, especially for the two cores Sgr B2(N) and Sgr B2(M).

By moving Sgr B2(N) along the line of sight behind the plane containing Sgr B2(M) as suggested by e.g. \citet{Goldsmith1990} (Model B), the emission from Sgr B2(N) gets attenuated by the envelope of Sgr B2(M) and an improved fit is possible. However, the general trend of the overestimation of the intensity at \unit[70]{\micron} and underestimation of the intensity at \unit[870]{\micron} remains, especially for Sgr B2(M). This is a sign of the dust column density being underestimated and the luminosity being overestimated.

In a first approach, we have reduced this effect by assuming that the envelopes of both, Sgr B2(N) and Sgr B2(M) are elongated along the line-of-sight, i.e. they basically look like a cigar (Model C). We achieved an improved fit by increasing the radii along the line-of-sight by factors of 1.4 and 2.3  for Sgr B2(N) and Sgr B2(M), respectively.  At \unit[70]{\micron} the dust becomes optically thick. So by increasing the dust column density along the line-of-sight, we are able to hide the contribution from the stars. On the other hand, this approach increases the intensity at \unit[870]{\micron}, because it increases the total dust column density, which is proportional to the intensity of the optically thin \unit[870]{\micron} emission.

Another solution would be to decrease the luminosity. The luminosity was calculated assuming a single ionized star in each \hii region. However, besides the Lyman continuum emission from the early-type star, additional UV photons could be emitted from accretion shocks  in the stellar neighborhood (Cesaroni et al. 2016, submitted) which would lead to assigning an earlier spectral type and thus overestimating the total luminosity. To test this possibility, we have multiplied the luminosity by an arbitrary factor of 0.5 (Model D). While doing so, we of course needed to increase the density to preserve the \unit[870]{\micron} fluxes. In this approach, there was no need for the elongated envelope, except for Sgr B2(M), where the envelope is still slightly elongated.  However, the observational and theoretical evidence of this Lyman continuum overluminosity seems to be mainly given for B-type stars \citep{Sanchez-Monge2013a, Smith2014b}. Thus it is unclear if O-type stars exhibit the same behavior. Furthermore, the effect of dust  inside the \hii regions has not been assessed. This  will be part of future work, see Sect.\ \ref{subs-outlook}.

We have kept the number of components in all these models fixed. All parameters of each component are listed in Table \ref{tab:sgrb2_power}. To facilitate the comparison of these different models, we have produced cuts of the intensity along the reversed right ascension axis at the constant declination of the components Sgr B2(NE), Sgr B2(N), Sgr B2(M) and Sgr B2(S). These cuts are shown in Fig.\ \ref{fig:cut_largeScale}. For comparison, we also show the observed and synthetic azimuthally averaged radial profile for all four models of the same components in Fig.\ \ref{fig:radProfiles_largeScale}. We present the ATLASGAL \unit[870]{\micron}, HiGAL-SPIRE 500, 350 and \unit[250]{\micron}, and the HiGAL-PACS \unit[70]{\micron} from left to right, i.e. the wavelength decreases from left to right.
These different models show that there is a degeneracy between the dust density, luminosity and the relative location of the different dust density centers and star clusters along the line of sight. So in summary Model A overestimates the \unit[70]{\micron} intensity for both, Sgr B2(N) and Sgr B2(M) while underestimating the intensity for both regions at \unit[870]{\micron}. Model B only improves the discrepancy for Sgr B2(N), but still gives a bad fit for Sgr B2(M). These two models are thus clearly unfavorable to proceed the analysis with them. Model D provides an improved fit, however it remains unclear to which degree the luminosity could be adjusted. We thus choose to use Model C for the following analysis, since it provides a reasonably good fit for Sgr B2(N) and Sgr B2(M) without assuming any modification of the luminosity.

The resulting large-scale single-dish maps of Model C are shown in Fig.\ \ref{fig:singleDishMaps} . The first row shows the observed maps, the second row the synthetic maps and the third row is a simple difference map between the observed data and synthetics maps. From left to right we have again the same wavelengths as in Fig.\ \ref{fig:cut_largeScale}: ATLASGAL 870\micron, SPIRE 500 \micron, SPIRE 350 \micron, SPIRE 250 \micron , and PACS 70 \micron. In  the observed maps, the two hot cores Sgr B2(N) and Sgr B2(M) are clearly visible. The extension to the north-east, i.e. Sgr B2(NE), and the one to the south, i.e. Sgr B2(S), are also distinguishable. 

In both the observed and the synthetic maps, the intensity of Sgr B2(M) is stronger than Sgr B2(N) except for \unit[870]{\micron} where the opposite is the case. So the model reproduces the general behaviour, but is currently incapable to reproduce the absolute intensities. Especially in Sgr B2(M), the model underestimates the flux at \unit[870]{\micron} and slightly overestimates the flux at \unit[70]{\micron}.  The observed maps and the intensity cuts (see Figs.\ \ref{fig:cut_largeScale} and \ref{fig:singleDishMaps}) furthermore show a clear asymmetry, i.e. the profile of the continuum emission has at all wavelengths a slope that is steeper on the east side of the peak than on the west side. To quantify the difference of these slopes, we have fitted the intensity profiles, $I \propto r^{-p}$, of Sgr B2(N), Sgr B2(M) and Sgr B2(S) for the wavelengths 870, 500, 350, and \unit[250]{\micron}. For the east side of the peak, we obtain an averaged $p$ of $1.0$ and for the west side of the peak we obtain an averaged $p$ of $0.4$. Looking at the slopes for the individual sources, we furthermore find that the steepness of the west side of the peak remains unchanged with declination, whereas the steepness of the east side of the peak changes significantly with declination. For Sgr B2(N) and Sgr B2(M), p is on average $1.2$, whereas it is $0.7$ for Sgr B2(S). This asymmetry is impossible to account for using a single spherically symmetric component. If the slope of the intensity profile, i.e. the column density distribution, is $p$, then the slope of the density profile should be $p+1(=\upeta)$. So based on the fitting of the intensity profile, the east wing should have a density exponent of $\upeta = 2.0$ and the west wing should have a density exponent of $\upeta = 1.4$.

In the model, we are able to approximate the asymmetry by using a superposition of two components with different exponents for each clump. The east wing is fitted with a density profile having an exponent of $\upeta = 2.5$, whereas the west wing is fitted with a density profile having a lower exponent of $\upeta = 1.8$. These exponents are in good agreement with the ones derived from the intensity profile fitting mentioned above. This clearly shows that our assumption of spherically symmetric clumps as basis limits the modeling effort.

In addition, it is noticeable that the peak of the intensity profiles is shifted to the west in the \unit[870]{\micron} ATLASGAL maps compared to all Hi-GAL maps by \unit[$\approx 0.2$]{pc}, corresponding to \unit[$\approx 5$]{\arcsec}. This is well within the relative pointing error of the ATLASGAL map (pointing accuracy \unit[$\sim$4]{\arcsec}) and the Hi-GAL maps (pointing accuracy \unit[$\sim$1]{\arcsec}). Thus we do not attempt to reproduce this shifting ot the peak with our model.

\subsection{Small scale dust and free-free continuum}

\begin{figure*}[t]
  \centering
     \includegraphics[width=0.98\textwidth]{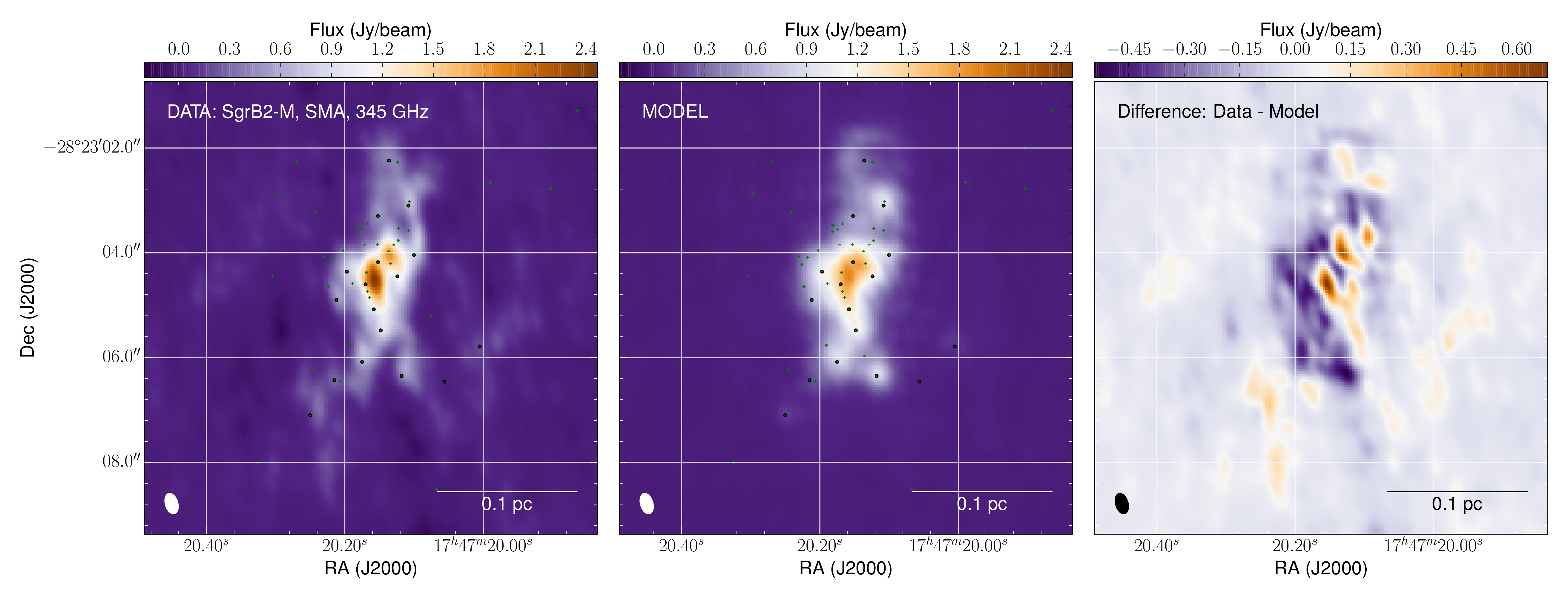}  \\
     \includegraphics[width=0.98\textwidth]{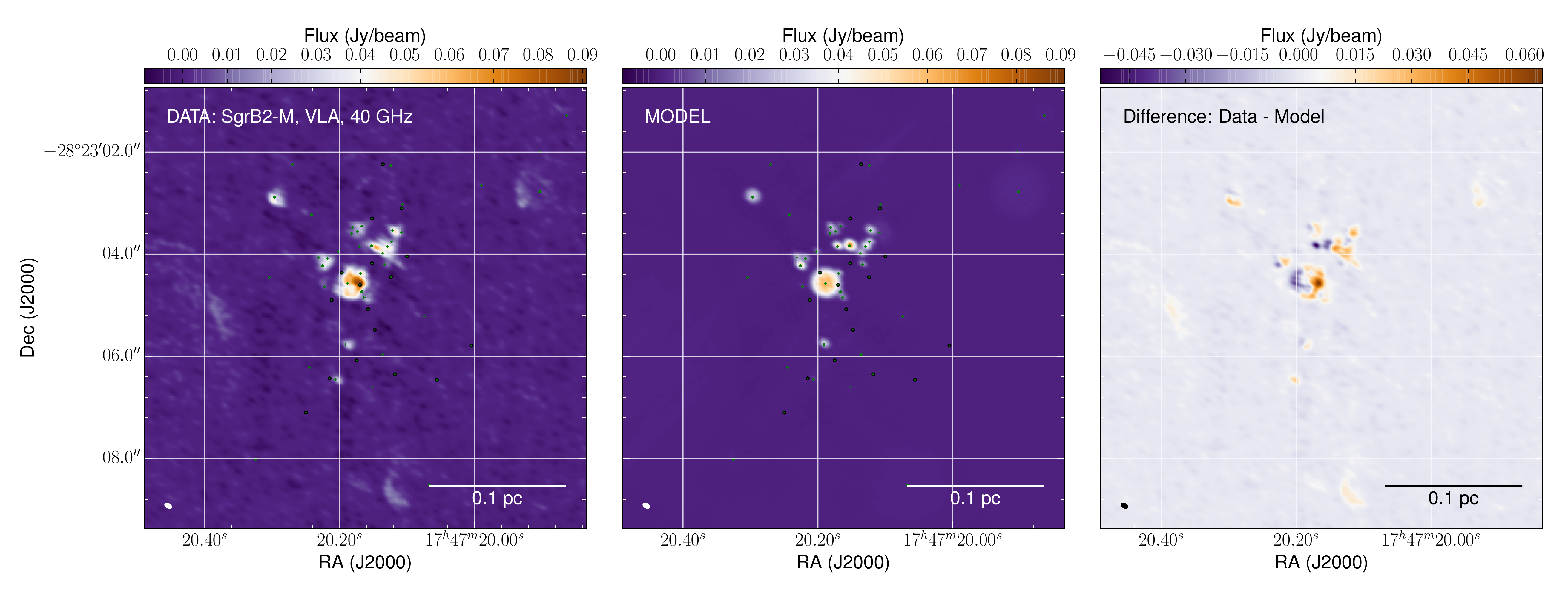}
\caption{Interferometric maps of Sgr B2(M). Top row: SMA dust continuum map at \unit[870]{\micron} (i.e. \unit[345]{GHz}). Bottom row: VLA free-free continuum map at \unit[7]{mm} (i.e. \unit[40]{GHz}). Left column: Observations. Middle column: Synthetic maps. Right column: Difference plot. The field of view is the same for both rows. The green pluses denote the center of the \hii regions and the black circles denote the center of the density cores.}\label{fig:interferometricMaps_M}
\end{figure*}

\begin{figure*}[t]
  \centering
     \includegraphics[width=0.98\textwidth]{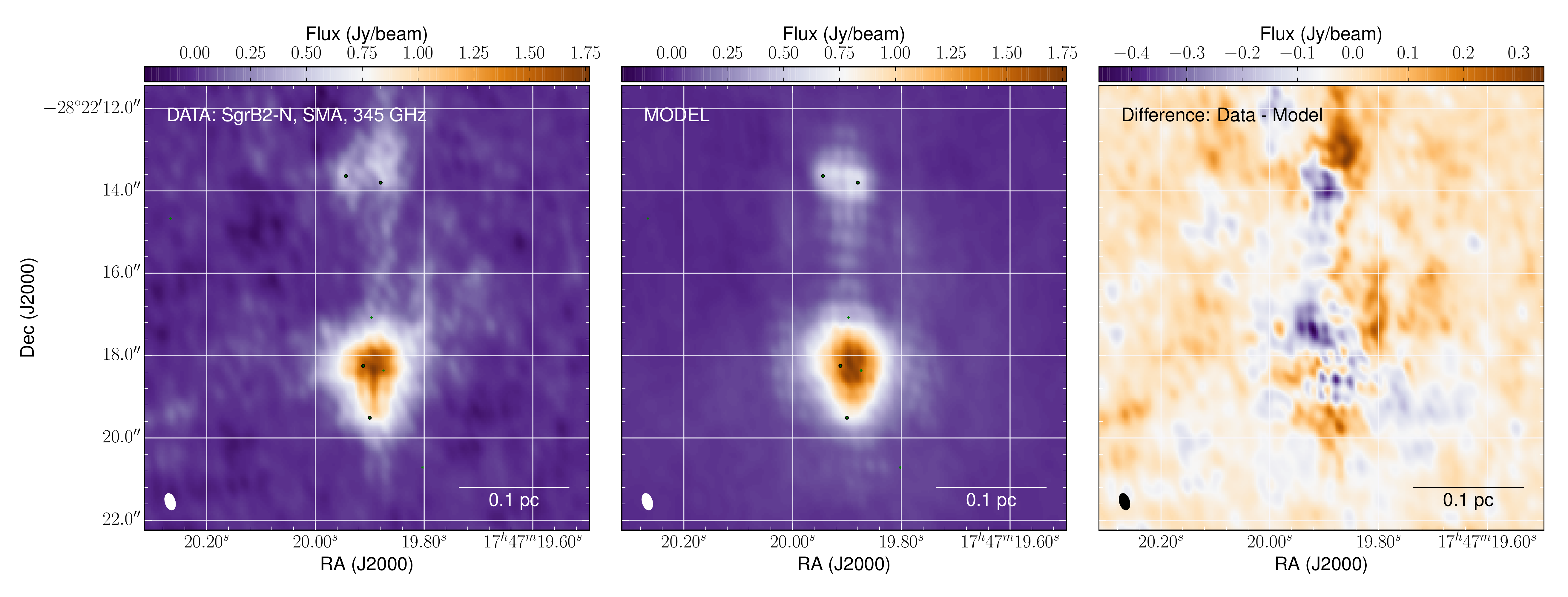}  \\
     \includegraphics[width=0.98\textwidth]{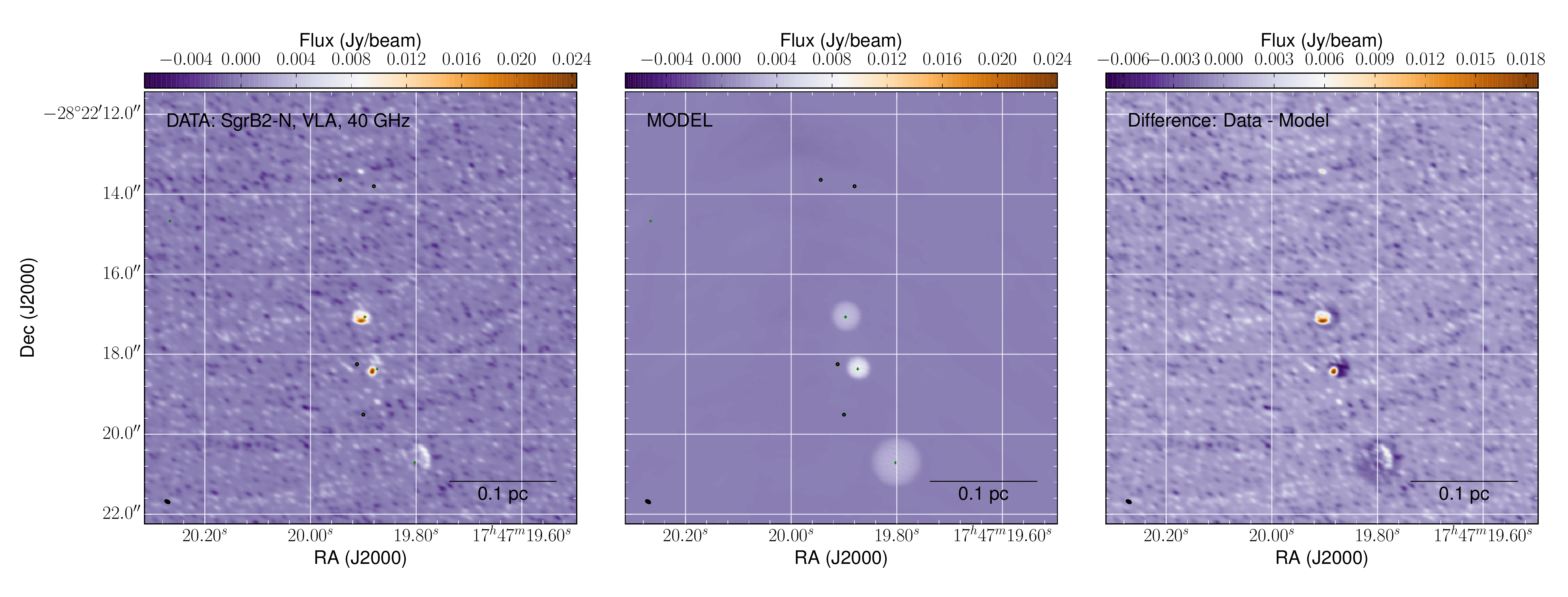}
\caption{Same as Fig.\ \ref{fig:interferometricMaps_M} for Sgr B2(N).}\label{fig:interferometricMaps_N}
\end{figure*}

\begin{figure}[t]
    \centering
    \includegraphics[width=0.45\textwidth]{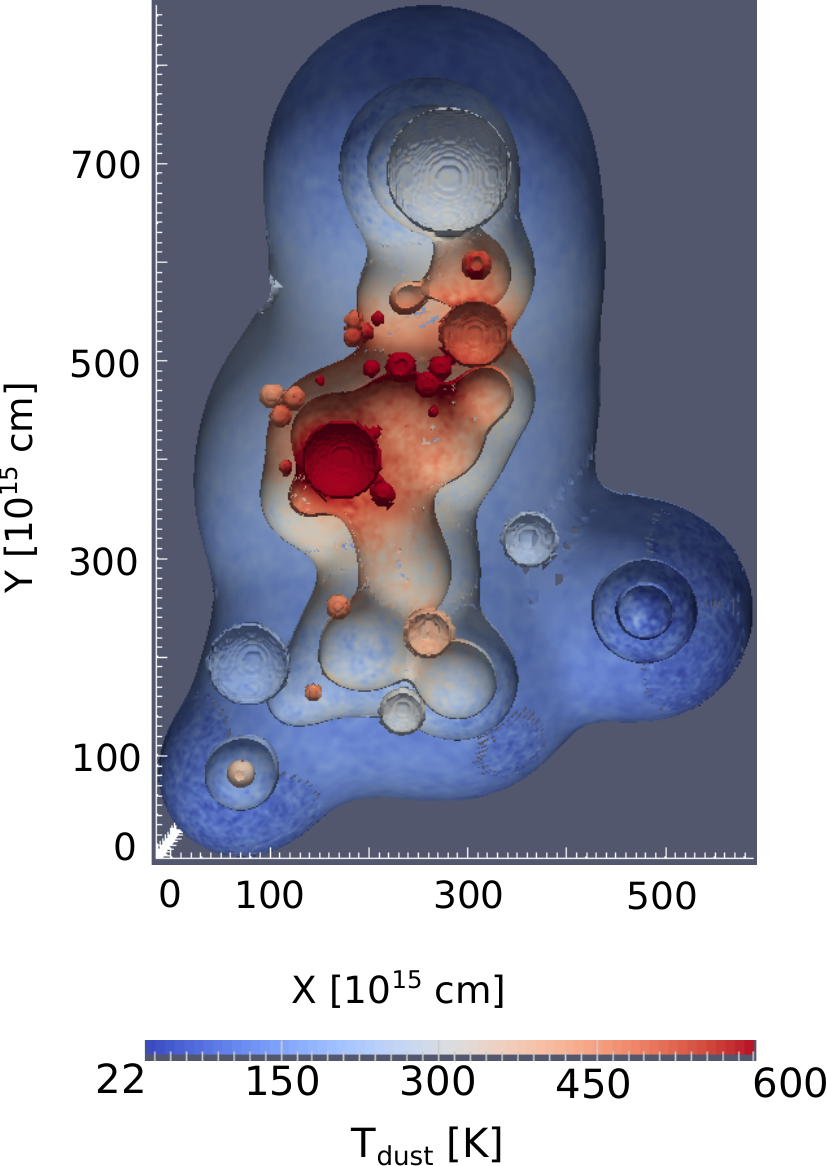}
    \caption{Impression of the 3d view of the dust density distribution in Sgr B2(M). Three density isocontours are presented, i.e. at densities of \unit[10$^{-17.5}$, 10$^{-18}$, and 10$^{-19}$]{g cm$^{-3}$}. They are colored using the dust temperature. The model is cut open at the model center to allow a view inside. The bubbles are the empty half-shells of the  distributed \hii regions, which are free of dust. The dust on the surface of these \hii regions is heated by the UV radiation field from the young stellar object embedded in the \hii region. This dust is thus very hot, exceeding dust temperatures of \unit[600]{K} (dark red spots). Visible here are mainly the \hii regions detected by \cite[][F1a -- F4d]{dePree1998}.}\label{fig:3dstructure}
\end{figure}

The interferometric maps covering spatial scales up to \unit[$\sim$ 0.25]{pc} are shown for Sgr B2(M) in Fig.\ \ref{fig:interferometricMaps_M} and for Sgr B2(N) in Fig.\ \ref{fig:interferometricMaps_N}. We refer to these spatial scales as the small-scale structure. In addition to the intensity maps at different wavelengths, we also show the physical setup of Sgr B2(M) in Fig. \ref{fig:3dstructure}. We plot three density isocontours colored using the dust temperature. To allow a view inside the model, we cut it open along the (x, y, 0)-plane. The bubbles visible there are the \hii regions that contain no dust. Their dust surface however gets heated by the UV radiation field from the embedded young stellar object. Dust temperatures as high as \unit[600]{K} are reached.

For the dust cores we find, similar to \citet[][for Sgr B2(N)]{Qin2011} and \citet[][for the hot molecular core G10.47+0.03]{Rolffs2011b}, a Plummer exponent of $\upeta = 5.0$ reproducing the small-scale density structure well. For a comparison, we plot the observed and synthetic azimuthally averaged radial profile of each component in Fig.\ \ref{fig:radProfiles_smallScale}. The parameters of all dust cores are listed in Table \ref{tab:sgrb2_submm}. Note that for some of the dust cores, we had to include an internal heating source by placing B-type stars at the center of the core. These could be e.g. stars that do (not yet) show signs of an \hii region. This is also documented in the above mentioned Table \ref{tab:sgrb2_submm}. We also had to shift a few of the many \hii regions in Sgr B2(M) behind the dust components to produce a good fit. Please note that at frequencies higher than \unit[100]{GHz} only one high resolution dataset is available in this modeling procedure. Thus our resulting density distribution is only one possible solution. Other density distributions, e.g. dust cores with a lower exponent and a smaller radius, might yield similar good fits.

However, when comparing Sgr B2(M) and Sgr B2(N) a clear difference is noticeable. Sgr B2(M) appears to be more fragmented compared to Sgr B2(N), which appears rather monolithic. This has already been noticed by \cite{Qin2011}. The model presented here allows us, however, to quantify this difference further. We will do this especially in subsection \ref{sub:SFE}. But first we have a look at the stellar population.

\begin{table*}
  \caption{Summary of the star clusters.}
  \label{tab:starCluster} 
  \centering
  \begin{tabular}{l r r r r r r r r r r}
  \hline
  \hline
	& radius & no. stars & no. stars & M$_\ast$ & M$_\ast$ & M$_\text{gas}$  & L   & L  & L  \tablefootmark{a}\Tstrut \\
	& [pc]	& & & [\msun]	& [10$^3$ \msun]& [\msun]  &[10$^6$ \lsun]  & [10$^6$ \lsun ] & [10$^6$  \lsun] \\
	& & (initial) & (all)& (initial) & (all) & & (initial) & (all) & \Bstrut \\
  \hline
    Sgr B2(NE) 	&  0.8	&  2	&   1282 &   52 &   1.2 & 7777	& 0.54	&   0.80 & ---\Tstrut\\
    Sgr B2(N)	&  0.4	&  6	&   2642 &  150 &   2.4	& 27897	& 1.38 	&   1.80 &  1.7\\
    Sgr B2(M)	&  0.5	& 60	&  22705 & 1295 &  20.7	& 9572	& 7.78	&  12.10 & 13.0\\
    Sgr B2(S)	&  0.35	&  2	&   1204 &   50 &   1.1	& 4472	& 0.45	&   0.66 & ---\\
    Envelope   	&  --	& 19 	&   8929 &  446 &   8.0	& ---	& 3.32	&   4.70 & ---\\
  \hline  
  \hline
    Sgr B2 	& 22.5	& 89 	&  36762 & 1993 &  33.4	& 8.0$\times$ 10$^6$ & 13.47 & 20.06 & ---\Tstrut \Bstrut \\
  \hline
 \end{tabular}
 \tablefoot{\tablefoottext{a}{\citet[their models C]{Lis1990}}}
 \end{table*}

\subsection{Heating sources}

\begin{figure}[t]
    \centering
    \includegraphics[width=0.48\textwidth]{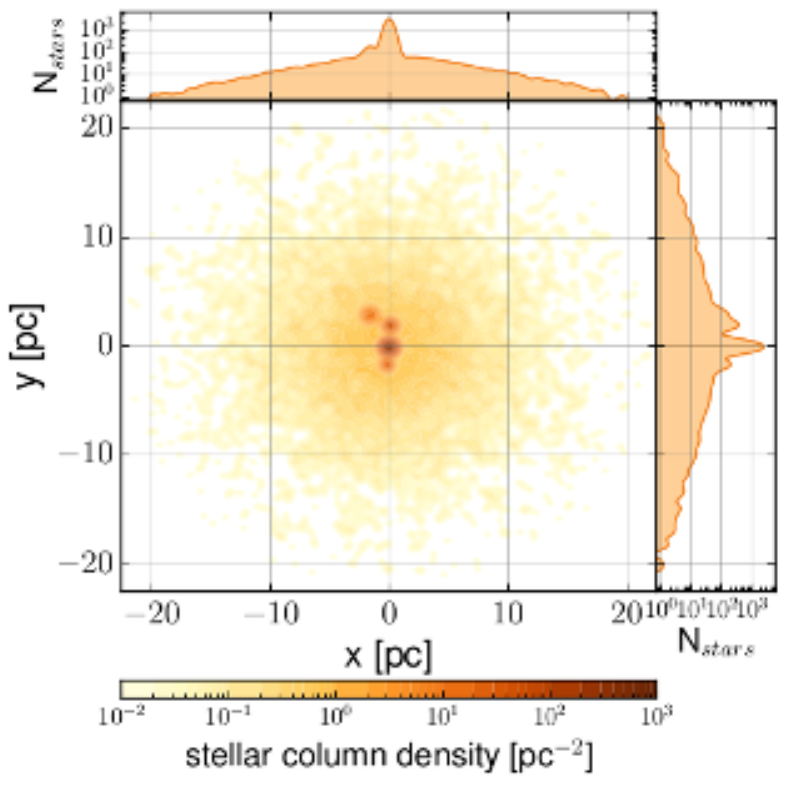}
    \caption{Center map: Stellar column density map of Sgr B2. The projected distribution of the stars along the x- and y-axis are shown in the right and top panel, respectively.}
    \label{fig:stellarDensity}
\end{figure}

\begin{figure}[t]
    \centering
    \includegraphics[width=0.49\textwidth]{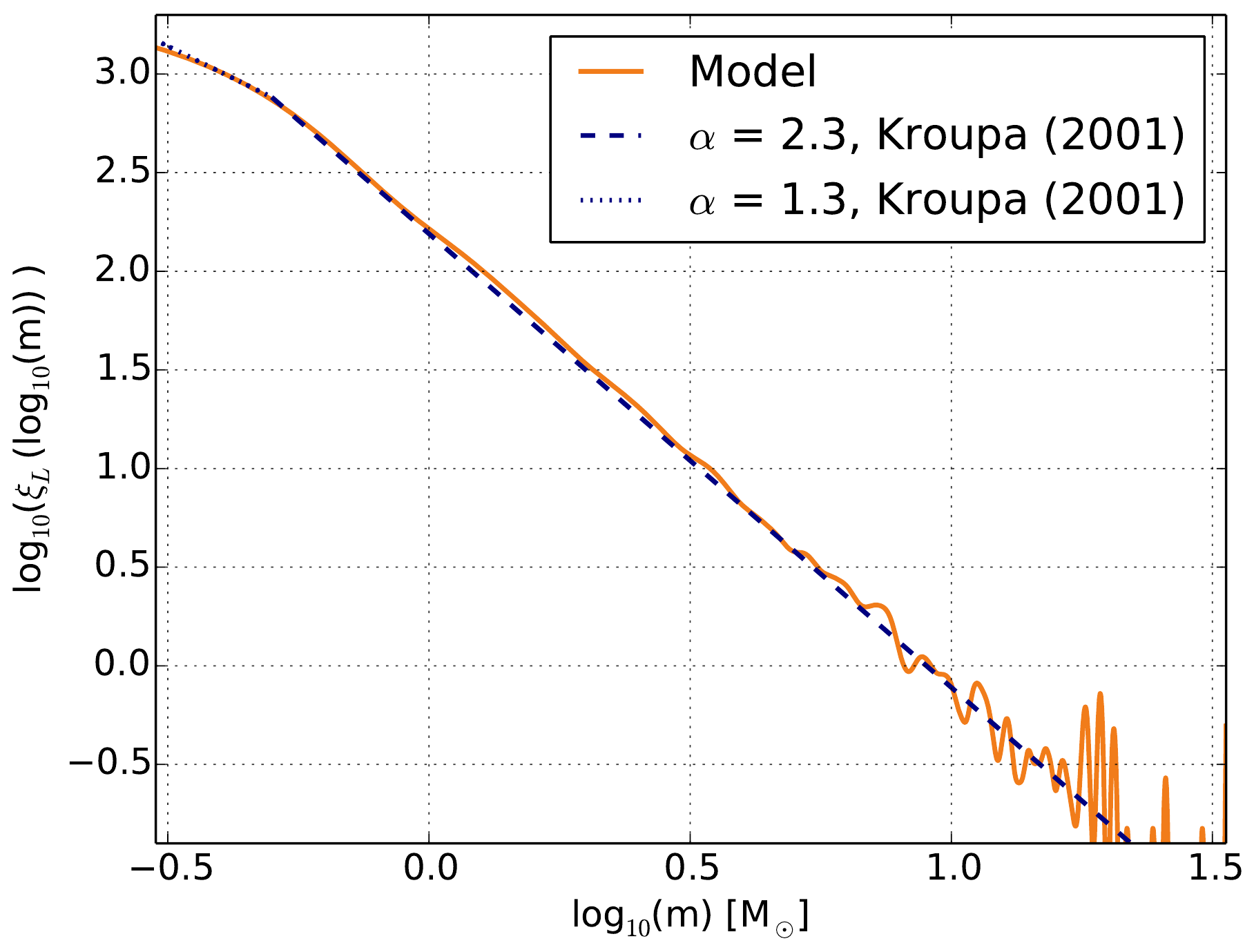}
    \caption{Mass distribution of the stellar population contained in the Sgr B2 complex, indicated by the orange line. The dotted and dashed blue lines indicate the corresponding segmented IMF from \cite{Kroupa2001}.}
    \label{fig:IMF}
\end{figure}

For all four models, we used the same stellar distribution. We account in total for four stellar clusters: Sgr B2(NE), Sgr B2(N), Sgr B2(M), and Sgr B2(S). In Fig.\ \ref{fig:starClusterRadius}, we plot for each of these clusters a radial distribution histogram of all high-mass stars placed in the model based on observed \hii regions (see Sect.\ \ref{subsec:hii}). We use these histograms to determine by-eye the star cluster radii of 0.8, 0.4, 0.5, and \unit[0.35]{pc} for Sgr B2(NE), Sgr B2(N), Sgr B2(M), and Sgr B2(S), respectively. Within these radii we sprinkle stars separately for each cluster (see Sect.\ \ref{ssec:heating}). We then treat Sgr B2 as a cluster itself and continue to sprinkle stars everywhere, except in the previously mentioned star clusters, until the calculated total stellar mass of Sgr B2 is reached. We call this part the envelope of Sgr B2. 
The stellar density distribution is plotted in Fig.\ \ref{fig:stellarDensity} and the mass distribution of the stellar mass produced with this approach (including the high-mass stars placed by hand) is shown in Fig.\ \ref{fig:IMF}. Instead of  plotting a histogram, we use a kernel density estimate. A histogram has several disadvantages. It is unsmooth and it depends on the end points as well as on the widths of the bins. By using a smooth kernel we obtain a smooth distribution that is independent of the choice of the end points of the bins and only depends on the chosen bandwidth. For our kernel density estimate a bandwidth of \unit[0.217]{\msun} has been calculated in linear space.

The results of applying this star sprinkling algorithm are listed in Table~\ref{tab:starCluster}. For each cluster we list the radius, the number of initial stars, the total number of stars,  the enclosed stellar mass, and the luminosity from the initial and all stars, respectively. Where available, we also list luminosities from \citet{Lis1990} as a reference. Additionally, this table lists the enclosed gas mass for each of the clusters. It is calculated from the dust density distribution assuming a gas-to-dust ratio of 100 \citep{Hildebrand1983}. We calculate the stellar luminosity to mass ratio as \unit[90]{\lsun \msun$^{-1}$}, \unit[71]{\lsun \msun$^{-1}$}, \unit[1176]{\lsun \msun$^{-1}$},  and \unit[120]{\lsun \msun$^{-1}$} for Sgr B2(NE). Sgr B2(N), Sgr B2(M), and Sgr B2(S), respectively.

In their appendix, \cite{Belloche2013} also estimate the stellar mass of the entire Sgr B2 region by extrapolating the already distributed stellar mass of all stars embedded in observed ultra-compact \hii regions using different IMFs. We apply the same method. However, our total stellar mass is an order of magnitude higher than their result. This difference is explained by the different initial parameters applied and different conversion tables for spectral type to stellar luminosity used.
\cite{Belloche2013} calculate a total stellar mass of \unit[675]{\msun} for the 41 ultra-compact \hii regions they account for. Their lowest mass star has a mass of \unit[11]{\msun} using the conversion table from \citet{Panagia1973}. Using Kroupa's IMF, they extrapolate a total stellar mass of \unit[$\sim$ 3900]{\msun} for the mass range of \unit[0.01 -- 120]{\msun}. In our study, we account for more than twice the number of \hii regions, of which the embedded lowest mass star has a mass of \unit[18]{\msun} using the conversion table from \citet{Vacca1996}.

\begin{figure*}[t]
    \centering
    \includegraphics[width=0.48\textwidth]{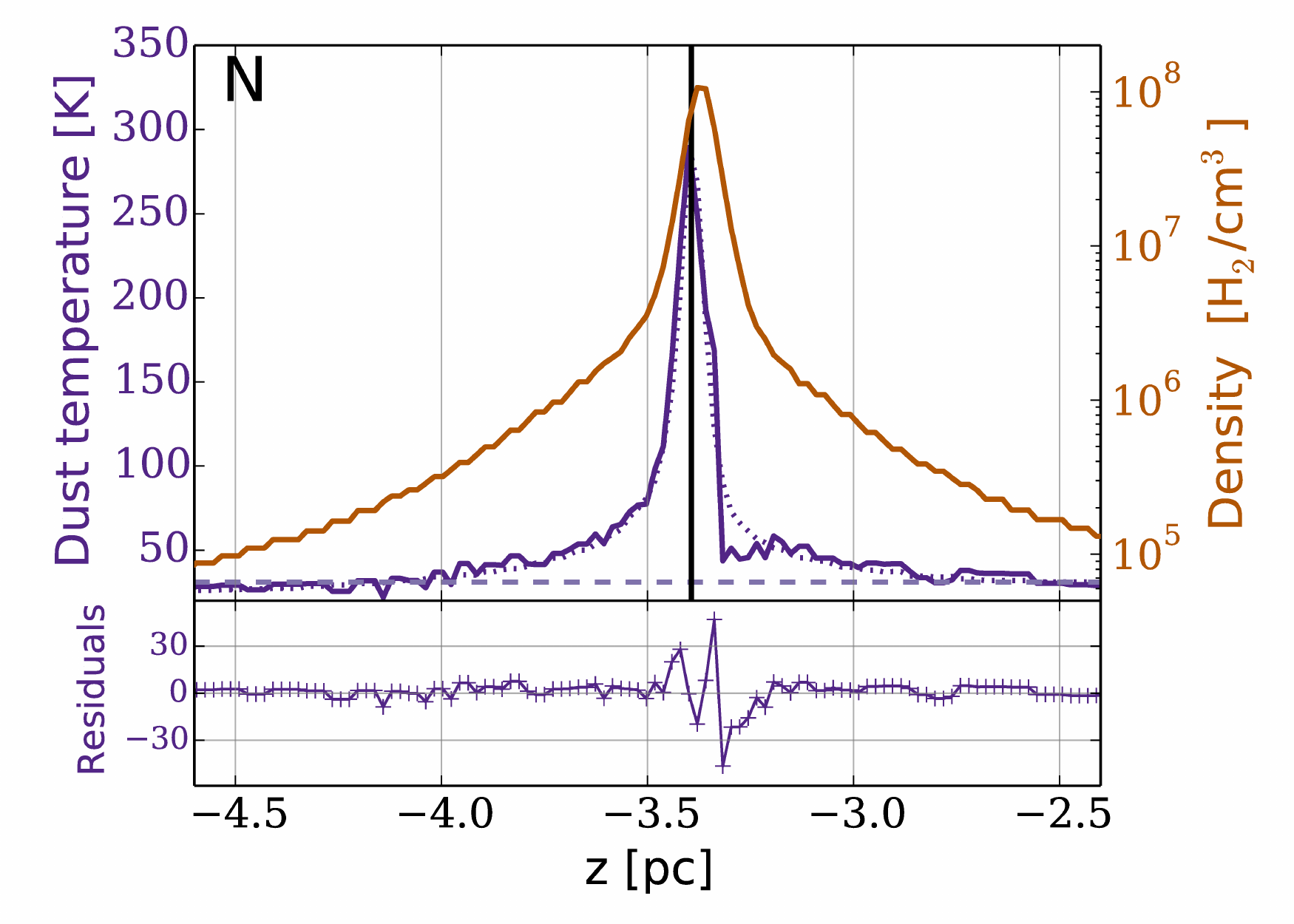}
    \includegraphics[width=0.48\textwidth]{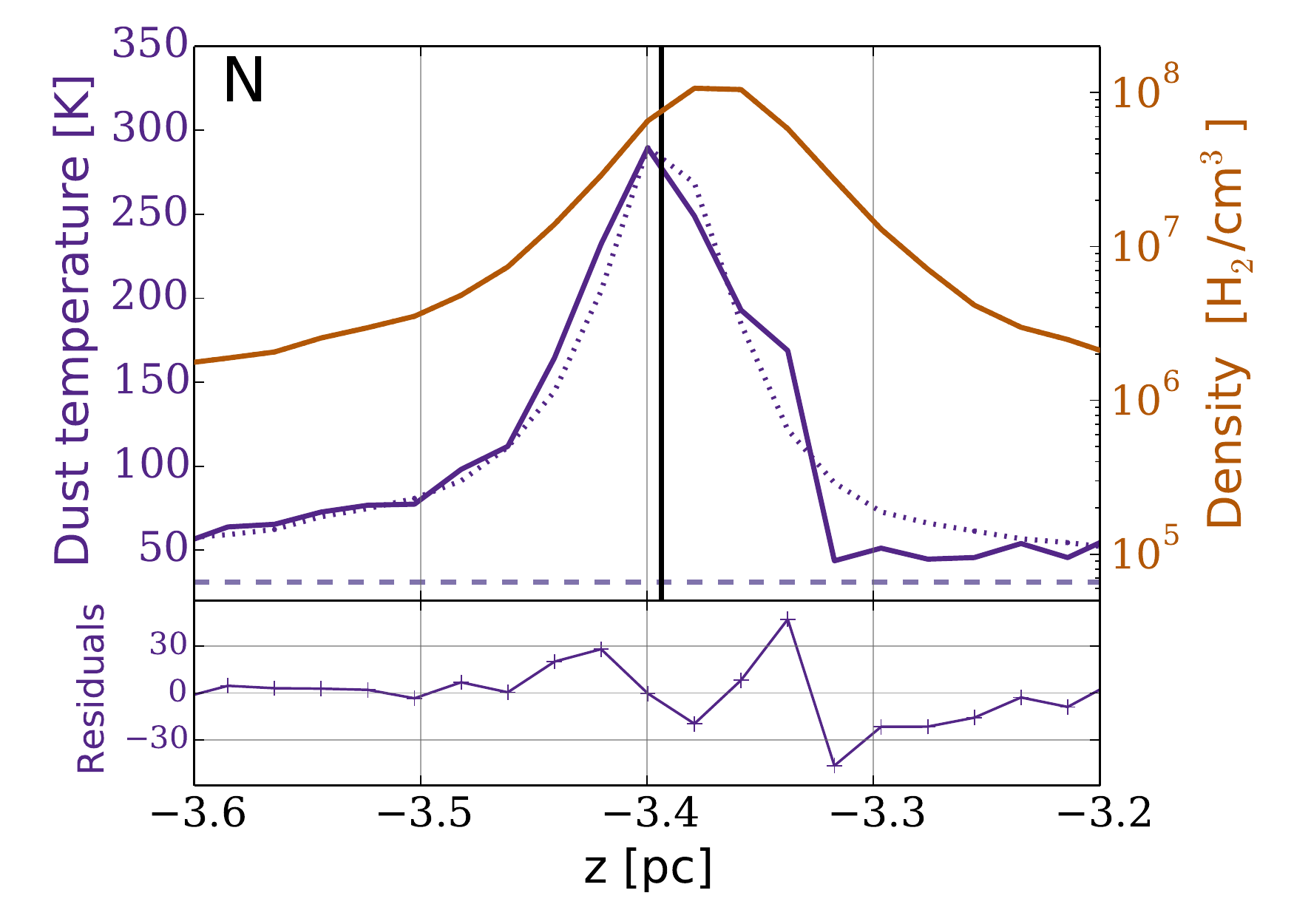}\\
    \includegraphics[width=0.48\textwidth]{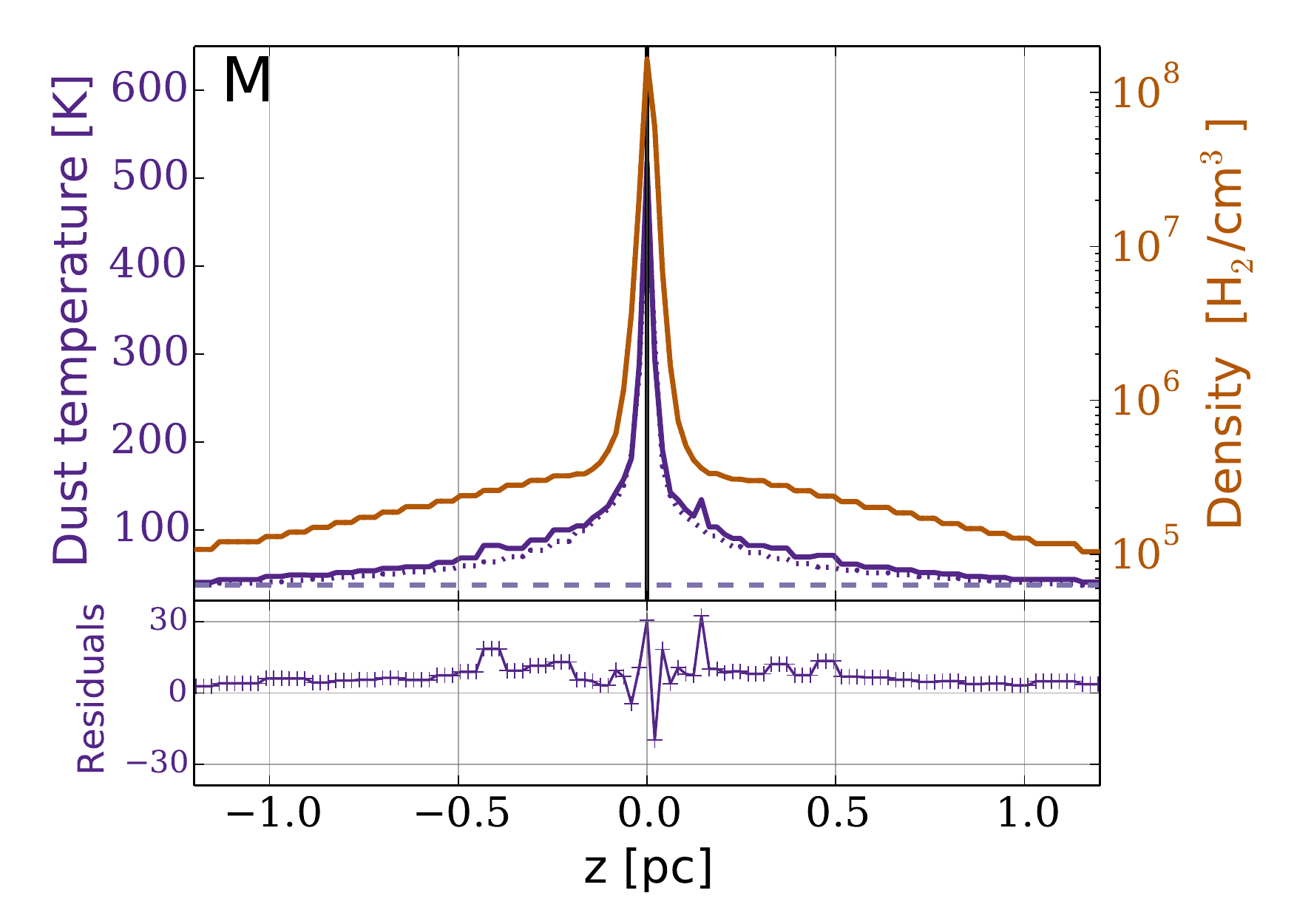}
    \includegraphics[width=0.48\textwidth]{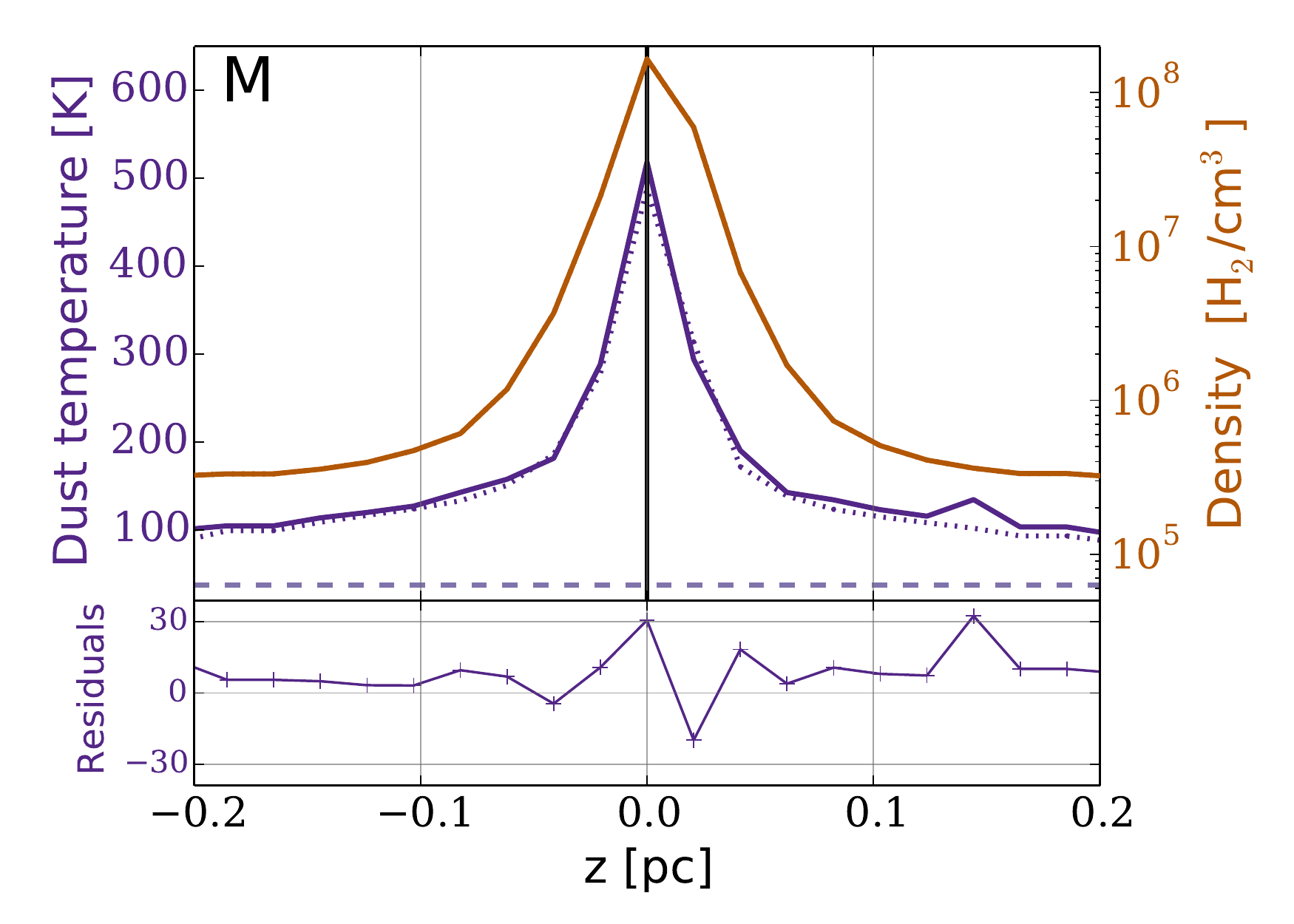}
    \caption{Density and dust temperature profile along the line-of-sight towards the two hot cores Sgr B2(N) (top) and Sgr B2(M) (bottom). This is obtained with a pixel resolution of \unit[0.5]{\arcsec}. The left panels show the profile along \unit[2]{pc} around the hot cores, whereas the panels on the right show a zoom-in view of the inner \unit[0.4]{pc}. The solid lines show the resulting distributions when including the star clusters. In addition, we show the temperature profile for the same model but without the extrapolated population of low- and intermediate-mass stars (dotted line). The bottom panel of each plot shows the residuals.}\label{fig:LOS}
\end{figure*}

We infer a large population of low- and intermediate mass stars. These low- and intermediate mass stars contain the major fraction of the stellar mass, but the high-mass stars contribute the major fraction of the luminosity. To evaluate the influence of the additional low- and intermediate mass stars, we have performed a simulation excluding these stellar populations. The intensity levels in the large-scale maps are unaffected. We then investigated the dust temperature profiles. Figure \ref{fig:LOS} shows the density and dust temperature profile along the line-of-sight towards the two hot cores. These line-of-sight profiles are obtained within a \unit[0.5]{arcsec} beam. Differences in dust temperature on the order of \unit[5-40]{K} (<10\%) are only visible on very small scales towards the cores. The envelope temperature is only affected marginally.

\subsection{Mass distribution and star formation efficiency}\label{sub:SFE}

\begin{figure}[t]
    \centering
    \includegraphics[width=0.48\textwidth]{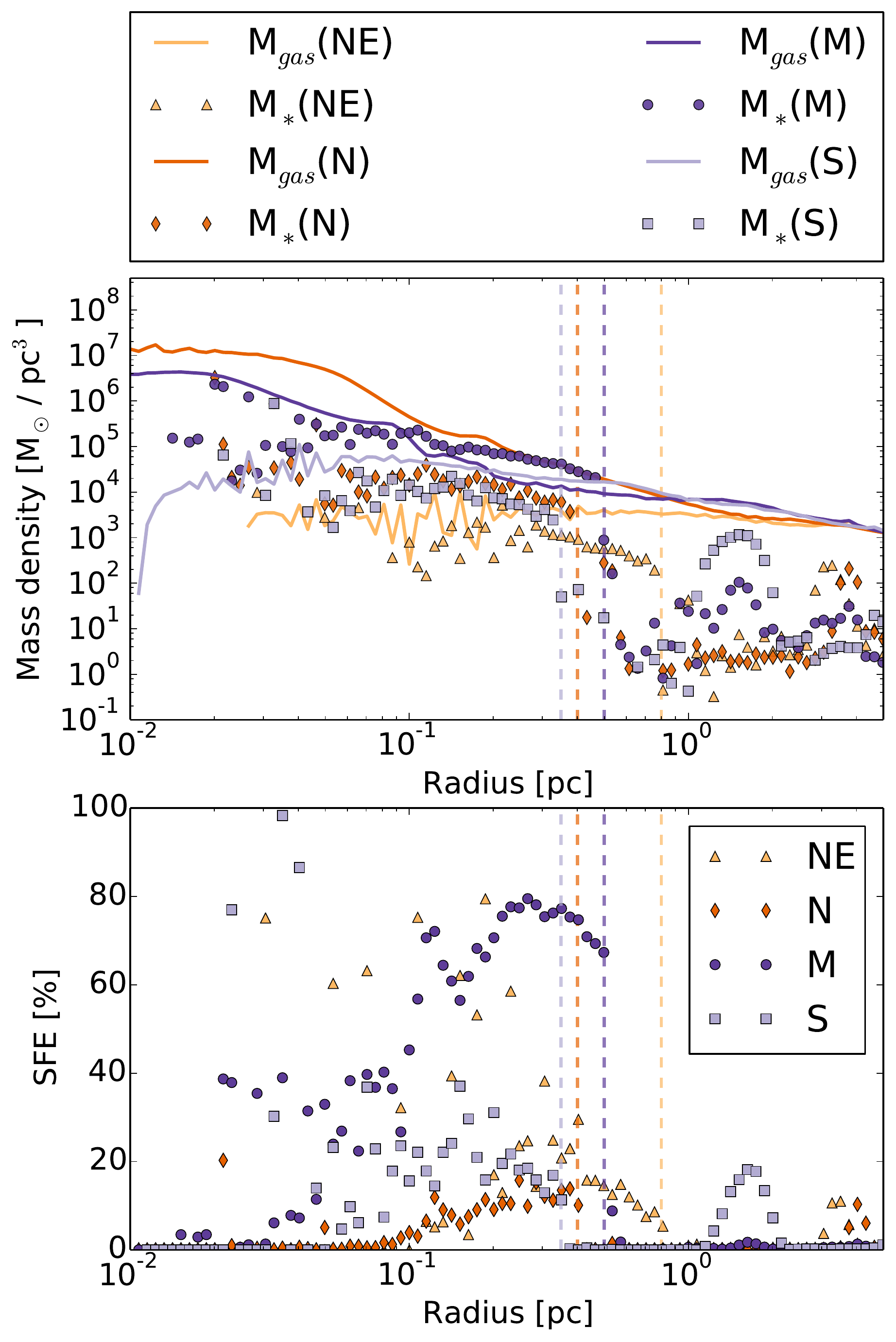}
    \caption{Top:Radial distribution of the gas and stellar mass density around Sgr B2(NE), Sgr B2(N), Sgr B2(M) and Sgr B2(S). Bottom: Calculated star formation efficiency for Sgr B2(NE), Sgr B2(N), Sgr B2(M), and Sgr B2(S). The vertical dashed lines mark the extent of each star cluster. The same colors belong to the same clusters. }\label{fig:enclosedMass}
\end{figure}

\begin{figure}
 \centering
 \includegraphics[width=.45\textwidth]{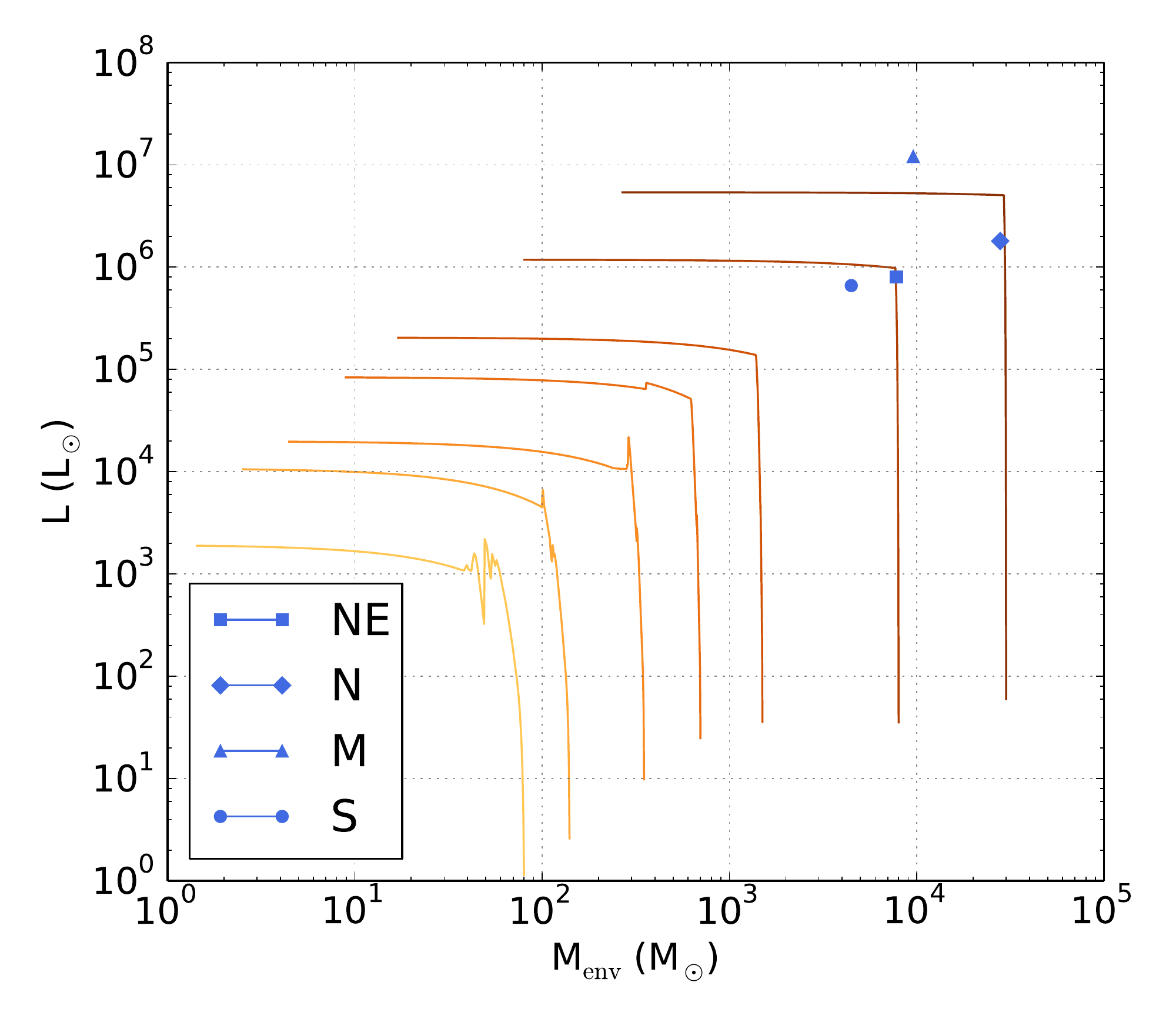}
 \caption{Luminosity-Mass diagram for the four clusters presented here. The five evolutionary tracks with the lowest initial envelope mass are taken from \citet{Molinari2008}. The two highest mass tracks are obtained by extending the evolutionary model from \citet{Molinari2008}.}\label{fig:massLuminosity}
\end{figure}

From the model we obtain the three-dimensional distribution of the stellar and gas mass. We can thus investigate the star formation efficiency (hereafter SFE), i.e. the fraction of gas that has been processed into stars. We calculate the SFE using the standard definition \citep[e.g.][]{Myers1986, Federrath2013}:
\begin{equation}
    \text{SFE} = \frac{M_\ast}{M_\ast + M_\text{gas}}
\end{equation}

\noindent where $M_\ast$ is the stellar mass and $M_\text{gas}$ is the gas mass. In Fig.\ \ref{fig:enclosedMass} we show the radial distribution of the stellar and gas mass around all four star clusters: Sgr B2(NE), Sgr B2(N), Sgr B2(M), and Sgr B2(S) as well as the derived radially resolved star formation efficiency. For Sgr B2(M) we derive star formation efficiencies of on average \unit[50]{\%}, for Sgr B2(N) the derived value is \unit[5]{\%} and thus significantly lower. 
\cite{Lada2003} suggest that the SFE of a cluster increases with time and can reach a maximum value of typically \unit[30]{\%} by the time the cluster emerges from its parental cloud core. This indicates, that Sgr B2(M) has already lost a significant portion of its gas mass through conversion into stars and dispersion. This stellar cluster is emerging from the parental cloud, whereas Sgr B2(N) is still forming stars. This agrees well with the findings by \citet{Vogel1987}. For Sgr B2(NE) as well as Sgr B2(S), we calculate rather high average star formation efficiencies of \unit[$\sim$ 30]{\%}.

These results have to be taken with caution. The SFE depends on the distributed gas and stellar mass. From the cross cuts in Fig. \ref{fig:cut_largeScale}, we think the derived gas masses on large scales are reasonable for all clusters. But while our star sprinkling algorithm will always distribute the same amount of stars if the number of initial stars and their mass is unchanged, the volume within which these stars are placed is set by the star cluster radius. A larger cluster radius effectively decreases the stellar mass density thus causing a lower SFE.

However, to interpret these results, we analyse the relationship between the gas mass of the clusters and their total luminosity. This allows us to study the evolution of the young stellar objects during the phase when they are still embedded in their parental cloud. We follow the same steps outlined in \citet{Molinari2008} and use their evolutionary tracks. In these evolutionary models, the fast accretion phase is stopped when the tracks approach the line where \hii regions are found (at a few thousands solar masses). To match the order of magnitude provided by the massive clusters studied here, we had to arbitrarily extend this line. Please note that these evolutionary models assume that a single massive star is forming, which is clearly not the case for at least Sgr B2(N) and Sgr B2(M). We present all tracks in Fig.\ \ref{fig:massLuminosity}. The five least massive tracks are the ones presented in \citet{Molinari2008}. To obtain the two most massive ones, we had to stop the accretion when the central star reaches 80 and \unit[220]{\msun}, respectively. Based on the instantaneous accretion rates along the tracks, we estimate that in reality the forming star reaches the ZAMS earlier than reaching the \hii line. In particular around \unit[50]{\msun} for the first track and \unit[75]{\msun} for the second. The track luminosity where this occurs is around \unit[4.8 $\times 10^5$]{\lsun} for the first track and around \unit[10$^6$]{\lsun} for the second track. If we assume a single massive star is forming, then these tracks suggest that Sgr B2(S) should be compatible with \hii regions (at least one \hii region is observed towards Sgr B2(S)), and Sgr B2(NE) as well as Sgr B2(N) are reaching the ZAMS now. For Sgr B2(M) we do not obtain any useful interpretation. Apart from the fact that a realistic calculation would use a stellar cluster rather than a single star, to reach this relation between gas mass and luminosity in full accretion would require producing a star of a very large mass; or else a star with much larger initial clump mass which would reach the location of M during the ZAMS evolution.

\citet{Kruijssen2015} presented an orbital structure of the Central Molecular Zone (CMZ), i.e. the central few \unit[100]{pc} of the Milky Way and derived an evolutionary timeline. According to their model, the orbital time between G0.253 (the Brick) and Sgr B2 is $\Delta t$ = \unit[0.43]{Myr}. Given the uncertainties in the ages and the age difference, we can only say that the approximate age is compatible with star formation beginning when Sgr B2 was at the position where G0.253 is now. However, the scenario is not able to shed light on the different development stages of the various sources in the region, because on the scales considered by the gaseous streams model all Sgr B2 sources are cospatial.

\subsection{Column density map and probability density function}

We convert the dust density to gas density by assuming a gas-to-dust ratio of 100 \citep{Hildebrand1983}, and that \unit[73]{\%} of the gas is in H$_2$ \citep{Cox2000}. An H$_2$ column density map of the whole cloud complex is then obtained by simply summing up the H$_2$ density distribution along the line-of-sight. The map of the full model with a resolution of \unit[0.1]{\arcsec} and zoom-ins to Sgr B2(N) and Sgr B2(M) are shown in Fig.\ \ref{fig:colDens}. The \hii regions are clearly visible in the zoom-in maps due to their lack of dust. 

The propability density function (hereafter: PDF) of the H$_2$ column density within molecular clouds is commonly used as a tool to investigate the influence of various competing star formation processes within them \citep[see e.g. ][]{Kainulainen2009, Schneider2013, Federrath2013}.

Unfortunately observational data covering the intermediate scales of Sgr B2 (see Fig.\ \ref{fig:freqScales}) are lacking. This affects the scales from \unit[$\sim$ 10 -- 20]{\arcsec}, corresponding to \unit[$\sim$ 0.4 -- 0.8]{pc}. So while the total flux on these scales is recovered in the Hi-GAL and ATLASGAL maps, its exact distribution is uncertain. We thus refrain from showing the PDF obtained from the H$_2$ column density map of the model. More observational data is needed.

\subsection{Fitting the spectral energy distribution (SED) towards N and M}\label{sec:bbfit}

\begin{figure}[t]
    \centering
    \includegraphics[width=0.48\textwidth]{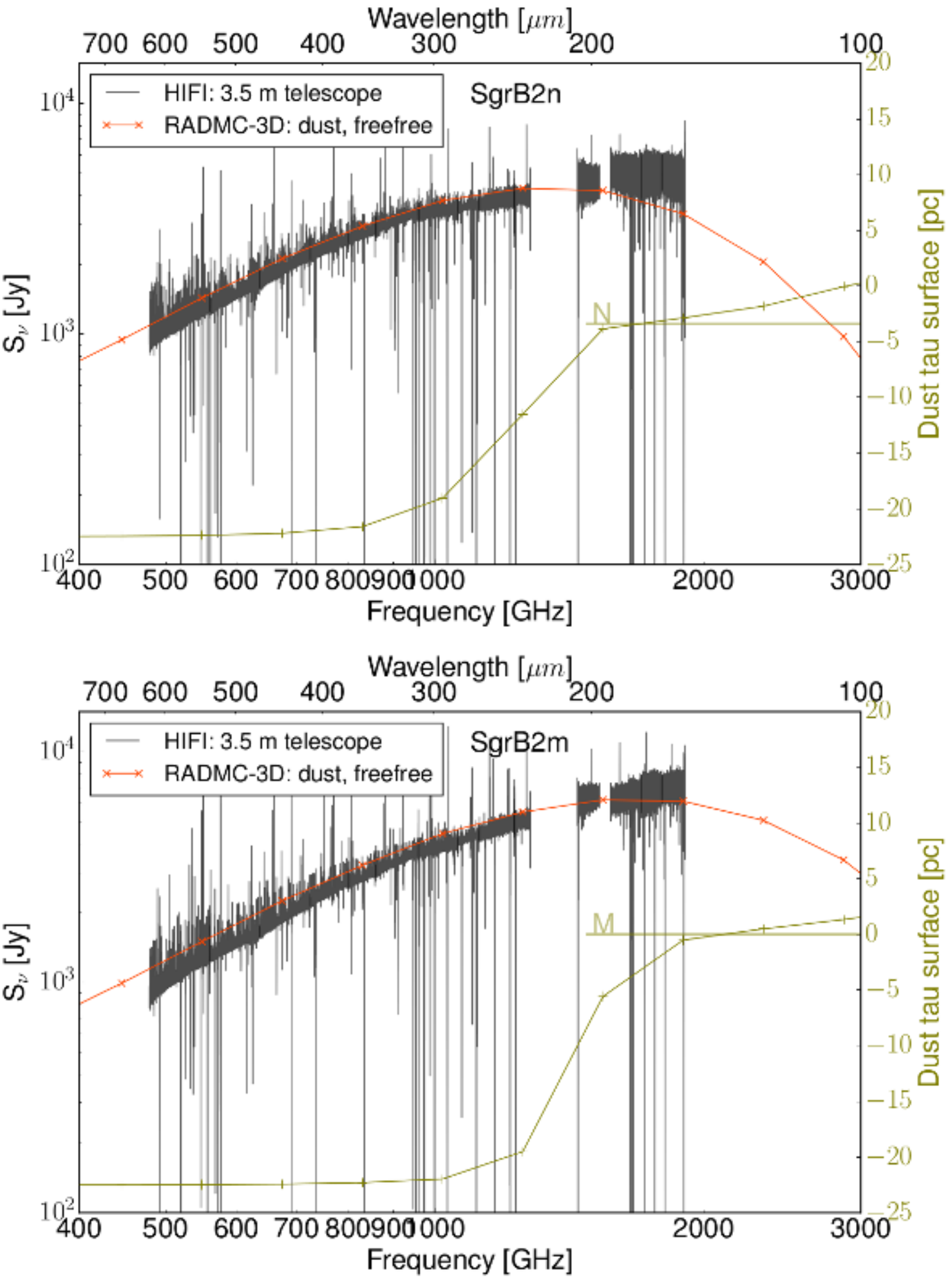}
    \caption{Spectral energy distribution of the HIFI data (in grey). The data is unsmoothed, what looks like noise are actually individual spectral lines. The best fit towards Sgr B2(M) and N are represented by the solid orange line. The synthetic intensity maps have been convolved to the frequency dependent beam of the \unit[3.5]{m} Herschel telescope. The fit includes dust and free-free emission. The surface, where the dust optical depth equals one is plotted in olive, the corresponding axis is shown on the right. The z-axis points towards the observer, Sgr B2(M) is located at z = 0 and Sgr B2(N) is located at z = \unit[7 $\times\, 10^5$]{au}. A tau surface datapoint at \unit[-22.5]{pc} indicates that the dust is optically thin at the corresponding frequency.}\label{fig:HifiSED}
\end{figure}

\begin{figure}[t]
    \centering
    \includegraphics[width=0.48\textwidth]{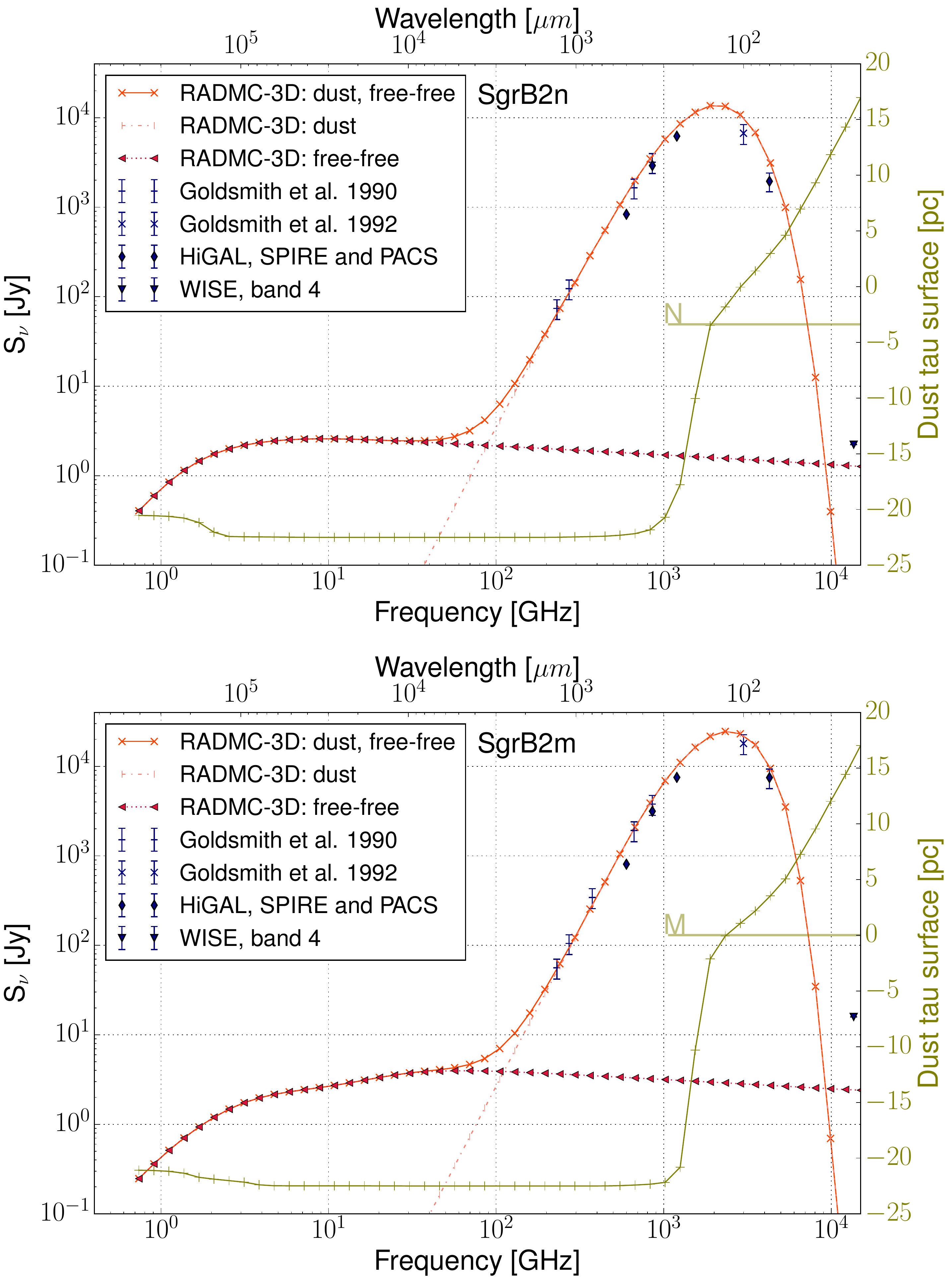}
    \caption{Spectral energy distribution plots using a fixed beamsize of \unit[30]{\arcsec}. The orange solid line is the RADMC-3D dust and free-free best fit, the orange dotted line represents the free-free contribution, the dashed-dotted line represents the contribution from dust emission. The different dark blue markers represent the observational results from \cite{Goldsmith1990, Goldsmith1992}, as well as from Hi-GAL and WISE maps. The surface, where the dust optical depth equals one is plotted in olive, the corresponding axis is shown on the right. The z-axis points towards the observer, Sgr B2(M) is located at z = 0 and Sgr B2(N) is located at z = \unit[7 $\times\, 10^5$]{au}. A tau surface datapoint at \unit[-22.5]{pc} indicates that the dust is optically thin at the corresponding frequency.}\label{fig:SEDfixedBeam}
\end{figure}

The fit of the spectral energy distribution towards Sgr B2(N) and Sgr B2(M) obtained with HIFI is shown in Fig.\ \ref{fig:HifiSED}. The fluxes from the simulated maps are convolved to the frequency-dependent beam of the 3.5 m Herschel telescope. This kind of fitting thus does not suffer from the necessary convolution to the worst spatial resolution required for modified blackbody fitting. For comparison with previous studies towards Sgr B2(M) and Sgr B2(N) \citep[e.g.][]{Goldsmith1992, Etxaluze2013}, we have convolved our continuum maps with a beam of \unit[30]{\arcsec} and extracted the flux towards both sources (see Fig.\ \ref{fig:SEDfixedBeam}). The agreement between observations and simulation in the ranges, where observational data is available is very good. We note however that the model is not able to reproduce fluxes at wavelengths shorter than \unit[70]{\micron}.

Full three-dimensional continuum radiative transfer modeling has so far been performed for disks, and $\rho$ Oph D \citep{Steinacker2005}. However, this has to our best knowledge not been attempted for Sgr B2, which limits our possibilities of comparison. So in order to compare our results to the work from other authors working on Sgr B2 \citep[e.g.][]{Lis1989, Goldsmith1990, Etxaluze2013}, we have applied the modified black body fitting technique to our synthetic maps. 

The intensity $I$ of a blackbody at frequency $\upnu$ is given by the Planck function
\begin{equation}
    I_\upnu = \frac{2\,h\,\upnu^3}{c^2} \,\left(e^{\frac{h\,\upnu}{k\,T}} -1 \right)^{-1},
\end{equation}
where $h$ is the Planck constant, $k$ is the Boltzmann constant, $c$ is the speed of light in vacuum and $T$ is the temperature of the blackbody.

This equation is multiplied by the factor $1-e^{-\uptau}$, where the optical depth $\uptau$ is given by
\begin{equation}
    \uptau = \upmu_{\text{H}_2}\, \text{m}_\text{H}\, \upkappa_\upnu\,\text{N}(\text{H}_2),
\end{equation}
$\upmu_{\text{H}_2}$ is the mean molecular weight of hydrogen, $\text{m}_\text{H}$ is the mass of hydrogen, $\text{N}(\text{H}_2)$ is the hydrogen column density. The dust opacity $\upkappa_\upnu$ is determined as a continuous function of frequency $\upnu$, by fitting a power-law to the dust opacities given by\\ \cite{Ossenkopf1994}:
\begin{equation}
    \upkappa_\upnu = \frac{\kappa_0}{\upchi_\text{d}}\,\left(\frac{\upnu}{\upnu_0}\right)^\upbeta,
\end{equation}
where $\kappa_0$ is the reference dust opacity and $\upchi_\text{d}$ is the gas-to-dust mass ratio.

For this, we have extracted the flux within a beam of \unit[40]{\arcsec} towards the SPIRE FTS pointing positions, given in \citet{Etxaluze2013}, namely \ra~=~\hms{17}{47}{20.00}, \dec~=~\ang{-28;22;17.44} (Sgr B2(N)) and \ra~=~\hms{17}{47}{20.30}, \dec~=~-\ang{28;23;04.1} (Sgr B2(M)).

\begin{table}
  \caption{Results from modified blackbody fitting of Sgr B2(N) in the upper part and Sgr B2(M) in the bottom part. For each core, we list results from three different approaches.}
  \label{table:modBBresults} 
  \centering
  \begin{tabular}{c c c c}
  \hline
  \hline
    Sgr B2(N) 					& T$_\text{d}$ 	& N(H$_2$) 		& $\upbeta$\Tstrut\\
						& [K] 	& [10$^{24}$ cm$^{-2}$] & \Bstrut\\
  \hline
    3d model\tablefootmark{a}			&  ---	& 2.6  	& ---  \Tstrut \\
    blackbody fitting\tablefootmark{b} 		& 31 	& 2.9  	& 1.8 \\
    \citet{Etxaluze2013}\tablefootmark{c}  	& 30 	& 7.0 	& 2.2 \\\\
  \hline 
  \hline
    Sgr B2(M) 					& T$_\text{d}$ 	& N(H$_2$) 		& $\upbeta$ \Tstrut\\
						& [K] 	& [10$^{24}$ cm$^{-2}$] & \Bstrut\\
  \hline
    3d model\tablefootmark{a}			&  ---	& 2.3 	& ---  \Tstrut \\
    blackbody fitting\tablefootmark{b} 		& 36 	& 2.5  	& 1.9 \\
    \citet{Etxaluze2013}\tablefootmark{c} 	& 37  	& 5.0	& 2.3 \\

 \end{tabular}
  \tablefoot{\\
	     \tablefoottext{a}{Hydrogen column density in a \unit[40]{\arcsec} beam directly taken from the model. This serves as the reference value.}\\
	     \tablefoottext{b}{Result from the modified black body fitting performed in the course of this paper.}\\
	     \tablefoottext{c}{Results obtained by \citet{Etxaluze2013}.}}
\end{table}

We apply MAGIX \citep{Moeller2013} using first the Genetic Algorithm to find the global best solution, followed by a Levenberg-Marquardt fit to obtain the local best fits for both cores, Sgr B2(M) and Sgr B2(N). The dust temperature $T_\text{d}$, the dust spectral index $\upbeta$ and the hydrogen column density $N$(H$_2$) where left as free parameters. The mean molecular weight $\upmu_{\text{H}_2}$ is assumed to be 2.8 \citep{Kauffmann2008}, the gas-to-dust mass ratio $\upchi_\text{d}$ is assumed to be 100 \citep{Hildebrand1983}, the reference dust opacity $\kappa_{500\mu m}$ is \unit[1.773]{cm$^2$g$^{-1}$} assuming no grain mantles and no coagulation \citep{Ossenkopf1994}.

\begin{figure}[t]
    \centering
    \includegraphics[width=0.48\textwidth]{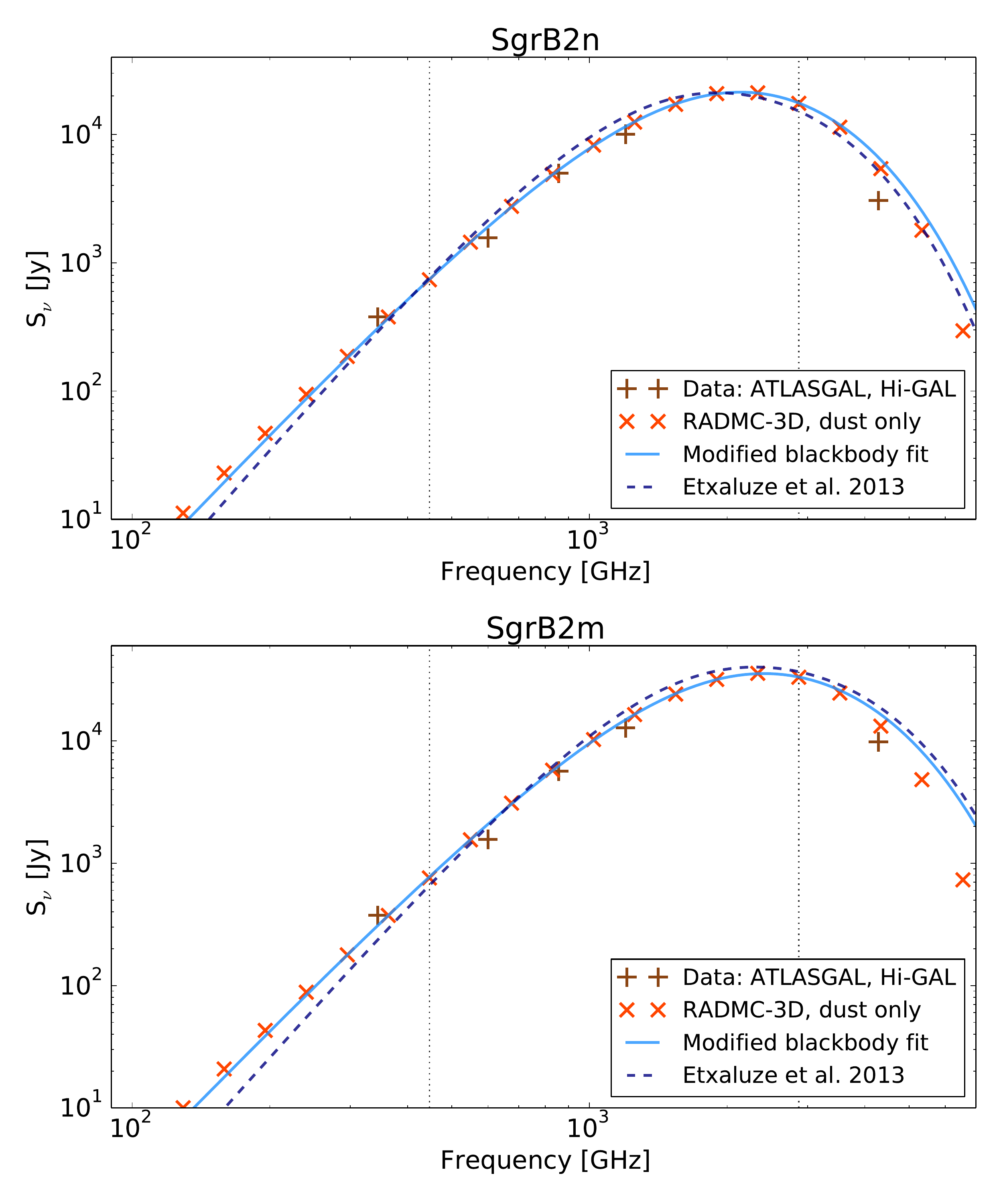}
    \caption{Spectral energy distribution from the modified blackbody fitting for a beamsize of \unit[40]{\arcsec}. The brown pluses represent the observed flux densities obtained from the ATLASGAL and Hi-GAL maps. The orange markers are the flux densities obtained with RADMC-3D including only dust. The solid blue line is the fitted SED and the dotted blue line is the SED from \cite{Etxaluze2013}. Top: Sgr B2(N), bottom: Sgr B2(M).}\label{fig:bbFit}
\end{figure}

We perform a fit covering the frequency range from \unit[450]{GHz} to \unit[2.87]{THz}, i.e. excluding the optically thick regime. The resulting SEDs are shown in Fig.\ \ref{fig:bbFit}. 
Our results are summarized in Table \ref{table:modBBresults}. For Sgr B2(N), we obtain a dust temperature $T_\text{d}$ of \unit[$\sim$ 31]{K}, a dust spectral index $\upbeta$ of 1.8 and a column density N(H$_2$) of \unit[2.9 \mal 10$^{24}$]{cm$^{-2}$}. For Sgr B2(M) we obtain a dust temperature $T_\text{d}$ of \unit[$\sim$ 36]{K}, a dust spectral index $\upbeta$ of 1.9 and a column density N(H$_2$) of \unit[2.5 \mal 10$^{24}$]{cm$^{-2}$}.

These values are the averaged values in a \unit[40]{\arcsec} beam and are thus the results for the envelopes of Sgr B2(N) and Sgr B2(M). The fitted dust temperature values correspond to radii of 2.0 and \unit[1.3]{pc} for Sgr B2(N) and Sgr B2(M), respectively (see Fig. \ref{fig:LOS}).

Recently \cite{Etxaluze2013} convolved the Herschel/SPIRE FTS spectral scan maps covering the wavelength range from \unit[194]{\micron} to \unit[671]{\micron} to \unit[40]{\arcsec} and performed  modified black body fits towards Sgr B2(N) and Sgr B2(M). They find spectral indices $\upbeta$ of 2.2 and 2.3, dust temperatures T$_\text{d}$ of \unit[30]{K} and \unit[37]{K}, and column densities N(H$_2$) of \unit[7 $\times$ 10$^{24}$]{cm$^{-2}$} and \unit[5 $\times$ 10$^{24}$]{cm$^{-2}$} for Sgr B2(N) and Sgr B2(M), respectively. Assuming a distance of \unit[8.5]{kpc}, a mean molecular weight $\upmu_{\text{H}_2}$ of 2.3, and a dust opacity of $\upkappa_{250\mu m}$ of \unit[5.17]{cm$^2$ g$^{-1}$} \citep{Li2001}, they then estimate dust masses of \unit[2500]{\msun} and \unit[2300]{\msun} and luminosities of \unit[1.1 $\times$ 10$^6$]{\lsun} and \unit[5 $\times$ 10$^6$]{\lsun} for Sgr B2(N) and Sgr B2(M), respectively. 

The derived dust temperatures from our study and the ones from \citet{Etxaluze2013} agree very well. But we obtain lower dust spectral indices for which we see two reasons. On the one hand our modeling setup is limited such that only a single dust species is considered. On the other hand different datasets are considered in both studies. We include the ATLASGAL \unit[870]{\micron} map, as well as the Hi-GAL 500, 350, 250, and \unit[70]{\micron} maps, all cross-calibrated with Planck data and in the case of the Hi-GAL map also cross-calibrated with IRAS data, whereas \citet{Etxaluze2013} only considered Hi-GAL data. This could introduce differences in the fluxes which may result in different dust spectral indices as well as different column densities. So this might also explain why we derive a factor of two difference in the dust column densities.

\begin{figure}[t]
    \centering
    \includegraphics[width=0.48\textwidth]{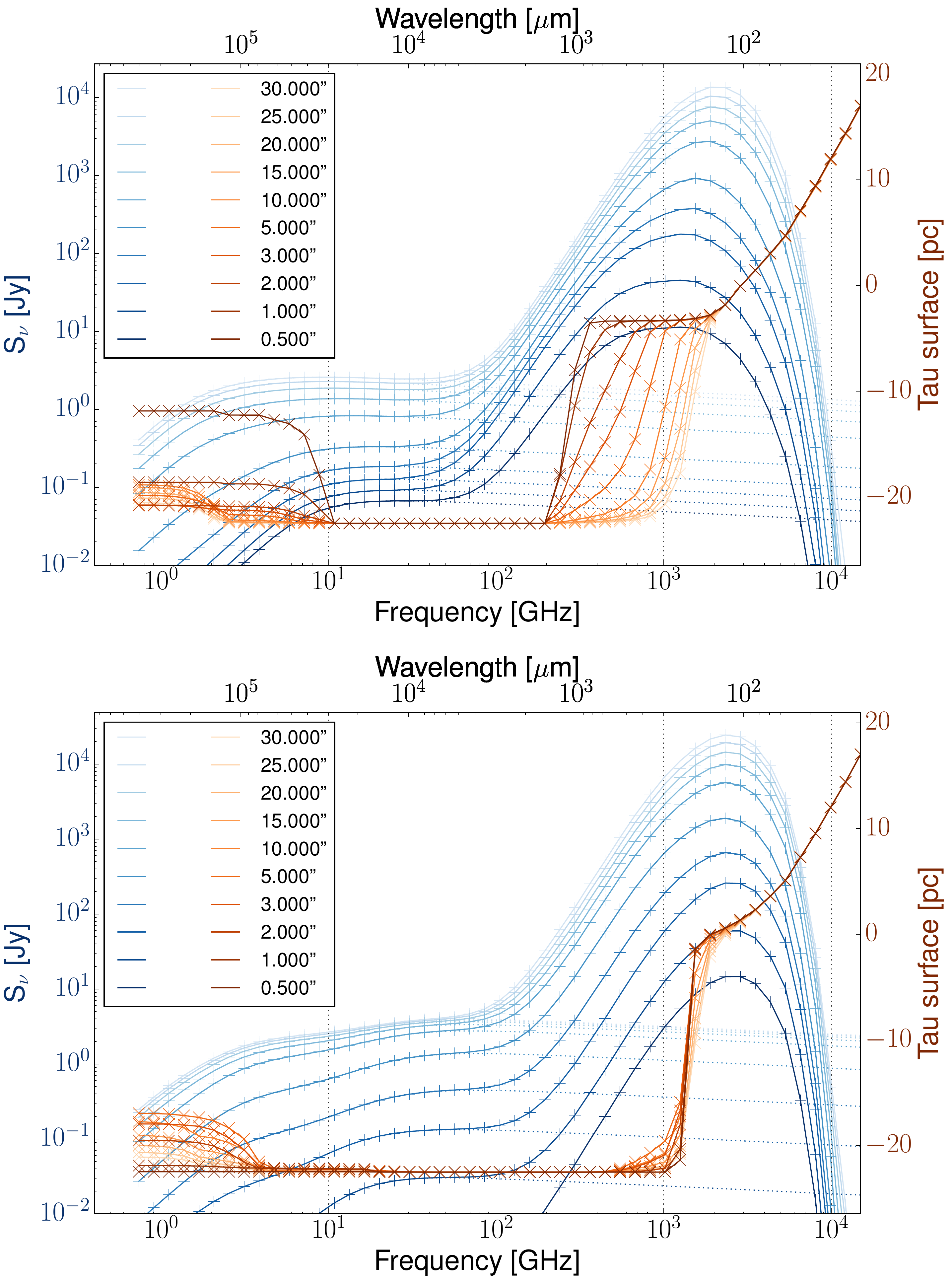}
    \caption{Top: Sgr B2(N), bottom: Sgr B2(M). The SED convolved with different beam sizes is plotted in blue. For each beam, the combined dust and free-free emission is indicated with the solid blue line, and the contributions from the free-free emission is indicated with the dotted blue line. The $\tau$ = 1 surface, i.e. the place along the line-of-sight where the optical depth $\tau$  equals one is plotted in orange. Darker colors indicate a smaller beam size.}\label{fig:optDepth}
\end{figure}

An important point is the resolution of the observation. We have calculated high resolution (i.e. the pixel size equals \unit[0.1]{\arcsec}) synthetic intensity maps covering the same frequency range as the SED plotted in Fig.\ \ref{fig:HifiSED} and Fig.\ \ref{fig:SEDfixedBeam}. We have then convolved these images with decreasing beamsizes from \unit[30]{\arcsec} down to \unit[0.5]{\arcsec}. We extract the SEDs towards the Herschel/HIFI positions of Sgr B2(N) and Sgr B2(M) as specified above for each beamsize and also extract the respective $\uptau$ = 1 surface at each wavelength. This is shown in Fig.\ \ref{fig:optDepth}. From these two SEDs, we can clearly see, that towards the position of Sgr B2(N) the free-free emission becomes optically thicker for smaller beamsizes. This indicates that an \hii region is located along the line-of-sight. Furthermore the dust also becomes optically thick towards Sgr B2(N) with decreasing beamsize, leaving a very narrow frequency range of \unit[100-200]{GHz}, where the dust is optically thin while still dominating over the free-free emission. The contribution of free-free emission at \unit[100]{GHz} is in the order of \unit[30]{\%}, which is still a significant fraction of the overall emission at this frequency. For Sgr B2(M) the dust remains optically thin for this specific position up to \unit[600]{GHz}. However, at \unit[100]{GHz}, the fraction of free-free emission is in the order of \unit[70]{\%} for Sgr B2(M), indicating that the free-free emission still dominates at this frequency. Looking at the short-frequency range it is interesting to note that the free-free is partly optically thick for intermediate beamsizes, but turns optically thin for small beamsizes. This shows that there is no \hii region located along the line-of-sight towards the position of the Herschel/HIFI beam on small scales. Given that there are a lot of \hii regions located along the line-of-sight towards Sgr B2(M) this shows that the pointing choice might be crucial for high-resolution observations. It furthermore shows that if instead of a smooth dust density distribution we would assume a clumpy distribution, this clumping would change the photon penetration depth at certain sightlines.

\section{Conclusions}

\label{s-conclusions}

We successfully reconstruct a possible three dimensional density distribution of Sgr B2, recovering the continuum structures covering a wide frequency range ($\upnu$ = \unit[40]{GHz} -- \unit[4]{THz}) on scales from \unit[100]{au} to \unit[45]{pc}.  We employ the publicly available three-dimensional radiative transfer program RADMC-3D and calculate the dust temperature self-consistently. 

\begin{itemize}
    \item We find that the density field of Sgr B2 can be reasonably well fitted by a superposition of spherical symmetric density cores with Plummer-like profiles. 
    \item To reproduce the spectral energy distribution, we locate Sgr B2(N) along the line of sight behind the plane containing Sgr B2(M).
    \item Sgr B2 comprises a total gas mass of \unit[8.0 $\times$ 10$^6$]{\msun} within a diameter of \unit[45]{pc}. This corresponds to an average gas density of \unit[170]{\msun pc$^{-3}$}. For Sgr B2(N) we find a stellar mass of \unit[2400]{\msun}, a luminosity of \unit[1.8~$\times$~10$^{6}$]{\lsun}, a H$_2$ column density of  \unit[2.9.7~$\times$~10$^{24}$]{cm$^{-2}$} in a \unit[40]{\arcsec} beam. For Sgr B2(M) we derive a stellar mass of \unit[20700]{\msun}, a luminosity of \unit[1.2~$\times$~10$^7$]{\lsun}, a H$_2$ column density of  \unit[2.5~$\times$~10$^{24}$]{cm$^{-2}$} in a \unit[40]{\arcsec} beam.  For Sgr B2(S) we find a stellar mass of \unit[1100]{\msun}, a luminosity of \unit[6.6~$\times$~10$^{5}$ ]{\lsun}, a H$_2$ column density of  \unit[2.2~$\times$~10$^{24}$]{cm$^{-2}$} in a \unit[40]{\arcsec} beam.
    \item The calculated star formation efficiency is very low for Sgr B2(N), \unit[$\sim$5]{\%}, and much higher for Sgr B2(M), \unit[$\sim$50]{\%}. This indicates that most of the gas in Sgr B2(M) has already been converted to stars or dispersed.
\end{itemize}

\section{Outlook}\label{subs-outlook}
 
 The setup of the density structure and temperature field presented here opens the stage of a plethora of different applications and improvements. We list some of them 

\begin{figure*}[t]
    \centering
    \includegraphics[height=0.87\textheight]{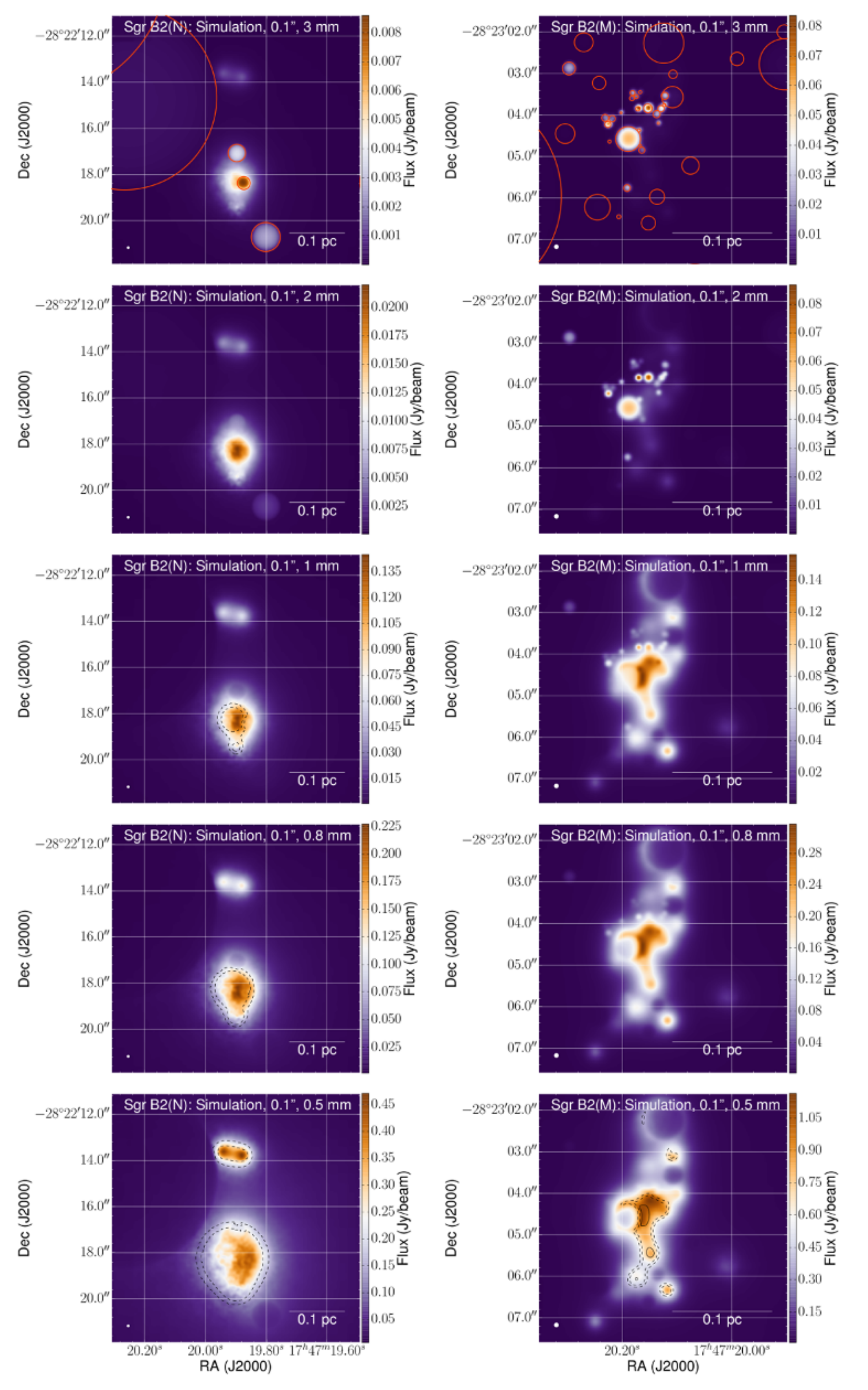}
    \caption{Simulation of high-resolution continuum maps towards Sgr B2(N), left,  and Sgr B2(M), right with a resolution of \unit[0.1]{\arcsec} at decreasing wavelengths from top to bottom. Observations like this are possible with ALMA. The red circles in the maps in the top row indicate the location and extent of the \hii regions. Maps having black dashed (and solid) contours are affected by optical thickness. The black dashed contours in maps of Sgr B2(N) indicate where the $\tau$ = 1 surface equals -3.50, and \unit[-3.39]{pc}, respectively. The black (dashed and solid) contours in maps of Sgr B2(M) show where the $\tau$ = 1 surface equals -1.0, -0.05, and \unit[0.0]{pc}, respectively. These maps indicate that, especially for Sgr B2(M), the free-free emission contributes significantly at \unit[3 and 2]{mm}. On the other hand, the dust becomes optically thick at \unit[$\sim$0.5]{mm} for Sgr B2(M), and already at \unit[$\sim$1]{mm} for Sgr B2(N).}\label{fig:alma}
\end{figure*}

\begin{itemize}
    \item \textbf{High resolution, envelope covering maps} With the current state of telescopes, it is now possible to efficiently map larger portions of the sky at high resolution. As indicated in Fig. \ref{fig:freqScales}, maps with high angular resolution of Sgr B2 are missing. However they are necessary to improve the setup of the envelope. Currently there is for example only one map available, that resolves dust structures down to \unit[$\sim$ 0.01]{pc} towards Sgr B2(N) and Sgr B2(M). To constrain the dust properties properly, at least another map covering similar scales at a different wavelength is needed. Studying spatial variations of the dust properties on scales smaller than \unit[10]{$''$} in the entire envelope is impossible with the data sets currently available. However, this could be partly achieved by using the Atacama Compact Array (ACA).
    
    \item \textbf{Physical setup:} The physical setup of the model presented here sets the stage for future improvements. By including the possibility to spatially vary the composition of dust, the influence of different compositions can be tested. The treatment of the \hii regions can be improved, eventually leading the way to also model radio recombination lines and thus getting access to the ionized gas content. Furthermore the effect of a clumpy density structure can be tested as well as the influence of dusty \hii regions.
    
    \item \textbf{Molecular line studies:} Ultimately, one wants to model molecular lines. The setup of the density and temperature structure presented here builds the foundation for such follow-up studies. By keeping the density structure fixed, the remaining main free parameters are the molecular abundance and the velocity field. By modeling a variety of different species, including their isotopologues, different surfaces where the opacity exceeds unity can be sampled, allowing a tomography of the source. Having one model for the whole region, also requires to find a solution of the velocity field that fits the plethora of available molecular line data.
    
    \item \textbf{(M)HD simulations:} While the work presented here focuses only on the modeling of a specific source - Sgr B2 - it builds a bridge between theoretical work, e.g. from (M)HD simulations, and observational work. 
    
    \item \textbf{Chemical models:} Combining the radiative transfer modeling efforts with sophisticated chemical modeling, would allow to fix the molecular abundances.
    
    \item \textbf{Planning observations:} The model enables us to examine the full parameter space, finding combinations of e.g. transitions, molecules, wavelengths that would allow to constrain a degenerate parameter observationally. Thus it is a useful tool to plan new observations.   In Fig.\ \ref{fig:alma} for example we show a prediction of high-resolution continuum maps at different wavelengths. They could be observed e.g. with ALMA. These maps indicate that in order to search for fragmentation and determine its level, it is necessary to obtain multi-wavelength observations as there is no single wavelength unaffected by optical thickness effects or free-free continuum contribution.
 
 \end{itemize}

\begin{acknowledgements}
We thank the anonymous referee for insightful comments that greatly improved this paper. We furthermore thank C. de Pree for providing the VLA continuum maps presented in \citet{Gaume1995, dePree1998}.

This research is carried out within the Colla\-bo\-ra\-tive Re\-search Centre 956, sub-project A6, funded by the Deutsche For\-schungs\-ge\-mein\-schaft (DFG). DCL acknowledges support for this work provided by NASA through an award issued by JPL/Caltech. SLQ is partly supported by the NSFC under grant Nos. 11373026, 11433004, by the Top Talents Program of Yunnan Province.

This research has made use of NASA's Astrophysics Data System, Astropy, a community-developed core Python package for Astronomy \citep{AstropyCollaboration2013}, APLpy, an open-source plotting package for Python hosted at http://aplpy.github.com, and the SIMBAD database, operated at CDS, Strasbourg, France. 

Herschel is an ESA space observatory with science instruments provided by European-led Principal Investigator consortia and with important participation from NASA. HIFI has been designed and built by a consortium of institutes and university departments from across Europe, Canada and the United States under the leadership of SRON Netherlands Institute for Space Research, Groningen, The Netherlands and with major contributions from Germany, France and the US. Consortium members are: Canada: CSA, U.Waterloo; France: CESR, LAB, LERMA, IRAM; Germany: KOSMA, MPIfR, MPS; Ireland, NUI Maynooth; Italy: ASI, IFSI-INAF, Osservatorio Astrofisico di Arcetri-INAF; Netherlands: SRON, TUD; Poland: CAMK, CBK; Spain: Observatorio Astron\'omico Nacional (IGN), Centro de Astrobiología (CSIC-INTA). Sweden: Chalmers University of Technology - MC2, RSS \& GARD; Onsala Space Observatory; Swedish National Space Board, Stockholm University - Stockholm Observatory; Switzerland: ETH Zurich, FHNW; USA: Caltech, JPL, NHSC.

The ATLASGAL project is a collaboration between the Max-Planck-Gesellschaft, the European Southern Observatory (ESO) and the Universidad de Chile. It includes projects E-181.C-0885, E-078.F-9040(A), M-079.C-9501(A), M-081.C-9501(A) plus Chilean data. 

The National Radio Astronomy Observatory is a facility of the National Science Foundation operated under cooperative agreement by Associated Universities, Inc.

The Submillimeter Array is a joint project between the Smithsonian Astrophysical Observatory and the Academia Sinica Institute of Astronomy and Astrophysics and is funded by the Smithsonian Institution and the Academia Sinica.

\end{acknowledgements}

\bibliographystyle{aa.bst}



\clearpage

\onecolumn
\begin{appendix}

%
%
\section{Figures}\label{s-figures}

\begin{figure*}[h]
    \centering
    \includegraphics[width=.78\textwidth]{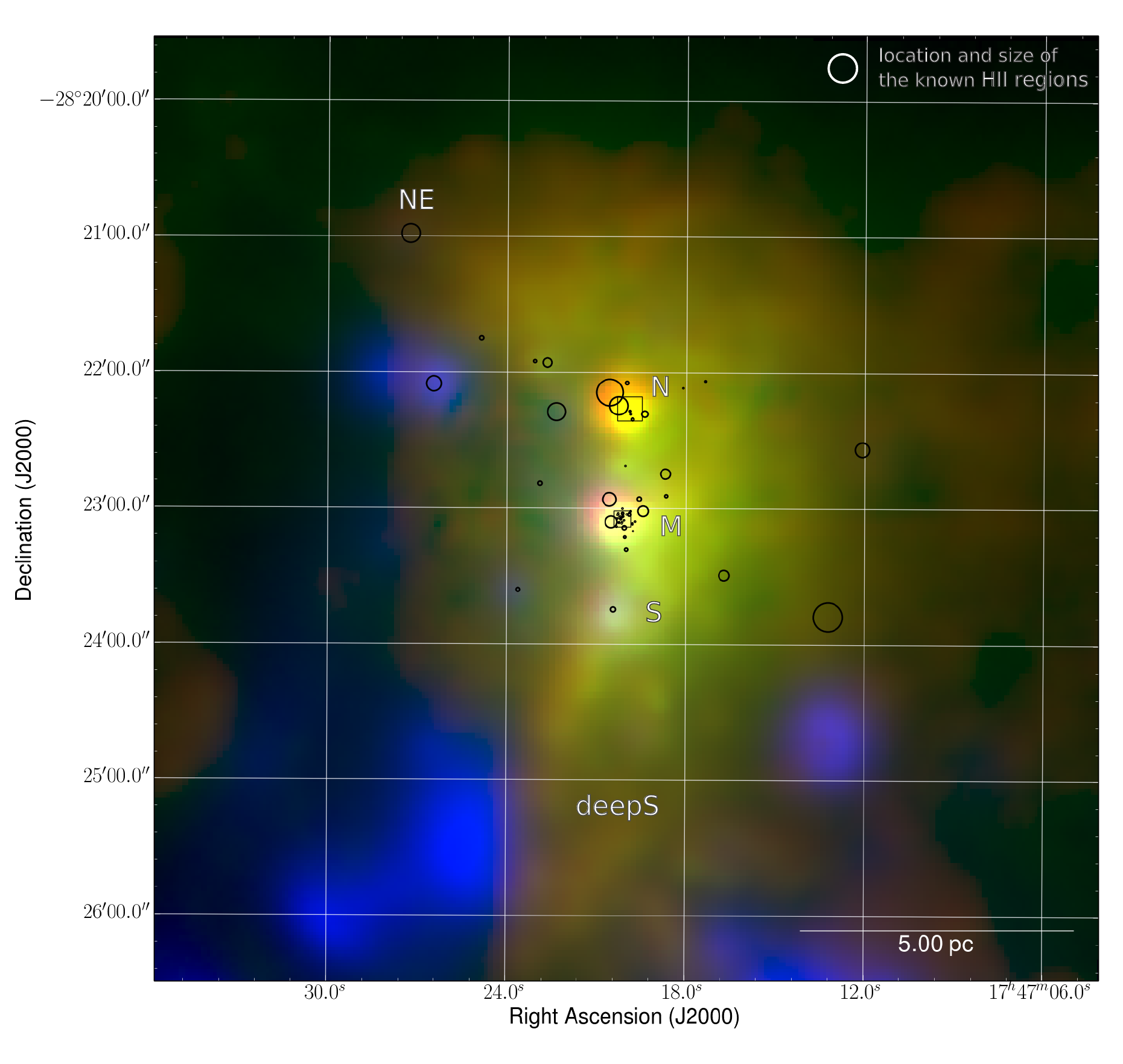}\\
    \includegraphics[width=.8\textwidth]{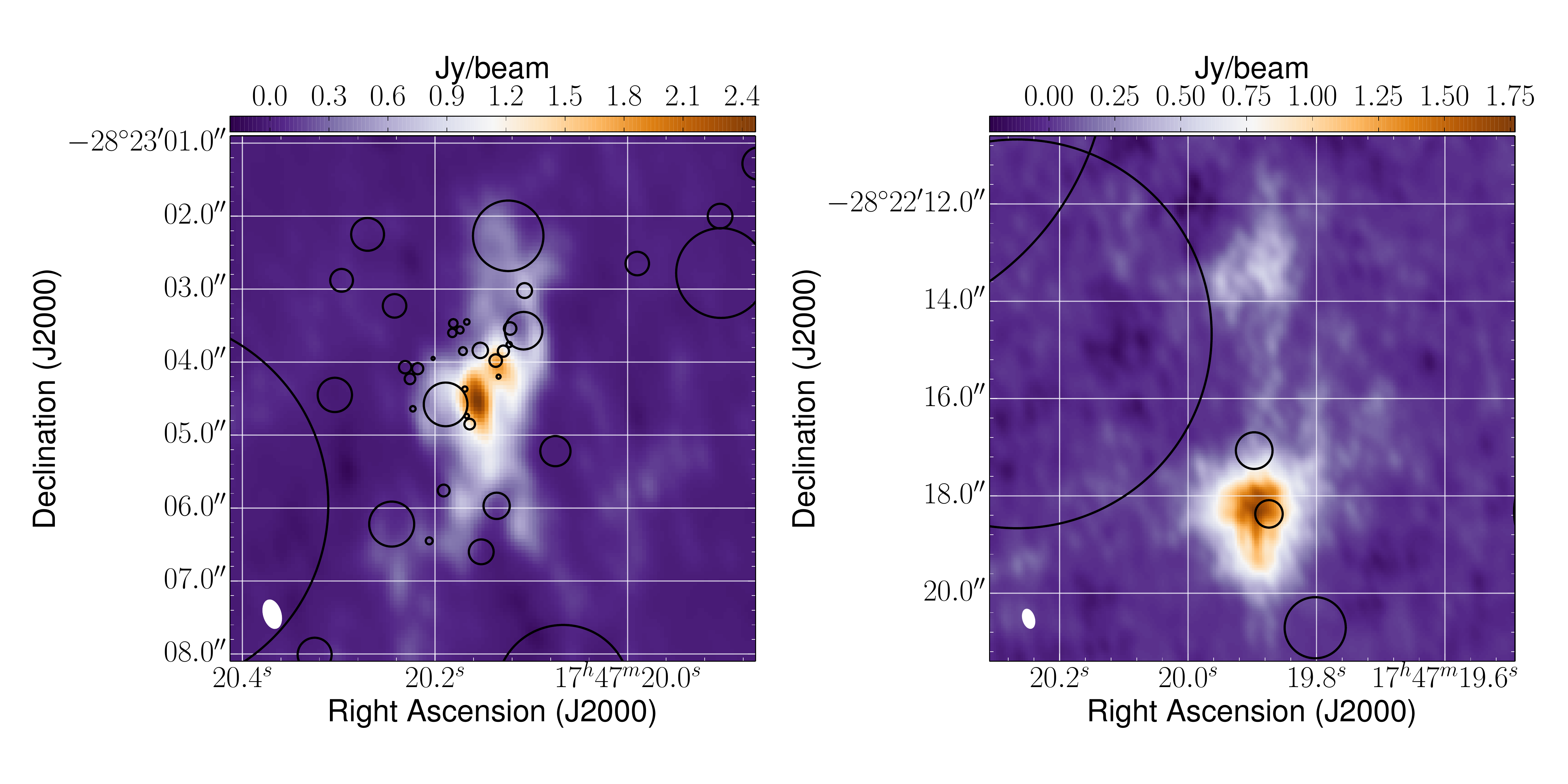}
    \caption{Top: Three color composite map of the large-scale structure of Sgr B2. Blue: JCMT - SCUBA \unit[850]{\micron}, green: CSO -- Sharc II  \unit[350]{\micron}, red: Herschel -- PACS \unit[70]{\micron}. Bottom left: Zoom-in to Sgr B2(M), SMA data. Bottom right: Zoom-in to Sgr B2(N), SMA data. The black circles mark the extent of the \hii regions. The rectangles in the upper map indicate the zoom-in region of the bottom maps.}\label{fig:largeScaleMap}
\end{figure*}

\begin{figure*}[h]
    \centering
    \includegraphics[width=.85\textwidth]{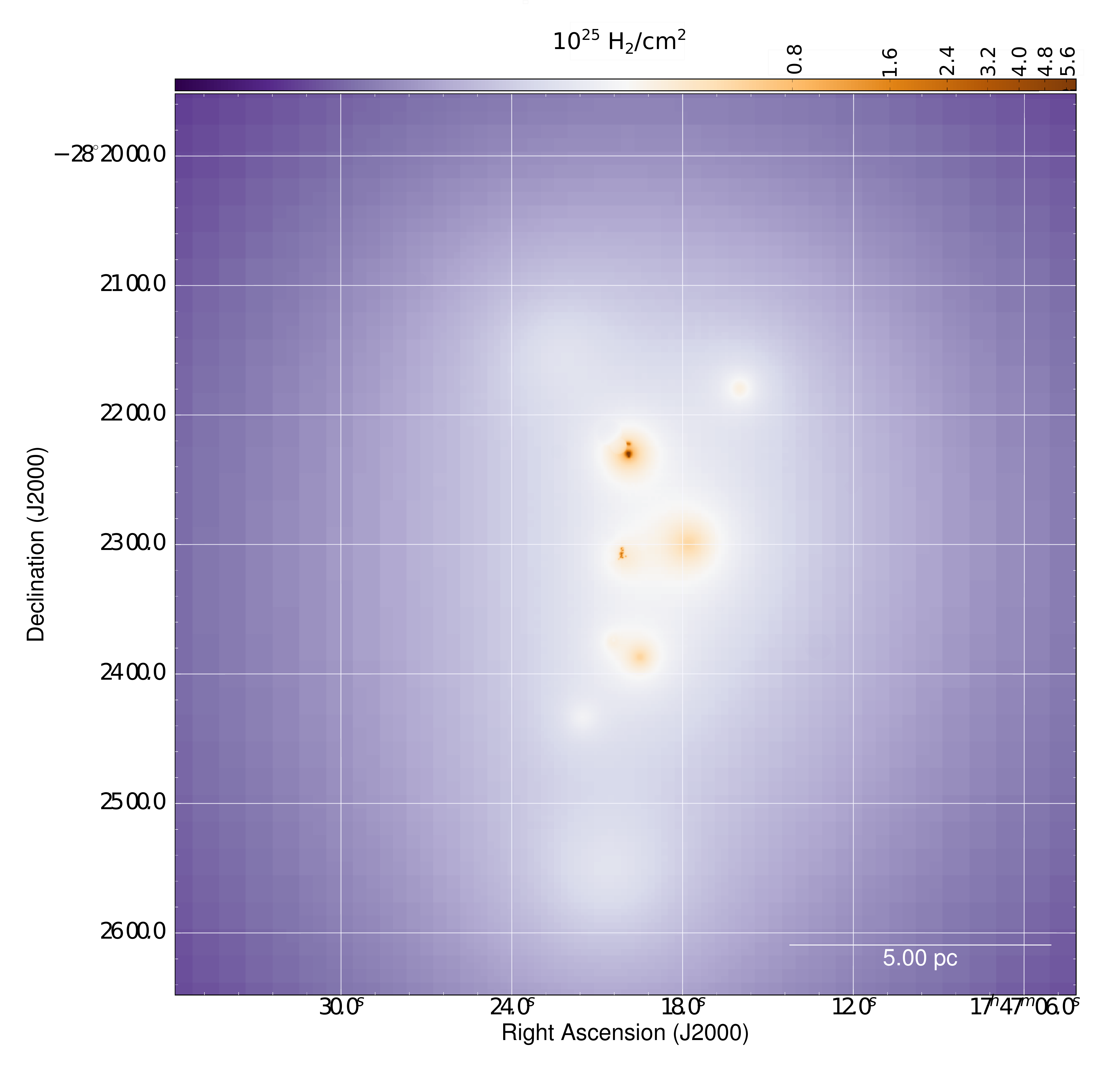}\\
    \includegraphics[width=.85\textwidth]{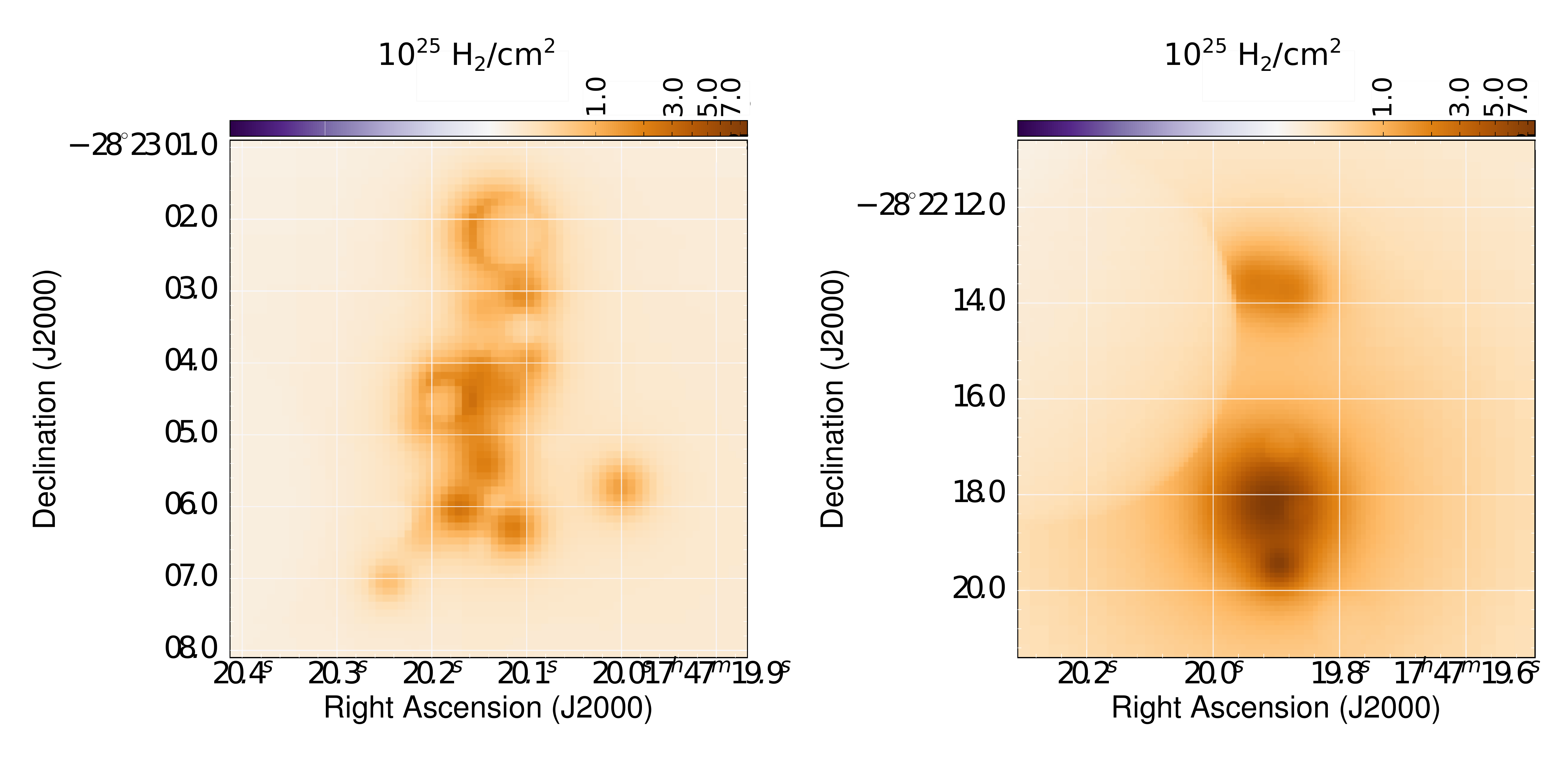}
    \caption{Top: Hydrogen column density map of the whole cloud complex. The pixel resolution is \unit[0.5]{arcsec}. Bottom left: Zoom-in to Sgr B2(M). Bottom right: Zoom-in to Sgr B2(N). The \hii regions are visible due to their lack of dust.}\label{fig:colDens}
\end{figure*}

\begin{figure*}[h]
    \centering
    \includegraphics[width=.9\textwidth]{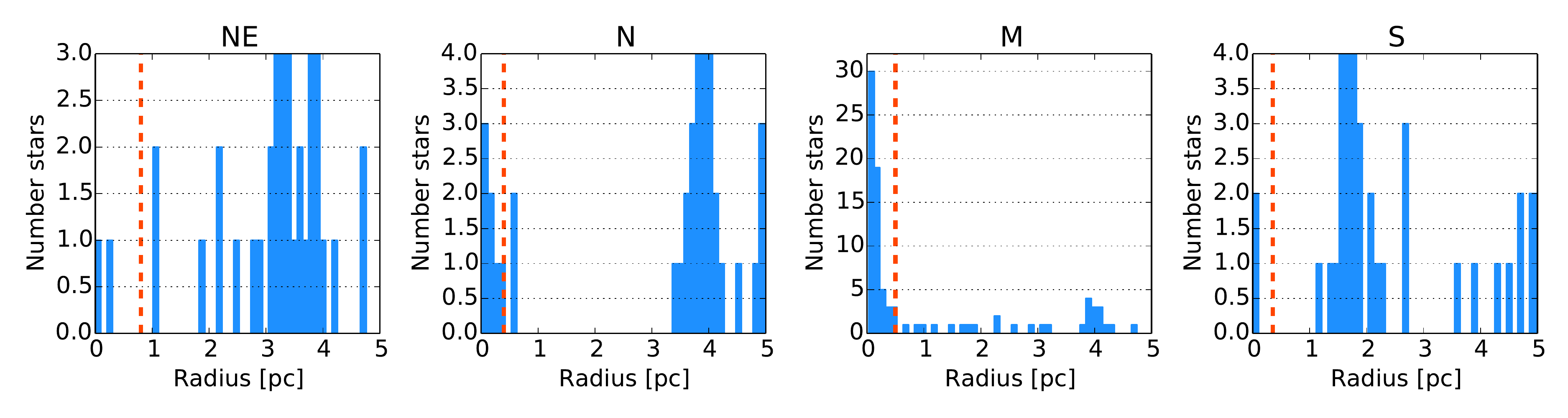}
    \caption{Histogram of the known high-mass stars around the star clusters to determine the star cluster radius. The dashed line indicates the star cluster radius determined based on a by-eye inspection of these histograms.}\label{fig:starClusterRadius}
\end{figure*}

\begin{figure*}[h]
    \centering
    \includegraphics[width=.98\textwidth]{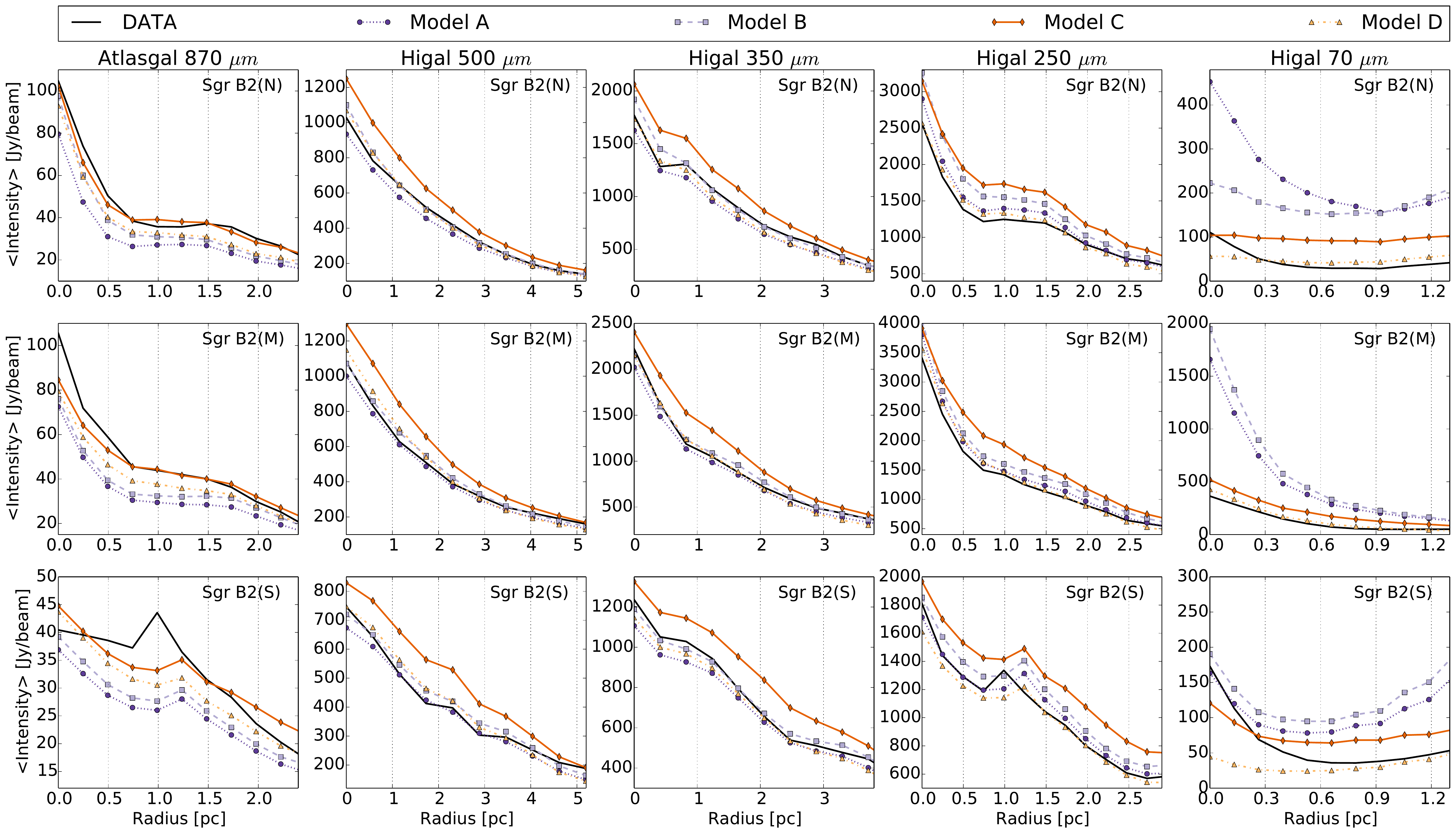}
    \caption{Azimuthally averaged radial profiles of the large-scale structure around the position of the envelope components of Sgr B2(N) (top row), Sgr B2(M), (middle row) and Sgr B2(S) (bottom row). The data is plotted in solid black. The wavelengths decreases from left to right: \unit[870]{$\upmu$m}, \unit[500]{$\upmu$m}, \unit[350]{$\upmu$m}, \unit[250]{$\upmu$m}, and \unit[70]{$\upmu$m}.}\label{fig:radProfiles_largeScale}
\end{figure*}

\begin{figure*}[th]
    \centering
    \includegraphics[width=.98\textwidth]{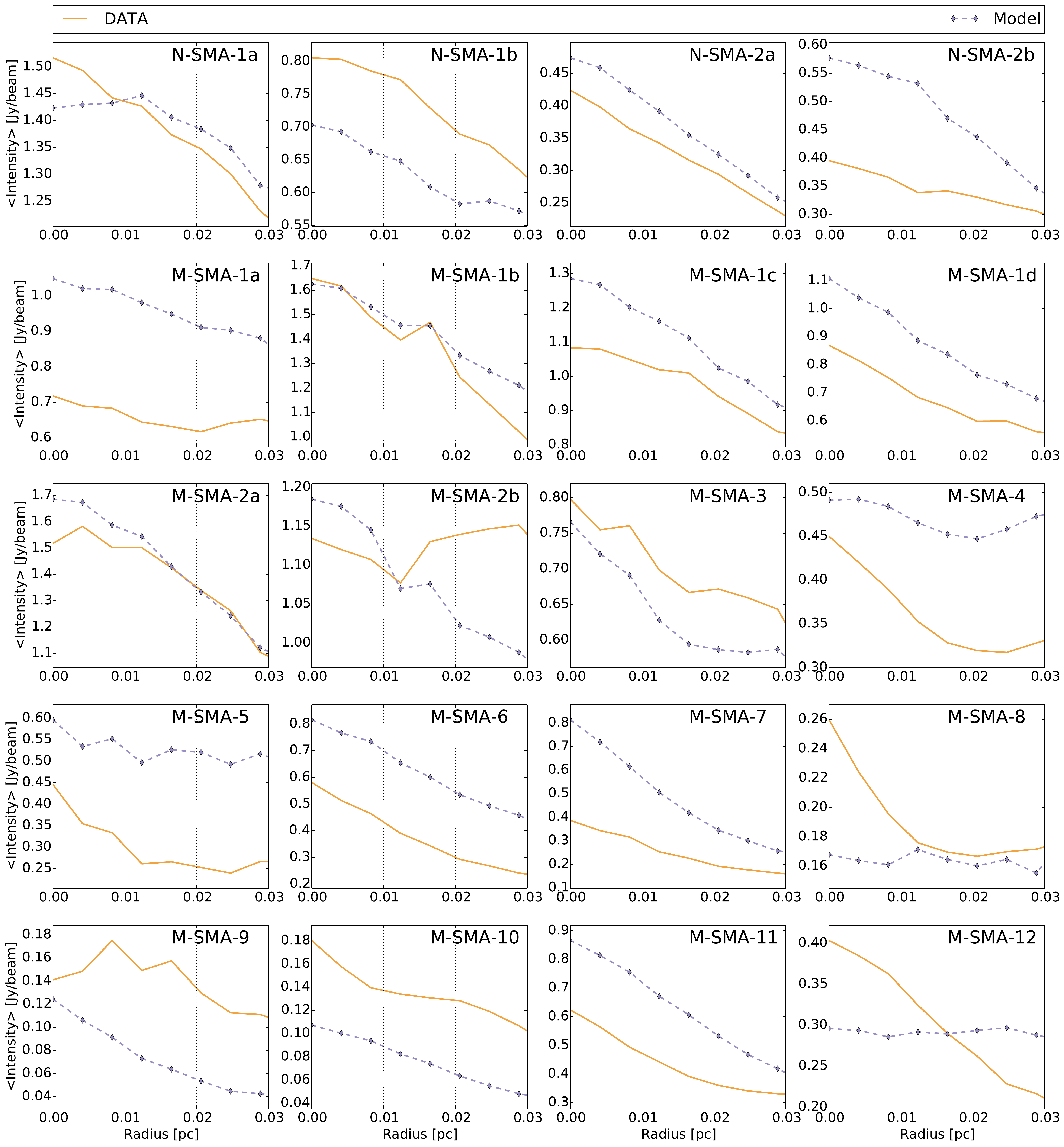}
    \caption{Radial profiles of the averaged intensity of the small-scale SMA structure around the position of the small-scale density components listed in Tab. \ref{tab:sgrb2_submm}. For each of the four different models (A -- D) described in this paper, the setup of the small-scale structure was fixed. The data is plotted in solid Orange, the model is shown in dashed blue. The identifier of the components are written in the upper right corner of each subplot.}\label{fig:radProfiles_smallScale}
\end{figure*}

\begin{figure*}[th]
    \centering
    \includegraphics[width=.9\textwidth]{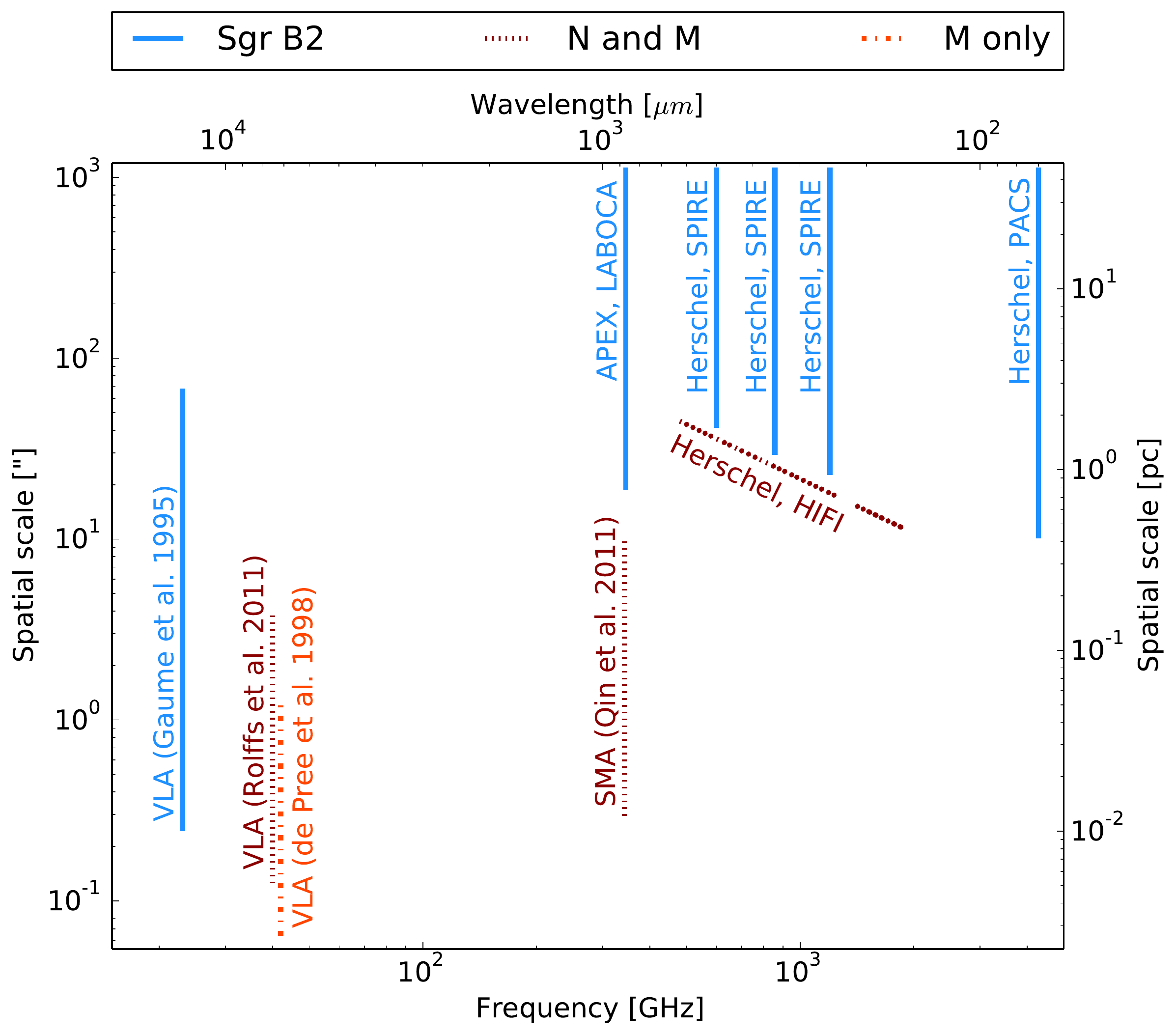}
    \caption{Covered spatial scales versus frequency of the different data employed in this study.}\label{fig:freqScales}
\end{figure*}
\clearpage

\onecolumn

\section {Tables}\label{s-tables}

\DTLloaddb{hiiDB}{tbl/sgrb2_hii.tab}

\begin{longtable}{llllrrrcrrrr}
    \caption{\label{tab:sgrb2_hii} Known \hii regions in the Sgr B2 region.\tablefootmark{a}}\\
   \hline
   \hline
    ID 						\tablefootmark{1} 	
    & Model  					\tablefootmark{2}
    & RA  	 				\tablefootmark{3}
    & DEC    					\tablefootmark{4}
    & dz    					\tablefootmark{5}
    & r$_\text{obs}$ 				\tablefootmark{6}
    & n$_e$	 				\tablefootmark{7}
    & ZAMS  					\tablefootmark{8}
    & EM 					\tablefootmark{9}
    & log($\mathrm{\dot N_\mathrm{i}}$) 	\tablefootmark{10} \Tstrut \\
	& & [h:m:s, J2000] & [d:m:s, J2000] & [10$^5$ au]  & [10$^3$ au] & [10$^4$ cm$^{-3}$] & type & [10$^7$ pc cm$^{-6}$] & [s$^{-1}$] \Bstrut \\
    \hline
    \endfirsthead
    \caption{(continued) Known \textsc{Hii} regions in Sgr B2.}\\
    \hline
    \hline
    ID  					\tablefootmark{1}	
    & Model 					\tablefootmark{2}
    & RA  	 				\tablefootmark{3}
    & DEC    					\tablefootmark{4}
    & dz    					\tablefootmark{5}
    & r$_\text{obs}$ 				\tablefootmark{6}
    & n$_e$	 				\tablefootmark{7}
    & ZAMS 					\tablefootmark{8}
    & EM 					\tablefootmark{9}
    & log($\mathrm{\dot N_\mathrm{i}}$) 	\tablefootmark{10} \Tstrut  \\
	& & [h:m:s, J2000] & [d:m:s, J2000] & [10$^5$ au]  & [10$^3$ au] & [10$^4$ cm$^{-3}$] & type & [10$^7$ pc cm$^{-6}$] & [s$^{-1}$] \Bstrut \\
    \hline
    \vspace{.3cm}
    \endhead
    \hline
    \endfoot
    
    \DTLforeach{hiiDB}{%
		\id=ID, \model=model, \ra=RA, \dec=DEC, \dz=dz, \ncen=ncen, \rin=rin, \zams=ZAMS, \em=EM, \nion=Nion}{%
		\\
		\id & \model & \ra & \dec & \dz & \rin & \ncen & \zams & \em & \nion }
\end{longtable}
\tablefoot{\tablefoottext{a}{All values listed in this table are the values used in the different models assuming \hii
			     regions are spherical symmetric regions of fully ionized gas with no dust and a single
			     ionizing source.}\\
	   \tablefoottext{b}{These regions are optically thick. Their electron density has been increased manually.}\\
	   \tablefoottext{1}{ID is the identifier used in the model. We use the same identifiers as \cite{Mehringer1993, Gaume1995, dePree1998}.}\\
	   \tablefoottext{2}{Model refers to one of the four model described in this paper.}\\
           \tablefoottext{3}{RA is the right ascension of the \hii region given in units of hours:minutes:seconds in the equatorial coordinate system.}\\	   
           \tablefoottext{4}{DEC is the declination of the \hii region given in units of degrees:arcminutes:arcseconds in the equatorial coordinate system.}\\
	   \tablefoottext{5}{dz is the displacement along the line of sight with respect to the model center. The z-axis is oriented such that it points towards the observer.}\\
	   \tablefoottext{6}{r$_\text{obs}$ is the observed radius of the \hii region.}\\
	   \tablefoottext{7}{n$_e$ is the number electron density.}\\
	   \tablefoottext{8}{ZAMS refers to the zero age main sequence star embedded in the \hii region.}\\
	   \tablefoottext{9}{EM is the emission measure.}\\
	   \tablefoottext{10}{log($\mathrm{\dot N_\mathrm{i}}$) is the logarithm of the number of Lyman continuum photons.}}  

\newpage
\DTLloaddb{plummerDB}{tbl/sgrb2_plummer.tab}

\begin{longtable}{llllrrrrr}
    \caption{\label{tab:sgrb2_submm} Small scale structure: Dust density components in Sgr B2.}\\
    \hline
    \hline
    ID 			\tablefootmark{1}
    & Model  		\tablefootmark{2} 	
    & RA 	 	\tablefootmark{3}
    & DEC  		\tablefootmark{4}
    & dz  		\tablefootmark{5}	
    & r$_0$  		\tablefootmark{6}
    & n$_\text{c}$ 	\tablefootmark{7}		 	
    & $\upeta$ 		\tablefootmark{8}
    & star 		\tablefootmark{9} \Tstrut \\
	& & [h:m:s, J2000] & [d:m:s, J2000] & [10$^5$ au] & [10$^3$ au] & [10$^7$ H$_2$ cm$^{-3}$] & & \Bstrut \\
    \hline
    \endfirsthead
    \caption{(continued) Dust density components in Sgr B2. Plummer density profile.}\\
    \hline\hline
    ID 			\tablefootmark{1}
    & Model 		\tablefootmark{2}
    & RA 		\tablefootmark{3}	 	
    & DEC 		\tablefootmark{4} 	
    & dz 		\tablefootmark{5} 		
    & r$_0$  		\tablefootmark{6}	
    & n$_\text{c}$ 	\tablefootmark{7}		 	
    & $\upeta$ 		\tablefootmark{8}
    & star 		\tablefootmark{9} \Tstrut \\		
	& & [h:m:s, J2000] & [d:m:s, J2000] & [10$^3$ au] & [10$^3$ au] & [10$^7$ H$_2$ cm$^{-3}$] & & \Bstrut \\    
    \hline
    \endhead
    \hline
    \endfoot
    
    \DTLforeach{plummerDB}{%
		\id=ID, \model=model, \ra=RA, \dec=DEC, \dz=dz, \nin=nin, \rin=rin, \exp=exp, \star=star}{%
		\\
		\id & \model &\ra & \dec & \dz & \rin & \nin & \exp & \star}
    
\end{longtable}
\tablefoot{\\
	   \tablefoottext{1}{ID is the identifier used in the model. These identifiers are identical to the ones introduced by \cite{Qin2011}. Note: Some of these objects identified by \cite{Qin2011} show an elongated intensity structure. We recover these objects with a superposition of several spherical symmetric clumps. We distinguish these components by adding additional lowercase letters to the identifier introduced by \cite{Qin2011}.}\\
	   \tablefoottext{2}{Model refers to one of the four model described in this paper.}\\
           \tablefoottext{3}{RA is the right ascension of the density component given in units of hours:minutes:seconds in the equatorial coordinate system.}\\	   
           \tablefoottext{4}{DEC is the declination of the density component given in units of degrees:arcminutes:arcseconds in the equatorial coordinate system.}\\
	   \tablefoottext{5}{dz is the displacement along the line of sight with respect to the model center. The z-axis is oriented such that it points towards the observer.}\\
	   \tablefoottext{6}{r$_\text{0}$ is the radius defining the component, as described in Eq. \ref{eq:plummer}.}\\
	   \tablefoottext{7}{n$_c$ is the central density.}\\
	   \tablefoottext{8}{$\upeta$ is the exponent of the dust density profile.}\\
	   \tablefoottext{9}{This column indicates whether an additional heating source had to be included inside the dust component. If this is the case, the spectral type of the star is given.}}

\newpage 
\DTLloaddb{powerDB}{tbl/sgrb2_envelope.tab}

\begin{longtable}{llllrrrrrrr}
    \caption{\label{tab:sgrb2_power} Large-scale structure: Dust density envelopes in Sgr B2.}\\
    \hline
    \hline
    ID		\tablefootmark{1} 
    & Model	\tablefootmark{2}
    & RA	\tablefootmark{3}
    & DEC	\tablefootmark{4}
    & dz 	\tablefootmark{5} 
    & r$_{0,x}$ \tablefootmark{6}
    & r$_{0,y}$ \tablefootmark{6}
    & r$_{0,z}$ \tablefootmark{6}
    & n$_0$ 	\tablefootmark{7}
    & $\upeta$ 	\tablefootmark{8}
    & star 	\tablefootmark{9} \Tstrut \\
	& & [h:m:s, J2000] & [d:m:s, J2000] & [10$^5$ au] & [10$^3$ au] & [10$^3$ au] & [10$^3$ au] & [10$^4$ H$_2$ cm$^{-3}$] & & \Bstrut \\
    \hline
    \endfirsthead
    
    \caption{(continued) Large-scale envelope in Sgr B2. Plummer-like density profile.}\\
    \hline
    \hline
    ID 		\tablefootmark{1} 
    & Model	\tablefootmark{2}
    & RA	\tablefootmark{3}
    & DEC 	\tablefootmark{4}
    & dz 	\tablefootmark{5} 
    & r$_{0,x}$	\tablefootmark{6}
    & r$_{0,y}$	\tablefootmark{6}
    & r$_{0,z}$	\tablefootmark{6}
    & n$_0$ 	\tablefootmark{7}
    & $\upeta$ 	\tablefootmark{8}
    & star 	\tablefootmark{9} \Tstrut \\
	& & [h:m:s, J2000] & [d:m:s, J2000] & [10$^5$ au] & [10$^3$ au] & [10$^3$ au] & [10$^3$ au] & [10$^4$ H$_2$ cm$^{-3}$] & & \Bstrut \\
    \hline
    \endhead
    \hline
    \endfoot
   
    \DTLforeach{powerDB}{%
		\id=ID, \model=model, \ra=RA, \dec=DEC, \dz=dz, \nin=nin, \rx=rx, \ry=ry, \rz=rz, \exp=exp, \star=star}{%
		\\
		\id & \model &\ra & \dec & \dz & \rx & \ry & \rz & \nin & \exp & \star}
		
\end{longtable}
    
\tablefoot{\\
	   \tablefoottext{1}{ID is the identifier used in the model. These identifiers follow the historic naming scheme explained in Sect. \ref{sec:intro}.}\\ 		  
	   \tablefoottext{2}{Model refers to one of the four model described in this paper.}\\
           \tablefoottext{3}{RA is the right ascension of the density component given in in units of hours:minutes:seconds in the equatorial system.}\\
	   \tablefoottext{4}{DEC is the declination. Both coordinates are given in units of degrees:arcminutes:arcseconds in the equatorial system.}\\
	   \tablefoottext{5}{dz is the displacement along the line of sight with respect to the model center. The z-axis is oriented such that it points towards the observer.}\\
	   \tablefoottext{6}{r$_x$, r$_y$, r$_z$ are the core radii in each principal direction, as described in Eq. \ref{eq:plummer}.}\\
	   \tablefoottext{7}{n$_c$ is the central density.}\\
	   \tablefoottext{8}{$\upeta$ is the exponent of the dust density profile.}\\
	   \tablefoottext{9}{This column indicates whether an additional heating source had to be included inside the dust component. If this is the case, the spectral type of the star is given.}}

\twocolumn
\section{Derivation of the physical parameters of \hii regions} \label{app:derivationHii}
 
\begin{figure}
  \centering
  \includegraphics[width=0.3\textwidth]{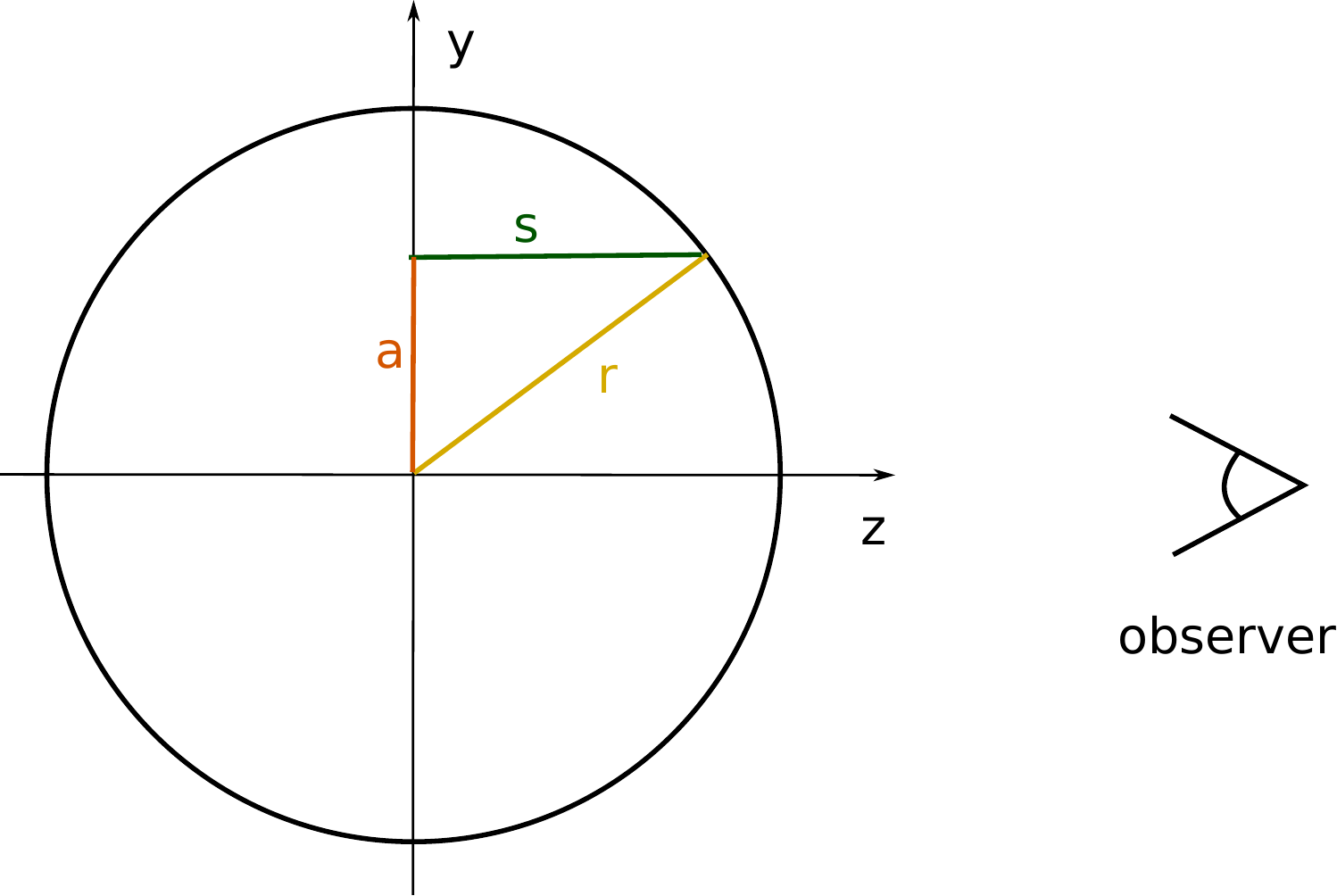}
  \caption{Sketch of the \hii region. Cut along the line of sight. The integration path along-the line of sight for the calculation of the emission measure is marked.}\label{fig:hii_EM_integration}
\end{figure}

We consider \hii regions as Str\"omgren spheres. They are fully ionized and contain no dust. Here we derive the emission measure, the electron density, and the total flux of ionizing photons for such a region. We assume an electron density distribution that equals $n_e$ within the radius $r_0$ of the \hii region and is zero elsewhere. We can then calculate the emission measure $EM$ by integrating the number of electrons and number of ions along the line-of-sight. The factor  accounts for both sides of the sphere.
\begin{equation}\label{eq:EM}
    EM = 2 \, \int_0^{z'} n_e n_i dz
\end{equation}
If we assume that the \hii regions only contain hydrogen, it follows that the electron density $n_e$ equals the ion density $n_i$. Since the electron density $n_e$ is uniform within the \hii region, we can rewrite Eq.\ \ref{eq:EM} taking the geometry given in Fig. \ref{fig:hii_EM_integration} into account.
\begin{align}
    EM 	 = {} & 2 \, n_e^2 \, \sqrt{r_0^2 - a^2}\\
	 = {} & 2 \, n_e^2 \, D \, \theta_0 \, \sqrt{1 - \left(\frac{\theta_a}{\theta_0}\right)^2}\\
	 = {} & 2 \, n_e^2 \, D \, \theta_0 \, \psi(\theta_a)
\end{align}
where $D$ is the distance to the source, $\uptheta_0 = r_0/D$ is the radius of the \hii region in angular units, and $\theta_a = \sqrt{x^2+y^2}/D$ is the angular position in the x-y plane ($z=0$). If we want to calculate the maximum emission measure, we set $a=0$ and the term $\sqrt{1 - \left(\frac{\theta_a}{\theta_0}\right)^2}$ vanishes. Introducing the angular source diameter $\theta_\text{source} = 2 \theta_0$, we obtain
\begin{equation}\label{eq:EM_intermediate}
    EM = n_e^2 \, D \, \theta_\text{source}
\end{equation}
However, neither the emission measure nor the electron density are known. So we need another equation to solve Eq. \ref{eq:EM_intermediate}. Integrating the specific intensity $I_\nu$ over the angular size of the source $\Omega_\text{source}$ yields the flux density $F_\nu$
\begin{equation}
    F_\nu = \int_{\Omega_\text{source}} I_\nu d\Omega
\end{equation}
The general solution of the radiative transfer equation for the specific continuum intensity is e.g. given by \citet{Rybicki1986} as
\begin{align}
    I_\nu = {} & I_0 \, e^{-\tau_\nu} + \int_0^{\tau_\nu} S_\nu \, e^{-(\tau_\nu-\tau_\nu')}  	
		  d\tau'\nonumber\\
	  = {} & I_0 \, e^{-\tau_\nu} + S_\nu \left(1 - e^{-\tau_\nu}\right)
\end{align}
where $S_\nu$ is the source function of the \hii region and $I_0$ is the background intensity. We assume an \hii region with uniform temperature and negligible background temperature, i.e. $I_\nu(0) \ll S_\nu$. The intensity $I_\nu$ is then given by 
\begin{equation}
    I_\nu = S_\nu \left(1 - e^{-\tau_\nu}\right)
\end{equation}
We have to distinguish two cases for the optical depth $\tau_\nu$:
\begin{equation}
  I_\nu = 
  \begin{cases}
   S_\nu \tau_\nu &\mbox{if } \tau_\nu \ll 1 \quad \text{(optical thin)}\\
   S_\nu &\mbox{if } \tau_\nu \gg 1 \quad \text{(optical thick)}
  \end{cases}
\end{equation}
The source function $S_\nu$ for the free-free radiation can be defined using the Planck function $B_\nu$ at an electron temperature $T_e$
\begin{align}
    S_\nu = {}       & B_\nu(T_e) \nonumber \\
	  \approx {} & \frac{2kT_e\nu^2}{c^2}
\end{align}
where $k$ is the Boltzman constant, $c$ is the speed of light in vacuum and $h$ is the Planck constant. The second step holds if $h\nu\ll kT$ (Rayleigh-Jeans approximation). The optical path length for free-free emission was derived by \citet{Oster1961}, an approximation is given by \citet{Altenhoff1960}.
\begin{equation}
    \tau_\text{Altenhoff} = 8.235 \times 10^{-2} \left(\frac{T_e}{\text{K}}\right)^{-1.35} \left(\frac{\nu}{\text{GHz}}\right)^{-2.1} \left(\frac{\text{EM}}{\text{pc cm}^{-6}}\right)
\end{equation}
This approximation deviates in the region of interest, i.e. $5\times 10^3 \le T_e \le 1.2\times 10^4\, \text{K}$ and $100 \text{MHz} \le \nu \le 35\, \text{GHz}$, less than 10\% \citep{Mezger1967}. 

Putting everything together, and assuming optical thin emission, we obtain
\begin{equation}
   \left(\frac{F_\nu}{Jy}\right)
	    = 2.525 \times 10^3 
	      \left(\frac{T_e}{\text{K}}\right)^{-0.35} 
	      \left(\frac{\nu}{\text{GHz}}\right)^{-0.1} 
	      \left(\frac{\text{EM}}{\text{pc cm}^{-6}}\right) 
	      \Omega_\text{source}
\end{equation}
Usually one fits a Gaussian to the data and extracts the flux density and the full width at half maximum from this fit. However, we determine the total flux in a given circular aperture. We thus calculate the solid angle for the given density distribution as follows.
\begin{align}
    \Omega_\text{source} 
    = {} & 2\pi \, \int_0^{\theta_0} \theta_a \psi(\theta_a) \, d\theta_a \nonumber \\
    = {} & \frac{\pi}{6} \theta_\text{source}^2 \nonumber \\
    = {} & 1.231 \times 10^{-11} \left(\frac{\theta_\text{source}}{\text{arcsec}}\right)^2
\end{align}
Now we obtain the following expression for the emission measure for a circular aperture
\begin{equation}
    \left(\frac{\text{EM}}{\text{pc cm}^{-6}}\right) 
    = 3.217 \times 10^{7} 
	\left(\frac{F_\nu}{\text{Jy}}\right)  
	\left(\frac{T_e}{\text{K}}\right)^{0.35} 
	\left(\frac{\nu}{\text{GHz}}\right)^{0.1}
	\left(\frac{\theta_\text{source}}{\text{arcsec}}\right)^{-2}
\end{equation}
From that we can derive an expression to calculate the electron density

\begin{align}\label{eq:electronDensity_appendix}
     \left(\frac{n_e}{\text{cm}^{-3}}\right)
     = {} & 2.576 \,\times\, 10^{6} \, 
	    \left(\frac{F_\nu}{\text{Jy}}\right)^{0.5}  
	    \left(\frac{T_e}{\text{K}}\right)^{0.175} 
	    \left(\frac{\nu}{\text{GHz}}\right)^{0.05} \nonumber \\
       & \times \left(\frac{\theta_\text{source}}{\text{arcsec}}\right)^{-1.5}
	    \left(\frac{D}{\text{pc}}\right)^{-0.5} 
\end{align}

To derive the flux of ionizing photons, $\dot N_\mathrm{i}$, we balance the number of recombinations and photoionizations within the \hii region. In a Str\"omgren sphere, this yields
\begin{equation}\label{eq:flux_ionPhotons}
  \dot N_\mathrm{i} = \int n_e^2 (\beta - \beta_1) dV
\end{equation}
where $\beta$ and $\beta_1$ are the rate coefficients for recombinations to all levels and to the ground state, respectively. Thus $(\beta - \beta_1)$ provides the recombination coefficient to level 2 or higher. \citet{Rubin1968} approximate the recombination coefficient given by \citet{Seaton1959} for electron temperatures $T_e$ generally found in \hii regions as
\begin{equation}\label{eq:recombinationCoefficient}
  \left(\frac{\beta - \beta_1}{\text{cm}^3\text{ s}^{-1}}\right) = 4.1 \times 10^{-10} \left(\frac{T_e}{\text{K}}\right)^{-0.8}
\end{equation}
Including Eq. \ref{eq:recombinationCoefficient} in Eq. \ref{eq:flux_ionPhotons} and solving the integral for a spherical symmetric clump yields:
\begin{equation}
    \left(\frac{\dot N_\mathrm{i}}{\text{s}^{-1}}\right)
    = \frac{4}{3}\pi
      \left(\frac{r_0}{\text{cm}}\right)^3 
      \left(\frac{n_e}{\text{cm}^{-3}}\right)^2 \times
       4.1 \times 10^{-10} \left(\frac{T_e}{\text{K}}\right)^{-0.8}
\end{equation}
Converting the radius $r_0$ to the angular diameter $\theta_\text{source}$ of the \hii region and using Eq. \ref{eq:electronDensity} yields an expression for the total flux of ionizing photons, given in practical units as
\begin{equation}
  \left(\frac{\dot N_\mathrm{i}}{\text{s}^{-1}}\right)
  = 4.771 \times 10^{42} \times
	    \left(\frac{F_\nu}{\text{Jy}}\right)  
	    \left(\frac{T_e}{\text{K}}\right)^{-0.45} 
	    \left(\frac{\nu}{\text{GHz}}\right)^{0.1}
	    \left(\frac{D}{\text{pc}}\right)^2
\end{equation}

\end{appendix}
\end{document}